\documentclass[sigconf, nonacm]{acmart}
\usepackage{subcaption}
\usepackage{listings}
\usepackage{xcolor}
\usepackage{microtype}
\usepackage{tcolorbox}
\usepackage{bo-macros}
\usepackage[linesnumbered,ruled,vlined]{algorithm2e}

\usepackage{mathabx}
\usepackage{booktabs}
\usepackage{balance}
\usepackage{paralist}
\usepackage{adjustbox}

\newcommand\vldbdoi{XX.XX/XXX.XX}
\newcommand\vldbpages{XXX-XXX}
\newcommand\vldbvolume{14}
\newcommand\vldbissue{1}
\newcommand\vldbyear{2020}
\newcommand\vldbauthors{\authors}
\newcommand\vldbtitle{\shorttitle} 
\newcommand\vldbavailabilityurl{URL_TO_YOUR_ARTIFACTS}
\newcommand\vldbpagestyle{plain} 

\newtheorem{problem}{Problem}

\makeatletter
\newcommand{\removelatexerror}{\let\@latex@error\@gobble}
\makeatother

\definecolor{mydarkred}{RGB}{200,0,0}

\SetCommentSty{mycommfont}
\definecolor{darkolivegreen}{rgb}{0.33, 0.42, 0.18}
\definecolor{cornellred}{rgb}{0.7, 0.11, 0.11}

\newcommand{\sys}{\textsc{Sharp}\xspace}
\newcommand{\mreg}{\texttt{\small MATCH\_RECOGNIZE}\xspace}

\begin{document}


\title{\sys: \underline{Shar}ed State Reduction for Efficient Matching of
Sequential \underline{P}atterns (Technical Report)}

\author{Cong Yu}
\authornote{Both authors have contributed equally.}
\affiliation{%
	\institution{Aalto University}
}
\email{cong.yu@aalto.fi}

\author{Tuo Shi}
\authornotemark[1]
\affiliation{%
	\institution{Aalto University}
}
\email{tuo.shi@aalto.fi}

\author{Matthias Weidlich}
\affiliation{%
	\institution{Humboldt-Universität zu Berlin}
}
\email{matthias.weidlich@hu-berlin.de}

\author{Bo Zhao}
\affiliation{%
	\institution{Aalto University}
}
\email{bo.zhao@aalto.fi}



\begin{abstract}
The detection of sequential patterns in data is a basic functionality of modern
data processing systems for complex event processing
(CEP), OLAP, and retrieval-augmented generation (RAG). 
To improve the results quality for downstream applications, pattern matching engines employ multiple shared patterns that collaboratively deliver more insights.
In practice, pattern
matching is
challenging, since common applications rely on a large set of patterns that shall
be evaluated with tight latency bounds. At the same time, matching needs to
maintain \emph{state}, \ie intermediate results, that grow exponentially in the input
size. 
Hence, systems turn to best-effort processing, striving for
maximal recall under a latency bound. Existing techniques, however, consider 
patterns in isolation, neglecting the optimization potential induced by state
sharing and corresponding interactions and interference across shared patterns. \looseness=-1

We describe \sys, a state management library that employs \emph{state reduction} for
efficient best-effort pattern matching in shared patterns.
To this end, \sys incorporates state sharing between patterns through a new
abstraction, coined pattern-sharing degree (PSD). At runtime, PSD
facilitates the categorization and indexing of partial pattern matches.
Once a latency bound is exceeded, \sys realizes best-effort
processing by using a cost model to select a subset of partial matches for further processing in
constant time.
In experiments with real-world data, \sys achieves a recall of
95\%, 93\% and 72\% for pattern matching in CEP, OLAP, and RAG applications,
under a bound of 50\% of the average processing latency.
\looseness=-1
\end{abstract}

\settopmatter{printfolios=true}
\maketitle
\pagestyle{\vldbpagestyle}
\begingroup\small\noindent\raggedright\textbf{PVLDB Reference Format:}\\
\vldbauthors. \vldbtitle. PVLDB, \vldbvolume(\vldbissue): \vldbpages, \vldbyear.
\href{https://doi.org/\vldbdoi}{doi:\vldbdoi}
\endgroup
\begingroup
\renewcommand\thefootnote{}\footnote{\noindent
This work is licensed under the Creative Commons BY-NC-ND 4.0 International License. Visit \url{https://creativecommons.org/licenses/by-nc-nd/4.0/} to view a copy of this license. For any use beyond those covered by this license, obtain permission by emailing \href{mailto:info@vldb.org}{info@vldb.org}. Copyright is held by the owner/author(s). Publication rights licensed to the VLDB Endowment. \\
\raggedright Proceedings of the VLDB Endowment, Vol. \vldbvolume, No. \vldbissue\ %
ISSN 2150-8097. \\
\href{https://doi.org/\vldbdoi}{doi:\vldbdoi} \\
}\addtocounter{footnote}{-1}\endgroup


\lstdefinelanguage{Cypher}{
    morekeywords={MATCH,RETURN,CREATE,WHERE,DELETE,DETACH,SET,MERGE,WITH,UNION,OPTIONAL, AS},
    sensitive=true, 
    morecomment=[l]{//},
    morestring=[b]' 
}

\lstset{
    keywordstyle=\color{blue}\bfseries, 
    commentstyle=\color{gray}\itshape, 
    stringstyle=\color{red}, 
    numbers=none,
    breaklines=true, 
    showstringspaces=false,
    basicstyle=\ttfamily\small, 
    frame=single 
}

\section{Introduction}\label{sec:intro}

The detection of sequential patterns with low latency is a data management
functionality with a wide range of applications:
Complex event
processing (CEP) engines detect user-defined patterns over high-velocity event
streams~\cite{cugola2012processing,10.1145/3654935,10.1145/3709682};
online analytical processing (OLAP) systems evaluate queries featuring the
\mreg operator that
takes a set of tuples as input and returns all matches of the given
pattern~\cite{iso19075-5,korber2021index,zhu2023high}; and graph databases evaluate patterns of regular path queries over knowledge graphs to facilitate
retrieval-augmented generation (RAG)~\cite{abiteboul1997regular,
angles2017foundations,pacaci2020regular}. In all these applications, patterns
define the order of data elements along with the correlation and
aggregation predicates over their attribute values.

Pattern matching is challenging, though: applications enforce strict
latency bounds~\cite{zhao2020load,
korber2021index, WadhwaPRBB19} as part of the service level
objectives (SLO)~\cite{liu2020resource}.
At the same time, the
evaluation is computationally hard, since it requires maintaining
\emph{state}, \ie partially matched patterns, which grow exponentially in
the size of the
input~\cite{zhang2014complexity, huang2023t, libkin2012regular,
pacaci2020regular}. Due to these challenges, exhaustive pattern evaluation
becomes infeasible, especially in short peak times of increased
computational load~\cite{tatbul2003load,he2013load}.
Systems therefore resort to best-effort
processing: they strive to maximize the number of
detected pattern matches, while satisfying a latency
bound~\cite{zhao2020load,chapnik2021darling, slo2023gspice, slo2019pspice,
hSPICE}.

In practice, the above mentioned challenges are amplified by the fact that
many applications require
the simultaneous evaluation of multiple \emph{shared patterns}. The reason
being that the result quality of many downstream
applications~\cite{DBLP:conf/iclr/LuoLHP24,kim-etal-2023-kg,10.1145/3709682,DBLP:journals/pvldb/GeorgeCW16}
 can directly be improved by evaluating a set of similar, yet different
patterns. We illustrate how shared patterns improve the quality of results
and the underlying performance challenges with an example of graph
retrieval-augmented generation (GraphRAG), as follows.

\begin{figure*}[t]
  \centering
  \begin{minipage}{0.69\textwidth}
    \begin{subfigure}{\linewidth}
      \centering
      \includegraphics[width=\linewidth]{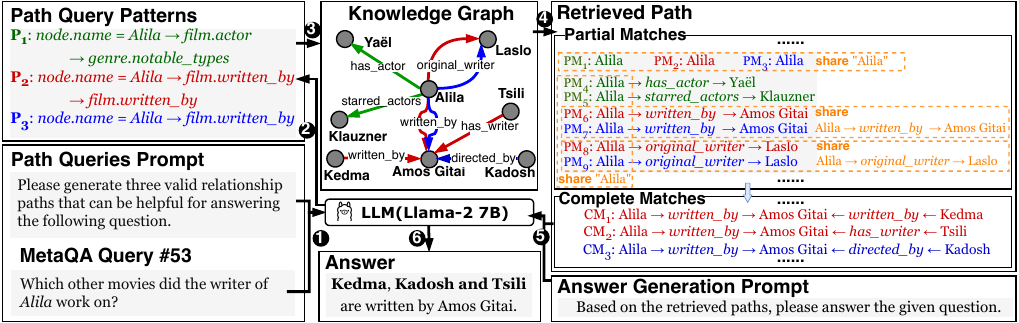}
      \caption{Workflow of a GraphRAG pipeline}
      \label{fig:basic_flow}
    \end{subfigure}
  \end{minipage}
  \begin{minipage}{0.29\textwidth}
    \begin{minipage}{\linewidth}
      \begin{subfigure}{0.49\linewidth}
        \centering
        \includegraphics[width=0.85\linewidth, height=1.5cm]{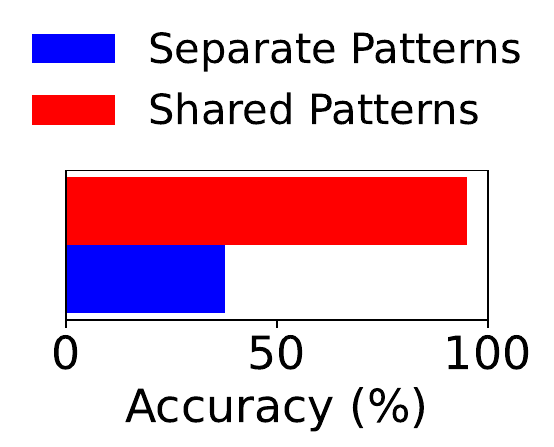}
        \caption{Result improvement}
        \label{fig:exm_acc_recall}
      \end{subfigure}%
      \begin{subfigure}{0.49\linewidth}
        \centering
        \includegraphics[width=0.85\linewidth, height=1.5cm]{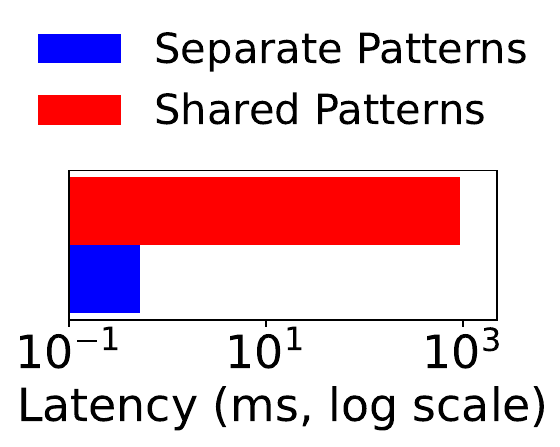}
        \caption{End-to-end latency}
        \label{fig:exm_lat}
      \end{subfigure}
    \end{minipage}

    \vspace{1.ex}

    \begin{minipage}{\linewidth}
      \begin{subfigure}{\linewidth}
        \centering
        \includegraphics[width=0.9\linewidth, height=1.7cm]{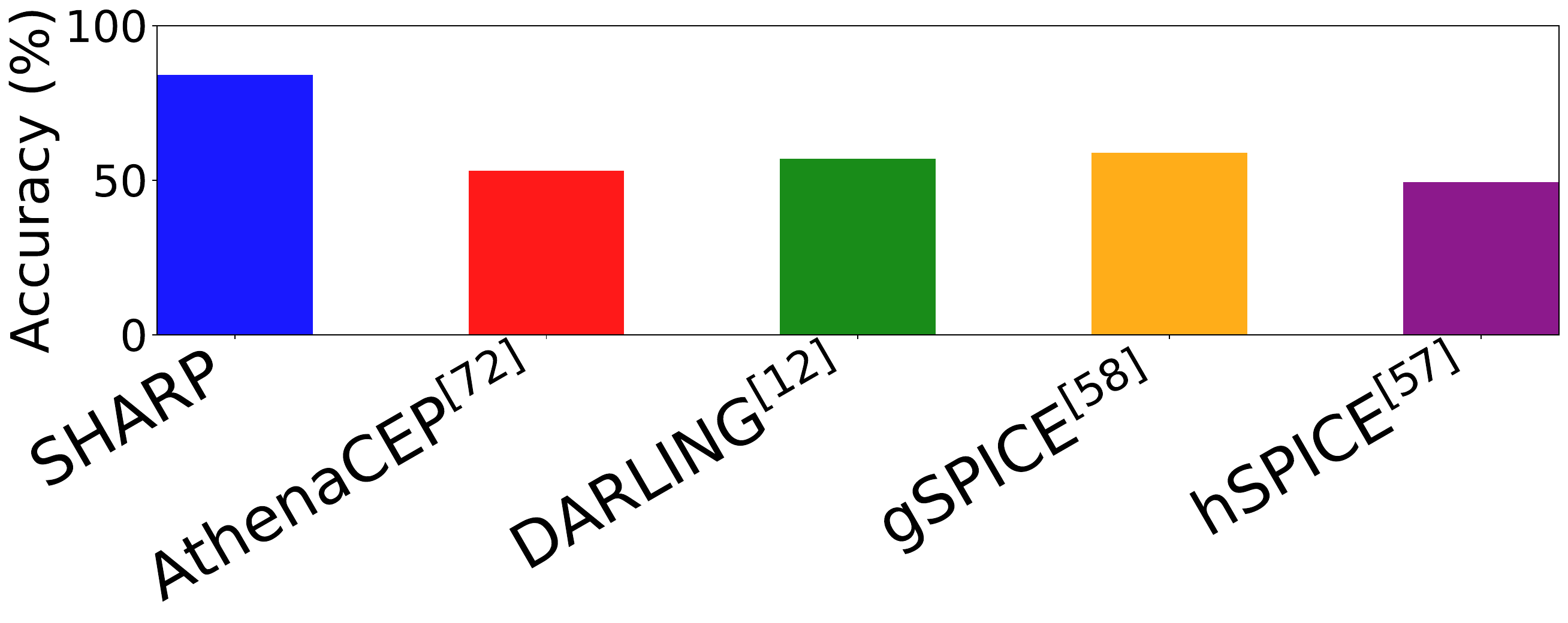}
        \caption{End-to-end accuracy of \sys and SOTA  at 300\unit{ms} latency bound of KG pattern matching}
        \label{fig:sota}
      \end{subfigure}
    \end{minipage}
  \end{minipage}

  \caption{ An example of how shared patterns enhance the end-to-end responses of GraphRAG and the underlying performance challenge}
  \label{fig:graphrag_full}
  \vspace{-1em}
\end{figure*}

 Consider an application in which the inference of a large language model (LLM) is augmented using an external knowledge graph~(KG), as
 shown in \F\ref{fig:basic_flow}: \myc{1} The application submits (i)
 a question for the LLM (from the MetaQA benchmark~\cite{Zhang_Dai_Kozareva_Smola_Song_2018}) and (ii) a query prompt to generate patterns of path queries for the KG.
 \myc{2} The patterns and prompts are sent to the LLM (fine-tuned on \code{WebQSP}~\cite{yih-etal-2016-value} and \code{CWQ}~\cite{Zhang_Dai_Kozareva_Smola_Song_2018}).
 The LLM then generates a set of path query patterns and selects the top-k
 ({\color{darkolivegreen}$P_1$}, {\color{cornellred}$P_2$} and
 {\color{blue}$P_3$}),
 e.g., using cosine similarity and beam-search~\cite{tan2025pathsovergraph}.
 \myc{3} The path queries are evaluated on the KG and \myc{4} the resulting
 patterns are \myc{5} integrated in an {answer generation prompt} for \myc{6} the final response.
 Here,  {\color{darkolivegreen}$P_1$}, {\color{cornellred}$P_2$} and {\color{blue}$P_3$} share sub-patterns and one snapshot of shared state is illustrated as \emph{partial matches} \code{$\left\{ PM_{1}, PM_{2}, \dots, PM_{9}, \dots \right\}$} in  \myc{4}.

Compared to the use of single, separate patterns, shared patterns improve
the result quality by 2.80$\times$ in accuracy and 2.51$\times$ in recall\footnote{Recall is the ratio of correct LLM responses to the number of ground-truth answers.} (over 14,872 questions
in MetaQA), see \F\ref{fig:exm_acc_recall}. 
For one question in \myc{1}, query \#53, the LLM only generates incorrect
answers (\ie LLM hallucination) if not using the shared patterns. Yet, the
improvement incurs a cost: Pattern matching latency in the KG \myc{4}
increases by \emph{four
orders of
	magnitudes} (
see \F\ref{fig:exm_lat})\footnote{Here, the
end-to-end processing latency comprises LLM generation
(325\unit{ms}) and KG pattern matching (1,079\unit{ms}); the
latter being the performance bottleneck.} compared to the use
of a single pattern, as 3,962$\times$ more partial matches are generated.
Even when adopting
optimizations for state
sharing~\cite{abul2016swarmguide,abul2017multiple}, the number of generated
partial matches still increases by 2,001$\times$. In
large-scale industrial applications, such as those reported for GraphRAG by
Microsoft~\cite{MicrosoftGraphRAG,
	MicrosoftPrivateData2024}, Amazon~\cite{AWSGraphRAG2024}, and
Siemens~\cite{SiemensKnowledgeGraph}, such computational challenges are
further amplified. \looseness=-1

Existing approaches for best-effort pattern matching, however, are limited
in their effectiveness as they treat each pattern in isolation.
Specifically, load shedding techniques~\cite{zhao2020load,
chapnik2021darling, slo2019pspice, hSPICE, slo2023gspice} that
discard input data (\code{DARLING}~\cite{chapnik2021darling},
\code{hSPICE}~\cite{hSPICE} and \code{gSPICE}~\cite{slo2023gspice}), partial
matches (\code{pSPICE}~\cite{slo2019pspice} and
\cite{DBLP:conf/icde/Zhao18}), or a combination of both
(\code{AthenaCEP}~\cite{zhao2020load}) take shedding decisions for each
pattern separately. While systems such as
\code{pSPICE}~\cite{slo2019pspice}, \code{hSPICE}~\cite{hSPICE} and
\code{gSPICE}~\cite{slo2023gspice} incorporate operators that are part of
multiple patterns, their utility assessment is done for each single pattern
in isolation. By ignoring the interaction among multiple patterns via their
shared state, these techniques neglect a significant optimization
potential. \looseness=-1

In this paper, we study how to realize best-effort
processing of pattern workloads, when incorporating state sharing in their
evaluation. This is difficult as the state (\ie partial matches) of pattern
matching affects the results and performance as follows:

\noindent (1) Partial matches differ in how they contribute to the result of a single pattern and in their computational
resources (\ie latency).

\noindent (2) Partial matches differ in their importance for shared multiple patterns, which may be captured further by a cost model.

\noindent (3) The relation between partial matches and patterns is subject to changes,
e.g., due to concept drift in data distributions,
and therefore, requires efficient indexing mechanisms to track this.

We describe \textbf{\sys}\footnote{publicly available at {\color{blue} \url{https://github.com/benyucong/SHARP}}}, a state management library for efficient
best-effort pattern matching with shared state. It addresses the above
challenges and overcomes the limitations of state-of-the-art approaches.
\F\ref{fig:sota} highlights that, under a latency bound of 300\unit{ms} (in KG pattern matching), the
state of the art achieves only an accuracy below $60\%$ for the
aforementioned pattern matching scenario.
In contrast, \sys incorporates optimizations for the shared state across
multiple patterns, boosting accuracy to around $85\%$.  These improvements
are facilitated by the following contributions:

\mypar{(1) Efficient pattern-sharing assessment (\S\ref{sec:clustering})}
\sys captures state sharing per partial match using a
new abstraction called \emph{pattern-sharing degree}~(PSD). PSD keeps track
of how different patterns share a partial match in terms of overlapping
sub-patterns and enables efficient lookup of this information for an
exponentially growing set of partial matches. The structure of the PSD is
derived from the pattern execution plan. At runtime, \sys relies on this
structure to cluster the generated partial matches and,
through a bitmap-based index, enables their retrieval in constant time. \looseness=-1

\tinyskip
\mypar{(2) Efficient cost model for state selection (\S\ref{sec:cost_model})} For each partial
match, \sys examines its contribution to all patterns and its computational
overhead, which determines the processing latency.
To achieve this, \sys maintains a cost model to estimate the number of
complete matches that each partial match may generate, as well as the runtime
and memory footprint caused by it.
\sys updates the cost model incrementally and facilitates a lookup of the
contribution and the overhead per partial match in constant time. \looseness = -1

\tinyskip
\mypar{(3) State reduction problem formulation (\S\ref{sec:problem}) and its
optimization (\S\ref{sec:shedding})}
We formulate the problem of satisfying latency bounds in pattern evaluation
as a \emph{state reduction} problem: Upon exhausting a latency bound,  \sys selects
a set of partial matches for further processing based on a multi-objective
optimization problem. \sys limits the overhead of solving this problem by a \emph{hierarchical selection}
lifting the pattern-sharing degree and the cost model to a coarse granularity by clustering, and by employing a greedy approximation strategy
for the multi-objective optimization space.

\tinyskip
We have implemented
\textbf{\sys} in C++
and Python with 3,500 LoC.
\sys has been evaluated (\S\ref{sec:experiments}) for three applications that rely on pattern
matching, i.e., complex event processing (CEP) over event streams, online
analytical processing (OLAP) with \mreg queries, and GraphRAG.
In a comprehensive experimental evaluation with real-world data, we observe
that \sys achieves a recall of
95\%, 93\%, and 72\% for shared-pattern matching in CEP, OLAP, and GraphRAG
applications, when enforcing a bound of 50\% of the average processing
latency. Compared to the existing state management strategies, \sys
significantly improves recall by $11.25 \times$ for CEP, $2.4\times$ for
OLAP, and $2.1\times$ for GraphRAG.

\section{Foundations of Pattern Matching}
\label{sec:background}
	\begin{figure*}[t]
		\centering
		\includegraphics[width =0.9\linewidth]{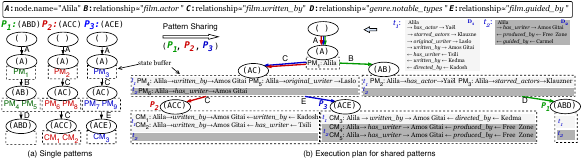}
		\caption{The execution plan DAG$^b$ of (a) separate single patterns and (b) multiple
			shared patterns } 
		\label{fig:pattern-graph}
		\vspace{-1.5em} 
	\end{figure*}

\subsection{Data and Execution Models}
\label{sec:model}
 
\myparr{Input data} of pattern matching is a sequence of data elements \code{$S=\langle d_1, d_2, \dots \rangle$}.
Each
\code{$d_i=\langle a_1, \dots, a_m
\rangle$}
 is an instance of a  schema
 defined by a finite sequence of
attributes.
We further define a prefix of the data
sequence $S$ up to index $k$,
\code{$S(\dots k) = \langle d_1, \dots, d_k \rangle$}
 and the suffix starting $k$ as
\code{$S(k \dots) = \langle d_k, d_{k+1}, \dots \rangle$}.

\myparr{Pattern} \code{$P$} defines a sequence of
data
that satisfy a set of predicates \code{$C$} (a time window, the sequential order, and value constraints over attributes).
For instance, {\color{darkolivegreen}$P_1$} in \F\ref{fig:basic_flow} \myc{3} defines the path sequence of three nodes in KG and the predicates, \eg \code{node.name=Alila}.
Pattern matching evaluates a
set of patterns \code{$\mathbf{P} = \{P_1, \dots, P_n\}$} over
\code{$S$},
ensuring \code{$P_i$} is detected within its
latency bound \code{$L_i$}
(based on SLO).

\myparr{Complete matches} are created by
evaluating \code{$P$} over
\code{$S$=$\langle d_1, d_2, \dots \rangle$}, each being a subsequence
\code{$\langle d'_1, \dots, d'_m \rangle$} of \code{$S$} that preserves the sequential order, \ie for \code{$d'_i = d_k$} and \code{$d'_j = d_l$}, it holds that \code{$i < j$} implies \code{$k < l$}, and satisfies all predicates in \code{$C$}.
We denote complete matches of \code{$P_i$} over \code{$S(\dots k)$} as \code{$\texttt{CM}_{P_i}(k)$}, and the complete matches for all patterns as \code{$\texttt{CM}(k) = \bigcup_{i} \texttt{CM}_{P_i}(k)$}.
\F\ref{fig:basic_flow} \myc{5} shows a snapshot of complete matches for {\color{cornellred} $P_2$}  (\code{$CM_{1}$} and \code{$CM_2$}) and {\color{blue} $P_3$} (\code{$CM_3$}).

\myparr{Partial matches} (PM) are sub-sequences of complete matches.
Their sequences of data elements strictly satisfy the time window and sequential order, but only satisfy a subset of predicates in \code{$C$}.
We write \code{$\texttt{PM}(k)$ = $\{\langle d_1, \dots , d_j \rangle, \dots, \langle d'_1, \dots, d'_l
\rangle \}$} for the set of PMs after evaluating patterns \code{$\mathbf{P}$} over \code{$S(\dots k)$}. \F\ref{fig:basic_flow} \myc{4} shows PMs for all patterns \code{$PM_{1-9}$} in corresponding colors.

\myparr{Execution plan} of a pattern \code{$P_i$} is represented as a directed acyclic graph with buffers (DAG$^b$), \code{$G_i$}, where the path from the \emph{starting vertex} to an \emph{ending vertex} constructs complete matches. \F\ref{fig:pattern-graph}a shows bespoke execution plans for {\color{darkolivegreen}$P_1$}, {\color{cornellred}$P_2$} and {\color{blue}$P_3$}.
Each node of \code{$G_i$} represents a sub-pattern \code{$SP$} of \code{$P_i$} and maintains a buffer to store the intermediate results, partial matches, in \code{$SP$}.
The edge \code{$(SP, SP')$} represents a transition between sub-patterns and is guarded by \code{$P_i$}'s predicates \code{$C_i$}.
When processing an input \code{$d_{k+1}$}, the pattern matching engine checks if \code{$d_{k+1}$} and all the partial matches stored in  \code{$SP$}, satisfy predicates  \code{$C_i$}. Then, it activates the corresponding state transitions and generates new partial matches and complete matches.   \looseness=-1

The DAG$^b$ model is general enough to capture bespoke automata- or tree-based execution models~\cite{wu2006high,mei2009zstream}
 for different applications (see \S\ref{sec:need4pattern}).
Formally, pattern execution is a \emph{function} that takes an  element \code{$d_{k+1}$}, the current
PMs \code{$\texttt{PM}(k)$},
and the execution plan \code{$G$} as input, and outputs
new PMs, \code{$\texttt{PM}(k+1)$} and complete matches
\code{$\texttt{CM}(k+1)$}, ensuring the latency below predefined bounds for all patterns: \looseness=-1
\tinyskip
\code{
	$f(d_{k+1}, \texttt{\footnotesize PM}(k), G)  \mapsto \{\texttt{\footnotesize PM}(k+1), \texttt{\footnotesize CM}(k+1)\} \textit{\footnotesize s.t.}\forall P_i \in \mathbf{P},\texttt{\footnotesize Latency}(P_i) \leq L_i$
}

\myparr{Pattern sharing} merges pattern execution plans to reuse overlapping
sub-patterns. As shown in \F\ref{fig:pattern-graph}b,
{\color{darkolivegreen}$P_1$}, {\color{cornellred}$P_2$} and
{\color{blue}$P_3$} share common sub-patterns \code{(A)}, while
{\color{cornellred}$P_2$} and {\color{blue}$P_3$} also share
sub-pattern \code{(AC)}.
The evaluation is similar to single patterns,
by checking state transition predicates between maintained PMs and the input data.
\code{(AC)} maintains two partial matches, \ie \code{$PM_4$} and \code{$PM_5$}, at time $t_1$.
When processing the two input data elements in $D_1$, \ie ``$\leftarrow$
\code{written\_by} $\leftarrow$ \code{Kedma}'' and ``$\leftarrow$
\code{has\_writer} $\leftarrow$ \code{Tsili}'', $PM_4$ derives two complete
matches (\code{$CM_1$}, \code{$CM_2$}) by satisfying the predicates in edges
\code{written\_by} and \code{has\_writer}. The state \code{(AC)} then
transitions to the state \code{(ACC)}.

\subsection{Need for Shared-State in Pattern Matching}
\label{sec:need4pattern}
\subsubsection{Pattern Matching Applications}
We target the fundamental problem of pattern matching, which is the backbone
of several categories of data processing systems: those for (i) complex
event
processing (CEP),
(ii) OLAP with the
\mreg operator,
and (iii) GraphRAG.
While each category induces different domain-specific applications, the data
and execution models defined in \S\ref{sec:model} are general enough for
these applications, as explained below.

\myparr{(i) CEP} detects predefined patterns in unbounded streams of
events with low latency~\cite{cugola2012processing,10.1145/3654935,10.1145/3709682}.
Each event is a tuple of attribute values.
 Patterns are defined as a combination of event types, operators,
predicates, and a time window. These operators include
conjunction, sequencing (\code{SEQ}) that enforces a temporal order of
events, Kleene closure (\code{KL}) that accepts one or more events
of a type, and negation (\code{NEG}) that requires the absence of specific-typed events.
\looseness=-1

Two execution models have been proposed for CEP. (1) In the automata-based model~\cite{wu2006high}, partial matches denote
partial runs of an automaton that encodes the required event occurrences. The state transitions are guarded by predicates
to check if partial runs advance
in the automaton.
(2) In the tree-based
model~\cite{mei2009zstream}, events are
inserted into a hierarchy of buffers that are guarded by predicates.
The evaluation then proceeds from the leaf buffers of the tree to the
root, filling operator buffers with derived partial matches.

Both automata- and tree-based models are covered by
 our 
 DAG$^b$ model (\S\ref{sec:model}). Automata and trees are specific forms of a DAG$^b$.
Fundamentally, the DAG$^b$ captures the essence of pattern matching: \emph{the state and state transitions}. CEP's \emph{selection and consumption policies} determine how many data elements can be skipped to select and if they can be reused~\cite{wu2006high,chakravarthy1994snoop,zimmer1999semantics,cugola2010tesla}.
In this technical report, we target all the combinations of selection and consumption policies.
\tinyskip
\myparr{(ii) OLAP} queries with \mreg clause~\cite{iso19075-5} perform pattern matching on the rows (\ie tuples) of a table
or view. Many platforms have supported this, including Oracle, Apache Flink,
Azure Streaming Analytics, Snowflake, and
Trino~\cite{findeisen2021, laker2017, paes2019,
	alves2019, snowflake2021}.
 \mreg  specifies types of rows
based on their attribute values and defines a pattern as a
regular expression over these types.
The pattern is then evaluated for a
certain partition of the input tuples.
OLAP systems also specify whether matches may overlap and how the result is constructed per match.
Common execution models for \mreg pattern matching are based on automata, which are covered by the DAG$^b$ model in \S\ref{sec:model}. Depending on the structure of the regular expression, a deterministic or non-deterministic automaton is
constructed, which is then used to process tuples while scanning the table
or view used as input.

\tinyskip
\myparr{(iii) GraphRAG} enhances the inference capabilities of RAG systems by integrating external knowledge graphs (KG),  particularly for rich relationships between entities~\cite{MicrosoftGraphRAG,MicrosoftPrivateData2024}.
As shown in \F\ref{fig:basic_flow}, this is achieved by evaluating the patterns of \emph{regular path queries} on the KG~\cite{DBLP:conf/iclr/SunXTW0GNSG24,DBLP:conf/iclr/LuoLHP24,nguyen-etal-2024-direct}.
Here, the input of patterns is the tuples of nodes representing entities and the edges that reflect semantic relationships between them.
A pattern specifies a sequence of edges \ie a regular
expression over edge labels, to retrieve structures of complex relations and contextual dependencies encoded in the KG.
Again, the common execution model is automata-based, searching the edges of the KG till accepting states in the automata. This is also captured by our DAG$^b$ model defined in \S\ref{sec:model}.

\subsubsection{Quality Enhancement via Shared Patterns}
The use of multiple shared patterns promises to significantly improve the query quality in CEP~\cite{ray2016scalable,poppe2021share, ma2022gloria}, OLAP~\cite{10.1007/978-3-030-27520-4_7,zhu2023high}, and GraphRAG~\cite{abul2017multiple,DBLP:conf/iclr/LuoLHP24,nguyen-etal-2024-direct,DBLP:conf/iclr/SunXTW0GNSG24}.
This is because sharing state allows several patterns to \emph{collaboratively}  generate more insights in higher-level semantics.

Consider the GraphRAG example in \F\ref{fig:graphrag_full}, the use of shared patterns enables the LLM to generate correct responses for all 14,872 questions in the MetaQA benchmark~\cite{Zhang_Dai_Kozareva_Smola_Song_2018}, improving the accuracy by 2.80$\times$.
In contrast, only using separate patterns in isolation failed to provide correct responses for 11,897 questions.

In particular, question \#53 requires multi-hop paths in the KG to first query the writer of \code{Alila}, \ie Amos Gitai, and then retrieve other movies that are directed or written by him.
Since the label of edges in the KG may be different than that specified in single patterns, for instance, \code{written\_by} and \code{has\_writer} are different labels but semantically equivalent. Shared patterns can leverage semantic similarity to evaluate a set of overlapping patterns (that only differ in the edge label) simultaneously. As a result, when using separate patterns, LLM reports ``\code{I apologize, but I don't think we discussed a movie called Alila or its scriptwriter}''.

 Although such enhancement can also be achieved by evaluating bespoke patterns first, buffering the matches, and aggregating them, the processing latency becomes unacceptable. In our experiments, it 
 takes 23 seconds (the shared pattern is finished in 1,079 ms). \looseness = -5

Similar quality enhancements are reported in industrial-scale use cases. Amazon employs GraphRAG
to improve response accuracy by 35\% in financial reports and vaccine documents
~\cite{AWSGraphRAG2024}.
Neo4j leverages GraphRAG to monitor and optimize supply chain management for a leading global automotive manufacturer~\cite{Neo4jGraphRAG2024, Neo4jAutomaker2024}.

\subsection{Challenges and the Design Space for Best-Effort Processing on Shared Patterns}
\label{sec:challenges}

\subsubsection{Challenges}
The evaluation of common pattern matching presents the challenge of exponential computational complexity.
Zhang \etal~\cite{zhang2014complexity} have proved exponential complexity in CEP patterns with Kleene closure operators.
In the same vein, Huang \etal~\cite{huang2023t} showed that \mreg presents exponential runtime complexity with Kleene closure.
Previous work~\cite{mendelzon1995finding, libkin2012regular, pacaci2020regular} has proven that the evaluation of a regular path query is NP-hard.
Therefore, pattern-matching systems turn to best-effort
processing to maximize the result quality,
\ie the number of complete matches,
while satisfying a latency bound. \looseness=-1

Existing best-effort approaches fall short in exploiting the optimization space of shared states across a set of patterns.
In particular, they discard selected input data~\cite{chapnik2021darling, zhao2020load, slo2023gspice, hSPICE} or partial matches~\cite{slo2019pspice, zhao2020load, DBLP:conf/icde/Zhao18} to reduce processing latency; however, they fail to capture the interactions and inferences of shared partial matches between shared patterns.
As a result, existing best-effort mechanisms cannot keep high quality within tight latency bounds.

\autoref{fig:sota} highlights such limitations.
At the 300\unit{ms} latency bound, input-based
 approaches achieve the accuracy of 56\% (\code{DARLING} \cite{chapnik2021darling}).
While the hybrid approach achieves 53\% (\code{AthenaCEP}~\cite{zhao2020load}).
  Their low accuracy performance is due to the focus on optimizing a single pattern.
  \code{pSPICE}~\cite{slo2019pspice}\footnote{We have not compared \texttt{pSPICE}~\cite{slo2019pspice} because its authors have demonstrated that  \texttt{hSPICE}~\cite{hSPICE}  outperforms it. Therefore, we report the performance of \texttt{hSPICE}.},
 \code{hSPICE}~\cite{hSPICE} and \code{gSPICE}~\cite{slo2023gspice} indeed consider a CEP operator in multiple pattern queries in which each query is assigned a weight in advance.
  Yet, they do not consider the interaction and interference of shared patterns.
  Instead, they calculate the utility of events to discard with respect to separate patterns in isolation, by multiplying the pattern's predefined weight. As a result, \code{hSPICE} and \code{gSPICE} only achieve 53\% and 60\% accuracy. \looseness=-1

\begin{figure*}[t]
	\centering
	\includegraphics[width=0.9\linewidth]{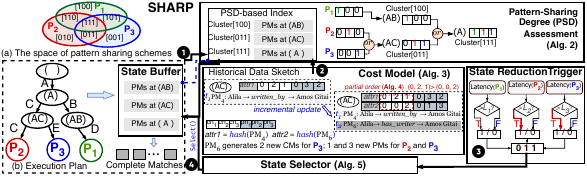}
	\caption{System architecture of \sys  (the exmaple execution plan is identical to \F\ref{fig:pattern-graph})} 
	\label{fig:sharp_flow} 
	\vspace{-1em} 
\end{figure*}

\subsubsection{Design Space}
We explore the design space of best effort pattern matching with shared state.
The fundamental problem is that the state of pattern matching,
\ie partial matches, has very different impact on the result quality and performance in \emph{three dimensions}: \looseness=-1

\textbf{(1)} \emph{For bespoke patterns}, partial matches contribute differently in constructing complete matches while consuming various computational resources (\ie increase the processing latency).
For instance, some partial matches generate more partial matches (longer sub-patterns) and consume lots of CPU cycles and memory footprint, but will not lead to complete matches due to the selection of query predicates.
\F\ref{fig:pattern-graph}b illustrates this dimension.
For pattern {\color{cornellred} $P_2$}, partial match \code{$PM_4$} (in state \code{(AC)}'s buffer) contributes two complete matches,
\code{$CM_1$} and \code{$CM_2$} (in state \code{(ACC)}'s buffer).
In contrast, \code{$PM_5$} has already consumed computational effort \ie state transitions till \code{(AC)}'s buffer, but does not generate any complete matches.

\textbf{(2)} \emph{An individual partial match} may have varying impacts on multiple shared patterns:
it may generate more complete matches for certain patterns
than others in the same set of shared patterns.
As demonstrated in \F\ref{fig:pattern-graph}b, \code{$PM_4$} generates two complete matches,
\code{$CM_1$} and \code{$CM_2$} for {\color{cornellred} $P_2$}, but only one complete match, \code{$CM_3$} for {\color{blue} $P_3$}.
Similarly, \code{$PM_{6}$} generates two complete matches, \code{$CM_4$} and \code{$CM_5$} for {\color{blue} $P_3$}, but none for {\color{cornellred} $P_2$}.
In addition, patterns may have various weights/utilities/priorities defined by user applications, which further complicates the impact of a single partial match on the final results.

\textbf{(3)} The impact between partial matches and shared patterns changes dynamically due to concept drift in payload value distribution, especially in data streams.
This means that pattern matching engines and optimization techniques shall adapt to such dynamics.
For instance, in \F\ref{fig:pattern-graph}b, at time point \code{$t_1$}, both state \code{(AC)} and \code{(AB)} have two PMs and \code{(ACC)} has two complete matches while \code{(ACE)} has one. That is equal resource consumption for PMs in \code{(AC)} and \code{(AB)}, but PMs contribute more to {\color{cornellred} $P_2$} than {\color{blue} $P_3$}.
However, when processing new data elements in \code{$D_2$}, at  \code{$t_2$}, one more PM is generated at \code{(AC)} while two more complete matches are generated at \code{(ACE)}.
Now, PMs in \code{(AC)} consume more computational resources than those in \code{(AB)} while PMs contribute more to {\color{blue} $P_3$}. \looseness=-1

\section{Research problem formulation}
\label{sec:problem}
In this paper,
we design a \emph{state reduction} mechanism that selects a subset of partial matches to process for shared patterns. The goal is to maximize results quality while satisfying bespoke latency bounds for all patterns.
We denote \code{CM$_{P_i}(j)$} and \code{CM$'_{P_i}(j)$} as the complete matches
without and with \emph{state reduction}.
Then, \code{$\delta_{P_i}(k) = \sum\nolimits_{1 \le j \le k} |\texttt{CM}_{P_i}(j)
\setminus \texttt{CM}'_{P_i}(j)|$} is the \textit{recall loss} for pattern
$P_i$.
We formulate the \emph{state reduction} as a {multi-objective
	optimization problem:} \looseness=-1
\begin{problem}
	\normalfont
	Given a prefix sequence \code{$S(\dots k)$}, an execution plan \code{$G$}, partial matches \code{$\texttt{PM}(k)$}, and the latency bounds \code{$L_i$} for each pattern \code{$P_i
	\in \mathbf{P}$},
	the state reduction problem for the best-effort pattern
	matching is to select a subset
	\code{$\texttt{PM'}(k) \subset \texttt{PM}(k)$}, such
	that {the recall loss \code{$\delta_{P_i}(k)$} is minimized for all \code{$P_i \in
		\mathbf{P}$}, respecting the latency bounds for all patterns, \code{$\texttt{Latency}(P_i) \leq
		L_i$}.} 
	\vspace{-5pt}
\end{problem}

\section{\sys Design}
\label{sec:design}

\begin{figure}[tb]
	\begin{minipage}[t]{1.0\linewidth}
		\removelatexerror 
		\begin{algorithm}[H]
			\begin{small}
				\caption{\sys workflow}
				\label{alg:work_flow}
				\KwIn{Input data element $d_{k+1}$ and sequence prefix $S(\dots k)$,
					execution plan $G$,
					patterns $\mathbf{P}$, complete matches
					$\texttt{CM}(k)$ and partial matches $\texttt{PM}(k)$.}
				\KwOut{A subset of partial matches \code{PM'(k)} to be
					processed.}
				$\mathbf{C} = \texttt{PSD}(G, \mathbf{P})$ \label{line:workflow_0}
				\tcp*{Assess Pattern-Sharing Degree (Alg.\ref{alg:cluster})}
				$\mathbf{Q} = \texttt{Cost}(\texttt{CM}(k),
				\mathbf{P}, \texttt{PM}(k))$ \tcp*{Calculate cost model (Alg.\ref{alg:cost_model})} \label{line:workflow_1}
				$\mathbf{C}.\texttt{Partial\_Order}(\mathbf{Q})$ \tcp*{Compute state partial order (Alg.\ref{alg:cost_order})} \label{line:workflow_2}
				\tcp{Scan input data and perform state reduction}
				\While{$d_{k+1}$ arrives}
				{
					\textbf{if} $\forall P_i \in \mathbf{P},~\exists l_{i,k+1} \ge L_i$ \tcp*{Trigger state reduction} \label{line:workflow_3}
						\textbf{then} \texttt{PM'(k)}=$\texttt{Select}(\mathbf{C}, \{P_i\}_{l_{i,k+1} \ge L_i})$ 	\tcp*{Select a subset of state for pattern matching (Alg.\ref{alg:pm_shedding})} \label{line:workflow_4} 
					\tcp{Incrementally update PSD indexing and the cost model}
					\textbf{while} \textit{a new match} $\rho'$ \textit{arrives} \textbf{do} $\mathbf{C}.\texttt{insert}(\rho')$, $\mathbf{Q}.\texttt{update}(\rho')$\; \label{line:workflow_5}
				}
				\textbf{return} \code{PM'(k)}\;
			\end{small}
		\end{algorithm}
	\end{minipage}
	\vspace{-2em}
\end{figure}

\sys's goal is to realize the \emph{state reduction} for best-effort processing in shared patterns.  
Its design is based on the analysis of the corresponding challenges and the design space to optimize the interactions and inferences of the shared state (\S\ref{sec:challenges}). 
To this end, \sys designs a new abstraction to capture the interactions and inferences of the shared state, and maintains a cost model to select partial matches for further processing to satisfy the latency bounds. 

\F\ref{fig:sharp_flow} illustrates the architecture of \sys, while Alg.\ref{alg:work_flow} explains its workflow.
First, \sys uses a new abstraction of \myc{1} \emph{Pattern-Sharing Degree} (PSD) to capture state sharing schemes across several patterns (line~\ref{line:workflow_0} in Alg.\ref{alg:work_flow}). 
To efficiently manage the dynamically changing state, \sys builds a bitmap-based indexing mechanism for PSD to cluster partial matches for fast lookup and updates. 

It then employs a \myc{2} \emph{cost model} to efficiently and effectively
 access partial matches' contribution to the final complete matches and their

  computational overhead
 (line~\ref{line:workflow_1} in Alg.\ref{alg:work_flow}). 
The cost model ensures that \sys always selects the most promising subset of partial matches to process (line~\ref{line:workflow_2} in Alg.\ref{alg:work_flow}), \ie state reduction. \looseness=-1

\sys monitors the processing latency for patterns, 
upon exceeding a latency bound (\ie we call this overload), 
\myc{3} the \emph{state reduction trigger} generates 
an \emph{overload label} encoded in a bitmap (line~\ref{line:workflow_3} 
in Alg.\ref{alg:work_flow}). 
After this,  \sys's \myc{4} \emph{state selector} uses this overload label and the PSD-based indexing to instantly locate state buffers that are related to violated latency bounds, and select a subset of partial matches to proceed with pattern matching (line~\ref{line:workflow_4} in Alg.\ref{alg:work_flow}).
Lastly, \sys incrementally updates the PSD and the cost model  (line~\ref{line:workflow_5} in Alg.\ref{alg:work_flow}). 
\sys repeats the above steps until the processing latency is below the latency bound.

Next, we explain the details of the pattern-sharing degree
(\autoref{sec:clustering}), the cost model
(\autoref{sec:cost_model}), and optimizations of the state selector  (\autoref{sec:shedding}).

\subsection{Pattern-Sharing Degree} 
\label{sec:clustering}

The \emph{pattern-sharing degree} (PSD) is \sys's abstraction to
capture how different patterns share the state through overlapping sub-patterns.
We design PSD to be expressive enough to cover the full space of pattern-sharing schemes.
For $n$ patterns the number of all possible sharing combinations is $2^n$ 
and \sys's PSD uses an $n$-bit vector to systematically encode 
the entire space of sharing schemes. \looseness=-1

\begin{example}
\normalfont
\F\ref{fig:sharp_flow}a shows the space of sharing schemes for patterns, {\color{darkolivegreen}$P_1$}, {\color{cornellred}$P_2$} and {\color{blue}$P_3$}: eight possible combinations in different colors.
\sys uses three bits to encode this space: 
\code{[111]} denotes the state shared by all patterns (\ie sharing degree of three), while \code{[011]} denotes those only shared by  {\color{cornellred}$P_2$} and {\color{blue}$P_3$}  (\ie sharing degree of two). \looseness=-1
\end{example}

Note that \sys takes an input execution plan for shared patterns. 
For a specific execution plan, the number of sharing schemes is already determined.  
The upper bound is linear---the \code{sum} of all pattern lengths (\ie no sharing at all).  
 \F\ref{fig:sharp_flow}b shows the execution plan with six states. Two of them contain shared 
states, \ie {\color{darkolivegreen}$P_1$}, {\color{cornellred}$P_2$} and {\color{blue}$P_3$} share \code{(A)}, 
while {\color{cornellred}$P_2$} and {\color{blue}$P_3$} share \code{(AC)}. \looseness=-1
 
To address the design space (see \S\ref{sec:challenges}) 
of state reduction for shared patterns, we design \sys's PSD to (i) incorporate how each partial match is shared across different patterns, 
(ii) efficiently retrieve all partial matches for bespoke shared patterns---based on corresponding sharing schemes, 
and (iii) efficiently adapt to dynamic changes between partial matches and shared patterns. 

To this end, we build bitmap structures to represent the sharing degree of each sub-pattern. This structure enables CPU bitwise instructions to efficiently manage PSD---reducing the system overhead of PSD.  
For a pattern \code{$P_i$}, the bitmap for each sub-pattern \code{$SP$} (and each PM $\rho$ in $SP$) is an $n$-bit array \code{\texttt{PSD}($SP$)} (and \code{\texttt{PSD}($\rho$)}), 
where the $i$-th bit is set to 1 if \code{$SP$} is a sub-pattern of \code{$P_i$}, and 0 otherwise.

\begin{figure}[tb]
	\begin{minipage}[t]{1.0\linewidth}
		\removelatexerror
\begin{algorithm}[H]
\begin{small}
    \caption{Pattern-sharing degree (PSD) assessment}
    \label{alg:cluster}
    \KwIn{ Evaluation Plan $G$, patterns $\mathbf{P}$ and partial matches \code{PM}($k$).}
    \KwOut{PSD-indexed partial match clusters $\mathbf{C}$.}
    \SetKwFunction{score}{score}
    \SetKwProg{Fn}{def}{:}{}

        \textbf{for} $P_i \in \mathbf{P}$ \textbf{do} \code{$\texttt{PSD}(P_i) \leftarrow 2^{n-i-1}$}\tcp*{ $\forall P_i $, set the $i^{th}$ bit to 1} \label{line:bitmap2}
        \tcp{PSD Assessment through Depth-First Traversal of $G$}
        
            \For{$P_i \in \mathbf{P}$ \textbf{and} \textit{there is a reachable state $SP$ in $G$}}  
            {\label{line:bitmap3}
                $\code{PSD}(SP) \leftarrow \code{PSD}(SP) \vee \code{PSD}(P_i)$ \tcp*{Update the PSD of SP} 
            }\label{line:bitmap5}
        \tcp{PSD-based Indexing}
        $\mathbf{C} \leftarrow$ \textit{Clustering sub-patterns with the same PSD}\;\label{line:bitmap6}
        { $\forall C(b') \in \mathbf{C},C(b') = \{\rho \in \code{PM}(k) | \texttt{PSD}(\rho) = b'\}$}\;  
          \label{line:bitmap7}
        \Return{$\mathbf{C}$}\; 
\end{small}
\end{algorithm}
\end{minipage}
\vspace{-2em}
\end{figure}

Algorithm \ref{alg:cluster} shows the process to construct PSD as follows:
First, \sys assigns each pattern $P_i$ an initial bitmap \code{PSD$(P_i) = 2^{n-i-1}$}, where only the $i$-th bit is 1 (line \ref{line:bitmap2}). 
\sys then traverses the execution plan $G$ in a depth-first manner to search all sub-patterns of each $P_i$
(line \ref{line:bitmap3}).
If there exists reachable state \code{$SP$} in \code{$G$},  \code{$SP$} is shared by $P_i$. 
\sys computes the bitmap for each sub-pattern \code{$SP$},  \code{$\texttt{PSD}(SP) = \bigvee_{SP' \in SP.\texttt{succ}} \texttt{PSD}(SP')$}. That is applying a bitwise \code{OR} operator across the bitmaps of its successor sub-patterns.

After constructing the PSD for the execution plan $G$, 
\sys organizes partial matches with the same bitmap $b'$ into a cluster 
\code{$C(b') = \{\rho \in \code{PM}(k) | \texttt{PSD}(\rho) = b'\}$} 
(lines \ref{line:bitmap6}-\ref{line:bitmap7}).
Whenever a new partial match is generated, \sys indexes it to the corresponding cluster.
Due to the efficiency of bitwise operating instructions, PSD indexing locates the partial matches via a bitmap in \code{$\mathcal{O}(1)$} time complexity.
Here, the number of clusters is bounded by the number of states (\ie nodes) in the execution plan \code{$G$}.
We illustrated the above process of PSD assessment ( (Alg. \ref{alg:cluster})) with following example
 
\vspace{-0.5em}
\begin{example} \label{example:2}
\normalfont
	For the execution plan in \F\ref{fig:sharp_flow}b \sys  (line \ref{line:bitmap2}) assigns each pattern an initial bitmap:
\code{PSD({\color{darkolivegreen}$P_1$})=[100]}, \code{PSD({\color{cornellred}$P_2$})=[010]}, and \code{PSD({\color{blue}$P_3$})=[001]}.
\sys then (line \ref{line:bitmap3}) computes the bitmap for each sub-pattern by performing \code{OR} across the bitmaps of its successor sub-patterns: 
\code{PSD(AB)=PSD({\color{darkolivegreen}$P_1$})}, 
\code{PSD(AC)=PSD({\color{cornellred}$P_2$})$\vee$  PSD({\color{blue}$P_3$})=[010]\\$\vee$[001]=[011]},
\code{PSD(A)=PSD(AC)$\vee$PSD(AB)=[100]$\vee$[011]=[111]}.
Finally (lines \ref{line:bitmap6}-\ref{line:bitmap7}), \sys groups partial matches into clusters \code{$C([100])$}, \code{$C([011])$}, and \code{$C([111])$} based on their bitmaps.

\end{example}

\subsection{Cost Model} \label{sec:cost_model}
The goal of the cost model is to assess (i) how ``promising'' a partial match is for the shared patterns---the number of complete matches it can contribute and
(ii) how ``expensive'' a partial match is for its entire lifespan---the computational overhead (resource consumption) it incurred.
In addition, (iii) the cost model calculation must be lightweight. 
Because the latency bounds have already been violated, the cost model shall not introduce extra overhead in the first place.

\subsubsection{Definition of the Cost Model} 
\label{subsubsection:pm_quality}
For a partial match \code{$\rho\in PM(k)$}, the cost model accesses its \textbf{\emph{contribution}} to the final results of shared patterns and the computational \textbf{\emph{overhead}} for final and intermediate results during pattern matching.

\mypar{Contribution}
A partial match \code{$\rho$} may result in
complete matches for multiple shared patterns.
\sys captures  $\rho$'s \emph{contribution} to
pattern $P_i$ as the number of complete matches that are
generated by $\rho$.
We define the contribution of $\rho$ up to a time point $k'$ as
\vspace{-.5em}
\code{
\begin{equation}\label{equ:contribution}
    \Delta_{P_i}^+(\rho) = |\{\rho' \in \texttt{CM}_{P_i}(k') | \text{$\rho$ \texttt{generates} $\rho'$} \}|,
\end{equation}
}where \code{$\texttt{CM}_{P_i}(k')$} is the complete matches of pattern \code{$P_i$} over \code{$S(\dots k')$}. 

\noindent
We define the contribution of $\rho$ to all patterns in $\mathbf{P}$ as a \emph{vector}

\centerline{$\Vec{\Delta}_{\mathbf{P}}^+(\rho) = [ \Delta_{P_1}^+(\rho), \dots, \Delta_{P_n}^+(\rho) ].$}
\mypar{Overhead}
\sys considers a partial match $\rho$'s computational overhead as the resource consumption caused by $\rho$ itself and all its derived partial matches and complete matches in the future. 
In addition, partial matches in different states consume different computational resources (\ie CPU cycles and memory footprint) due to predicates' complexity and the size of the partial match itself (\ie the length of a sub-pattern).
We capture this through a function \code{$\varTheta (\rho) \mapsto r$, $r \in \mathbf{N}^+$} that maps $\rho$'s computational overhead to a real number.
Users can materialize \code{$\varTheta$} based on their specific applications.
The computational overhead of \code{$\rho$} to pattern \code{$P_i$} is 
\code{
\begin{equation}\label{equ:overhead}
    \Delta_{P_i}^-(\rho) = \sum\nolimits_{\rho' \in \bigcup\texttt{PM}_{P_i}(k') \wedge \text{$\rho$ \texttt{generates} $\rho'$}} \varTheta(\rho'),
\end{equation}
}
\code{$\texttt{PM}_{P_i}(k')$} is the set of partial matches of \code{$P_i$} over \code{$S(...k')$}. 
We define the computational overhead of $\rho$ to all patterns in $\mathbf{P}$ as a \emph{vector}
\code{
\centerline{$\Vec{\Delta}_{\mathbf{P}}^-(\rho) = [ \Delta_{P_1}^-(\rho), \dots, \Delta_{P_n}^-(\rho) ].$}
}

\vspace{-0.5em}
\begin{example} \label{example:3}
\normalfont
For the partial match \code{$PM_4$}
	in \F\ref{fig:sharp_flow} \myc{2}, 
	its contribution and overhead to \code{{\color{darkolivegreen} $P_1$}$,$ {\color{cornellred} $P_2$}$,$ {\color{blue}$P_3$}} at time $t_1$ are:
\code{$\Vec{\Delta}_{\mathbf{P}}^+({PM}_4) = [0, 2, 1]$} and \code{$\Vec{\Delta}_{\mathbf{P}}^-({PM}_4) = [0, 3, 2]$}, as it generates two complete matches (\code{$CM_1$}, \code{$CM_2$}) for {\color{cornellred} $P_2$} and one (\code{$CM_3$}) for {\color{blue}$P_3$}, while generating three (\code{$PM_4$}, \code{$CM_1$}, \code{$CM_2$}) and two (\code{$PM_4$}, \code{$CM_3$}) partial/complete matches for {\color{cornellred}$P_2$} and {\color{blue}$P_3$}, respectively. (Here, for simplicity, we assume \code{$\varTheta(\rho)=1$}.)
\end{example}
\vspace{-1em}

\begin{figure}[tb]
	\begin{minipage}[t]{1.0\linewidth}
		\removelatexerror
\begin{algorithm}[H]
    \begin{small}
        \caption{Cost model estimation}
        \label{alg:cost_model}
        \KwIn{Patterns $\mathbf{P}$ 
        and partial matches $\texttt{PM}(k)$.} 
        \KwOut{The contribution $\Delta_{P_i}^+(\rho)$ and the computational overhead $\Delta_{P_i}^-(\rho)$ of each $\rho \in \texttt{PM}(k)$ to all patterns $P_i \in \mathbf{P}$.}

        \tcp{Incremental Update of the Historical Data Sketch}
        
        \If{a newly generated match $\rho'$ arrives} 
        {\label{line:cost_0}
            \tcp{Update the complete/partial match counting}
            \For{$\rho$ that generates $\rho'$}
            {\label{line:cost_1-2}
                $attr= \texttt{hash}(\rho)$\tcp*{Get the attribute key}
                \textbf{if} $\rho'$ is a complete match \textbf{then} $\texttt{Sketch}[attr].cn_i$++\; 
                \textbf{if} $\rho'$ is a partial match \textbf{then} $\texttt{Sketch}[attr].pn_i$++\; \label{line:cost_1-3}
            }
        }
        \tcp{Look-up contribution and computational overhead}
        \While{evaluating $\rho \in \texttt{PM}(k)$}
        {\label{line:cost_1-6}

            $attr= \texttt{hash}(\rho)$\tcp*{Get the attribute key}
            $\forall P_i \in \mathbf{P}, \Delta_{P_i}^+(\rho) = \texttt{Sketch}[attr].cn_i$ \tcp*{Contribution}
            $\forall P_i \in \mathbf{P}, \Delta_{P_i}^-(\rho) = \texttt{Sketch}[attr].pn_i \times \varTheta(\rho)$ \tcp*{Overhead}
        }\label{line:cost_1-7}
        \Return $[\Delta_{P_i}^+(\rho)]_{P_i \in \mathbf{P},~\rho \in \texttt{PM}(k)}, [\Delta_{P_i}^-(\rho))]_{P_i \in \mathbf{P},~\rho \in \texttt{PM}(k)}$\;
    \end{small}
\end{algorithm}
\end{minipage}
\vspace{-1em}
\end{figure}

\subsubsection{Efficient Estimation of the Cost Model} 
\label{sec:historical_data}

The materialization of the cost model, 
 $\Vec{\Delta}_{\mathbf{P}}^+$ and $\Vec{\Delta}_{\mathbf{P}}^-$,
requires \emph{future} statistics (\eg the number of generated partial matches in runtime), which can only be computed in retrospect.
To address this issue,
\sys designs an efficient estimation for the cost model based on historical statistics of pattern matching. 
That is, the number of partial and complete matches generated by each partial match in the previous time window or time slice~\cite{zhao2020load}. 
The rationale is that the fresh history of runtime statistics may predict the counterparts in the near future~\cite{zhao2020load,DBLP:journals/pvldb/LiG16a}. \looseness=-1

\sys maintains a \textit{historical data sketch} for each state buffer in the execution plan to efficiently monitor and estimate the number of future complete and partial matches generated per partial match.
The sketch contains multiple entries, each summarizing the number of partial matches with the same attribute values.
\sys represents each entry as a vector \code{$[attr, cn_1, \dots, cn_n,$ $pn_1, \dots, pn_n]$} where $attr$ is the \textit{attribute key} value.
\sys uses a hash function to derive the attribute key of  \code{$\rho$},  \ie \code{$attr = \code{hash}(\rho)$}. 
\code{$pn_i$} and \code{$cn_i$} denote the number of partial and complete matches of pattern \code{$P_i$} generated from all partial matches associated with \code{$attr$}.
\code{$PM_4$} in Example \ref{example:3} 
	is hashed to \code{$attr_1$}: {\small [$attr_1$, $cn_1$=0, $cn_2$=2, $cn_3$=1, $pn_1$=0, $pn_2$=3, $pn_2$=2]}.
 
For efficiency, \sys \emph{incrementally} updates the historical data sketch, as illustrated in Alg.\ref{alg:cost_model}, lines \ref{line:cost_0}-\ref{line:cost_1-3}. When a new match $\rho'$ of pattern $P_i$ is generated, \sys updates the corresponding entries based on the hashed attribute key.
 Specifically, (i) if $\rho'$ is a complete match, \sys updates the counter {\small $cn_i$=$cn_i$ + 1}, 
 and (ii) if $\rho'$ is a partial match, it increases the partial match counter  {\small $pn_i$ = $pn_i$ + 1}.
For instance, in \F\ref{fig:sharp_flow} \myc{2}, when
 	 \code{$PM_6$} is created
 	   at $t_2$, 
 a new entry 
  \code{$attr_2$} is constrcuted:
 {\small [$attr_2$, $cn_1$=0, $cn_2$=0, $cn_3$=2, $pn_1$= 0, $pn_2$= 1, $pn_2$=3]}.
  $cn_i$ and $pn_i$ are updated accordingly.

\sys incrementally updates the historical data sketche: new matches refresh statistics while stale estimations fade—either via a sliding time window (\eg CEP) or updated statistics by newer observations (\eg GraphRAG).
This design enables lightweight maintenance of data skeetfhes and daptiveness to concept drifts.

 \sys estimates the cost model using the updated historical data sketch (Alg.~\ref{alg:cost_model}, lines~\ref{line:cost_1-6}--\ref{line:cost_1-7}).
 For a partial match $\rho$,
 \sys first hashes the \textit{attribute key}, \ie \code{$attr = \texttt{hash}(\rho)$}, and then looks up the historical data sketch to locate the entry \code{$[attr, cn_1, \dots, cn_n, pn_1, \dots, pn_n]$}. 
 The contribution and computational overhead of $\rho$ to pattern $P_i$ are estimated as
 \code{$\Delta_{P_i}^+ = cn_i$} and \code{$\Delta_{P_i}^- = pn_i \times \varTheta(\rho).$}

The design of \emph{historical data sketch} enables the estimation and the lookup in $\mathcal{O}(1)$ time complexity. In practice, users may also customize the hash function (\eg based on application-specific value distribution) to avoid hash collisions for the efficacy of the cost model estimation.

\begin{figure}[tb]
	\begin{minipage}[t]{1.0\linewidth}
		\removelatexerror
		\begin{algorithm}[H]
			\begin{small}
				\caption{Cost model-based partial order of state}
				\label{alg:cost_order}
				\KwIn{Two partial matches $\rho$ and $\rho'$, and their contributions, $\Delta_{P_i}^+(\rho)$ and $\Delta_{P_i}^+(\rho')$, to all patterns $P_i \in \mathbf{P}$.}
				\KwOut{The partial order between $\rho$ and $\rho'$.}
				\textbf{if }~$\forall P_i \in \mathbf{P},~\Delta^+_{P_i}(\rho) > \Delta^+_{P_i}(\rho')$ \textbf{then} \Return $\rho > \rho'$\; \label{line:shedding5}
				\textbf{else if}~$\sum\nolimits_i \Delta^+_{P_i}(\rho) >  \sum\nolimits_i \Delta^+_{P_i}(\rho')$ \textbf{then} \Return $\rho > \rho'$\;\label{line:shedding6}
				\textbf{else} \Return $\rho \leq \rho'$\; \label{line:shedding7}
				
			\end{small}
		\end{algorithm}
	\end{minipage}
	\vspace{-2em}
\end{figure}

\subsubsection{Partial Order of Partial Matches}
\label{sec:partial_order}
To facilitate efficient state reduction  in \S\ref{sec:shedding}, \sys organizes partial matches in a partial order based on the cost model.
Such a partial order allows \sys to effeciently and approxiamtely select partial matches for state reduction (see \S\ref{sec:appSelect})

The partial order must consider that a partial match may contribute differently to multiple shared patterns. 
In \F\ref{fig:sharp_flow} \myc{2}, \code{$PM_4$} contributes to {\color{cornellred}$P_2$} and {\color{blue}$P_3$} with contributions {\small [$cn_1$=0, $cn_2$=2, $cn_3$=1]}.
	In contrast, \code{$PM_6$} only contributes to {\color{blue}$P_3$} with {\small [$cn_1$=0, $cn_2$=0, $cn_3$=2]}. \looseness=-1

Algorithm~\ref {alg:cost_order} outlines how \sys computes such a partial order.
First (line \ref{line:shedding5}), \sys compares the contribution of two partial matches, $\rho$ and $\rho'$, on each pattern $P_i$. 
If \code{$\forall P_i \in \mathbf{P},~\Delta^+_{P_i}(\rho) > \Delta^+_{P_i}(\rho')$}, $\rho$ is better than $\rho'$. 
If that does not hold (line \ref{line:shedding6}), \sys compares the total contribution of $\rho$ and $\rho'$. If \code{$\sum\nolimits_i \Delta^+_{P_i}(\rho) >  \sum\nolimits_i \Delta^+_{P_i}(\rho')$},  $\rho$ is better than $\rho'$. Otherwise (line \ref{line:shedding7}), $\rho$ is worse than $\rho'$.
\vspace{-0.5em}
\begin{example} \label{example:4}
\normalfont
For \code{$PM_4$} and \code{$PM_6$} in \F\ref{fig:sharp_flow} \myc{2}, since \code{$PM_4$}.$cn_2$ $>$ \code{$PM_6$}.$cn_2$
and \code{$PM_4$}.$cn_3$ $<$ \code{$PM_6$}.$cn_3$, \sys cannot determine their order in line \ref{line:shedding5} of Alg. \ref{alg:cost_order}.
\sys  then compares their total contributions (line \ref{line:shedding6}).
The total contribution of \code{$PM_4$} is \code{$PM_4$}.$cn_2$ + \code{$PM_4$}.$cn_3$ = 3, which is greater than that of \code{$PM_6$}, \ie \code{$PM_6$}.$cn_3$ = 2.
Thus, \code{$PM_4$} $>$ \code{$PM_6$}.
\end{example}
\vspace{-0.5em}

For efficient implementation of the partial order, \sys maintains partial matches in a max-heap structure~\cite{heap} for state buffers. 
It is constructed during the initialization of PSD.
Updating the max-heap is sublinear time complexity~\cite{heap}, which is negligible compared to the overhead of the evaluation of pattern matching itself. \looseness=-1

\subsection{State Selector}
\label{sec:shedding}

\sys's \emph{state selector} carefully chooses a subset of partial matches to process, when the \emph{state reduction trigger} is activated by overloading, \ie processing latency exceeds the pre-defined bound (\eg SLO).

\subsubsection{Problem Formulation}
\label{sec:PMselectProb}

We formulate the partial match selection as a \textit{multi-objective optimization problem}
 to decide \emph{what} and \emph{how many} partial matches to select, based on \sys's \emph{cost model}. 
 
 The goal is to select a set of partial matches \code{$\rho\in \code{PM'(k)}$} with the highest total contribution. 
 That is to maximize {\small $\sum\nolimits_{\rho \in \texttt{PM'(k)}} \Vec{\Delta}_{\mathbf{P}}^+(\rho) = [\varDelta^+_{P_1}, \dots, \varDelta^+_{P_n}]$}.
At the same time, for each pattern \code{$P_i$}, the processing latency incurred by \code{PM'(k)}, \code{Latency($P_i$)}, must be below the latency bound \code{LatencyBound$_i$}.
 
Note that the processing latency  \code{Latency($P_i$)} is caused by the computational overhead of partial matches, \code{$\Delta_{P_i}^-$}.
Reducing the overhead results in lower latency.
This means that the latency bound indicates the upper-bound capacity of computational overhead that is allowed in pattern matching. 
Based  thereon, 
 \code{PM'(k)}'s total computational overhead must be below 
{\small $\frac{\texttt{LatencyBound}_i }{\texttt{Latency}(P_i)} \Delta_{P_i}^-$}.

Therefore, we formulate the state selection in shared patterns as the following multi-objective optimization problem.
\code{
	\begin{align}\label{equ:shedding_constraints}
		&\max\quad \text{\small $\sum\nolimits_{\rho \in \texttt{PM'(k)}} \Vec{\Delta}_{\mathbf{P}}^+(\rho) = [\varDelta^+_{P_1}, \dots, \varDelta^+_{P_n}]$} \\
		\textit{s.t.}~&\text{\small $\sum\nolimits_{\rho \in \texttt{PM'}(k)} \texttt{PSD}(\rho)[i] \cdot \Delta_{P_i}^-(\rho)\le$} \text{\footnotesize $\frac{\texttt{LatencyBound}_i}{{\texttt{Latency}(P_{i})}}$} \text{\small $ \Delta_{P_i}^-, \forall P_i \in \mathbf{P}$} \nonumber
	\end{align}
} 

\noindent \code{PSD}($\rho$)[$i$] represents the pattern sharing degree: if $\rho$  is shared by \code{$P_i$}, \code{PSD}($\rho$)[$i$]$\ge$1, indicting that $\rho$'s computational overhead should be counted under $P_i$.
The objective function, \code{$\sum\nolimits_{\rho \in \texttt{PM'(k)}} \Vec{\Delta}_{\mathbf{P}}^+(\rho)$}, is a vector of contributions for patterns in $\mathbf{P}$, with various or even conflicting utilities and therefore, different optimization goals. \ie multi-objective optimization.  \looseness=-1

The above problem is NP-hard via reduction from the multi-dimensional multi-objective knapsack problem~\cite{lust2012multiobjective,da2008core}, where each element corresponds to a partial match and the knapsack capacity corresponds to the computational overhead.
It is impractical to solve it online in the already overloaded pattern-matching engine with violated latency bounds.
Therefore, we design an efficient implementation for \sys's state selector to approximate the solution (\S\ref{sec:appSelect}). \looseness=-1

\subsubsection{Efficient Implementation in \sys}
\label{sec:appSelect}

\begin{figure}[tb]
	\begin{minipage}[t]{1.0\linewidth}
		\removelatexerror
		\begin{algorithm}[H]
			\begin{small}
				\caption{State selection}
				\label{alg:pm_shedding}
				\KwIn{Overload label $b_{OL}$, PSD-indexed partial match clusters $\mathbf{C}$.}
				\KwOut{A set of partial matches $\texttt{PM}'(k)$ to be processed.}
				\tcp{Select partial matches related to non-overloaded patterns}
				$\texttt{PM}'(k) \leftarrow \bigcup\nolimits_{\texttt{PSD'}:~\texttt{PSD' (bitmap index)}~\texttt{AND}~b_{OL} = 0} C(\texttt{PSD'})$\;\label{line:shedding1}
				\tcp{Select partial matches related to overloaded patterns}
				\While{the count constraint in Eq. (\ref{equ:shedding_constraints}) is holding}
				{\label{line:shedding2}
					\tcp{Select the highest quality PM based on the PSD}
                    \code{PSD} = \code{$\max$\{PSD: PSD (bitmap index)~\texttt{\footnotesize AND}~$b_{OL}\neq$ 0\}}\; \label{line:shedding2-2} 
                    $\rho = C(\texttt{PSD}).pop()$, $\texttt{PM}'(k)\leftarrow \texttt{PM}'(k) \cup \{\rho\}$\; \label{line:shedding2-3}
                    
				}\label{line:shedding3}
				\Return $\texttt{PM}'(k)$\;
			\end{small}
		\end{algorithm}
	\end{minipage}
	\vspace{-2em}
\end{figure}

The design of the state selector is based on the idea of \emph{hierarchical selection}. 
Specifically, \sys must (i) first select all partial matches from state buffers that are not associated with patterns with violated latency bounds using PSD (\ie not causing overload),
(ii) efficiently locate the affected state buffers and retrieve the partial matches using PSD, and (iii) efficiently select the partial matches from affected buffers based on both PSD and the partial order of the cost model. 

To achieve (i) and (ii), \sys employs the bitmap-based \emph{overload label} in the state reduction trigger (\S\ref{sec:design}, \F\ref{fig:sharp_flow} \myc{3}) to connect PSD's bitmap index to instantly locate the unaffected state buffers (\ie not related to latency violation) while retrieving partial matches from the affected buffers.
 Specifically, we define the overload label \code{$b_{OL}$} as an $n$-bit array (\ie bitmap). 
Whenever a pattern \code{$P_i \in \mathbf{P}$} violates its latency bound, 
\sys's state reduction trigger sets  \code{$b_{OL}$}'s $i$-th bit to 1.\looseness=-1

The overload label and PSD bitmap index are both n-bitmaps. 
\sys performs the bitwise \code{AND} operator between the overload label and the PSD index to locate unaffected state buffers (\ie the result is a zero bitmap). 
This is because the $i$-th bit in PSD index means if the state is shared by the $i$-th pattern \code{$P_i$}.
The state selector then selects all partial matches in unaffected state buffers without computing Eq.~(\ref{equ:shedding_constraints}), as shown in Alg.~\ref{alg:pm_shedding}, line~\ref{line:shedding1}.
\sys then locates the affected state buffers in constant time, based on non-zero results from the \code{AND} operation. \looseness=-1

To realize (iii), select partial matches in affected state buffers, \sys uses a \emph{greedy} approach (Alg.~\ref{alg:pm_shedding}, line~\ref{line:shedding2-2}) based on the partial order of the cost model and the PSD of partial matches.  
In particular, \sys selects partial matches in \emph{the order of PSD values}. 
A larger PSD value means the state is shared by more patterns and conveys a larger contribution value. 

Within a state buffer (Alg.~\ref{alg:pm_shedding}, line~\ref{line:shedding2-3}), \sys greedily selects partial matches based on the partial order of the cost model. 
That is, \sys always prioritizes the selection of partial matches with the highest contribution value, until reaching the upper bounds of the overhead capacity of partial matches (defined in Eq.\ref{equ:shedding_constraints}),
resulting
in the linear complexity (compared to the NP-hard in Eq.\ref{equ:shedding_constraints}).
In addition, such selection is performed at the attribute cluster level, instead of the PM instance level, which significantly further reduces the search space. \looseness=-1

\vspace{-0.5em}
\begin{example}
\normalfont
When {\color{cornellred}$P_2$} and {\color{blue}$P_3$} violate their latency bounds,
\sys configures the overload label, \code{$b_{OL}$} = \code{[011]} (see \F\ref{fig:sharp_flow} \myc{3}).
\sys then performs an \code{AND} between \code{$b_{OL}$} and instantly identify affected states via the PSD index.
Here, \code{(AB)} is unaffected because it is not shared by  {\color{cornellred}$P_2$} and {\color{blue}$P_3$}, \ie \code{PSD(AB)} \code{AND} \code{$b_{OL}$} = \code{[100]} \code{AND} \code{[011]} = \code{[000]}.
In contrast, \code{(AC)} and \code{(A)} are indeed affected due to the sharing  by both patterns. 
As a result,
\sys first selects partial matches in \code{(AB)}, then in affected states \code{(AC)} and \code{(A)}.
For the selection between \code{(AC)} and \code{(A)},  \sys prioritizes selection in \code{(A)} due to its larger PSD value (\code{[111]} vs \code{[011]}).
For efficiency, \sys approximtaely selects partial matches
at each state buffer, based on the partial order (selects \code{$PM_4$} rather than \code{$PM_6$} in Example~\ref{example:4}).  
This process continues until reaching the overhead bound  in Eq.~(\ref{equ:shedding_constraints}).
\end{example}
\vspace{-0.5em}

The state selector is lightweight. In (i) and (ii), the PSD-based index enables $\mathcal{O}(1)$ time to locate state buffers and retrieve partial matches. 
In (iii), the partial order enables $\mathcal{O}(n)$ search time in the number of clusters within a single state buffer--significantly small $n$ in practice. 
Furthermore, this searching overhead can be hidden, since scanning partial matches is inherently the core processing of pattern matching---overlapping the searching at $d_k$ and pattern match evaluation at $d_{k+1}$. 
As a result, the state selector only introduces negligible incremental system overhead. 

\section{Evaluation}
\label{sec:experiments}

We evaluated the effectiveness and efficiency of \sys in various scenarios.
After outlining the experimental setup in \S\ref{sec:exe_setting}, our experimental evaluation answers the following questions:

\tinyskip

\noindent (1) What are the overall effectiveness and efficiency of \sys? (\S\ref{subsec: overall_effect})

\noindent (2) How sensitive is \sys to pattern properties including
pattern selectivity, pattern length, and the time window?
(\S\ref{sec:exe_sens})

\noindent (3) How do pattern sharing mechanisms impact \sys? (\S\ref{sec:exe_sharingScheme})
 
\noindent (4) How does \sys adapt to concept drifts of input data? (\S\ref{sec:exe_adaptivity})

\noindent (5) How do resource constraints impact \sys?
(\S\ref{sec:exe_resourceCon})

\noindent (6) How does \sys adapt to complex pattern interactions? (\S\ref{sec:diff_bounds})

\noindent (7) What is the scalability performance of \sys? (\S \ref{sec:scalability})

\noindent (8) How do the selection/consumption policies impact \sys (\S \ref{sec:selection_consumption_policy})

\noindent (9) \sys' performance compared to SOTA GraphRAG systems.\looseness=-1 (\S\ref{sec:neo4j})

\noindent (10) \sys's state selection compared to the optimal solution. (\S\ref{sec:optimal}) \looseness=-1

\subsection{Experimental Setup}
\label{sec:exe_setting}

Our experiments have the following setup:

\mypar{Testbeds} We conduct experiments on two clusters. \textbf{(1)} LUMI supercomputer~\cite{lumi}--each node features two AMD EPYC 7763 CPUs (128 cores) and 512\unit{GB} RAM, running SUSE Linux
Enterprise Server 15 SP5. 
\textbf{(2)} A GPU cluster~\cite{triton}--each node being equipped with four NVIDIA H100 80GB GPUs, two Intel Xeon Platinum 8468 CPUs (96 cores), and 1.5\unit{TB} RAM, running Red Hat Enterprise 9.5.\looseness=-1

\begin{figure*}[t]
  \centering
  \begin{minipage}{0.48\linewidth}
    \centering
    \begin{subfigure}[b]{0.49\linewidth}
      \centering
      \includegraphics[width=\linewidth]{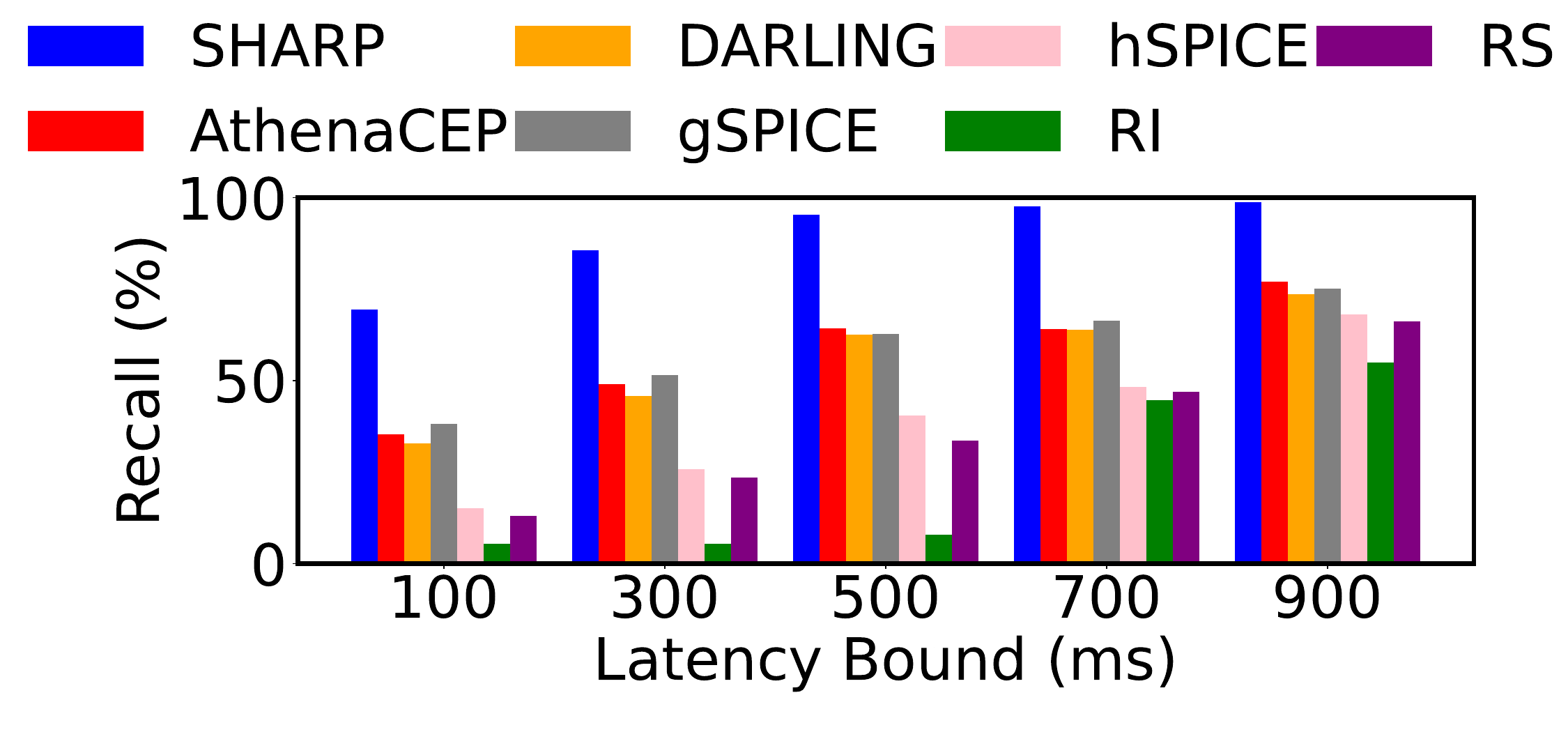}
      \caption{Recall}
      \label{fig:recall}
    \end{subfigure}
    \hfill
    \begin{subfigure}[b]{0.49\linewidth}
      \centering
      \includegraphics[width=\linewidth]{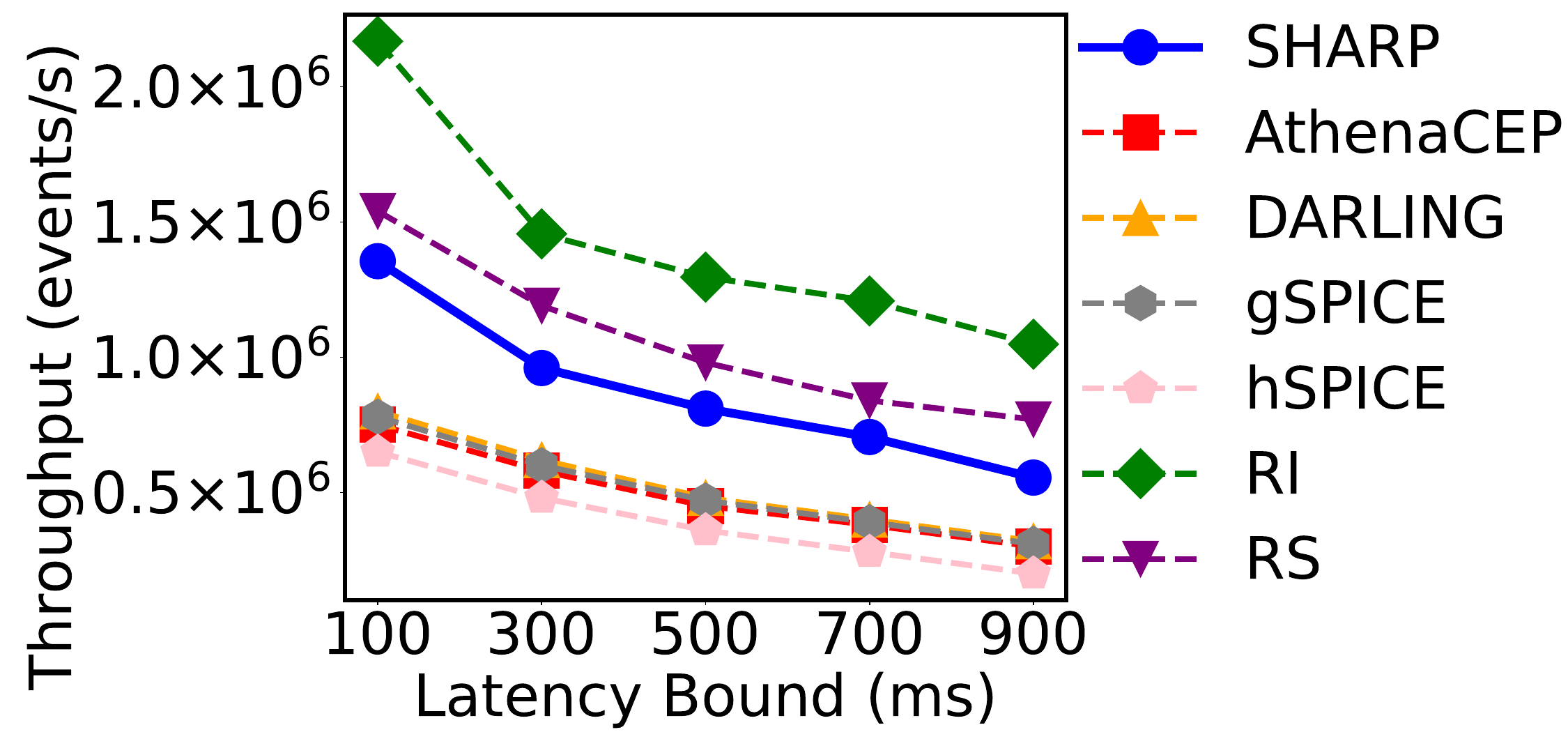}
      \caption{Throughput}
      \label{fig:throughput}
    \end{subfigure}
    \caption{The overall performance of shared CEP patterns \code{P$_{3}$-P$_4$} over \code{DS1} at different latency bounds
    	}
    \label{fig:overall}
  \end{minipage}
  \hfill
  \begin{minipage}{0.48\linewidth}
    \centering
    \begin{subfigure}[b]{0.49\linewidth}
      \centering
      \includegraphics[width=\linewidth]{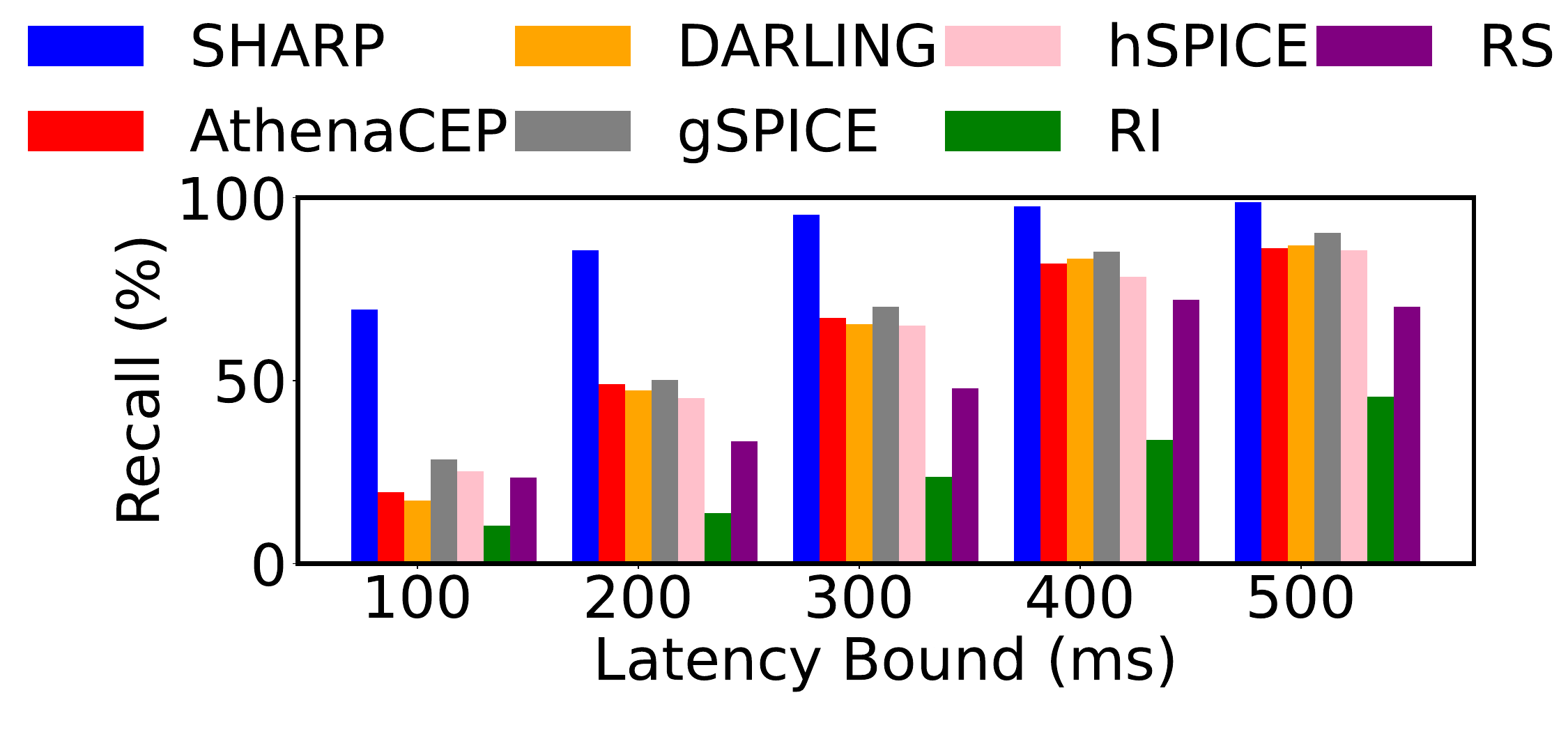}
      \caption{Recall}
      \label{fig:citi_recall}
    \end{subfigure}
    \hfill
    \begin{subfigure}[b]{0.49\linewidth}
      \centering
      \includegraphics[width=\linewidth]{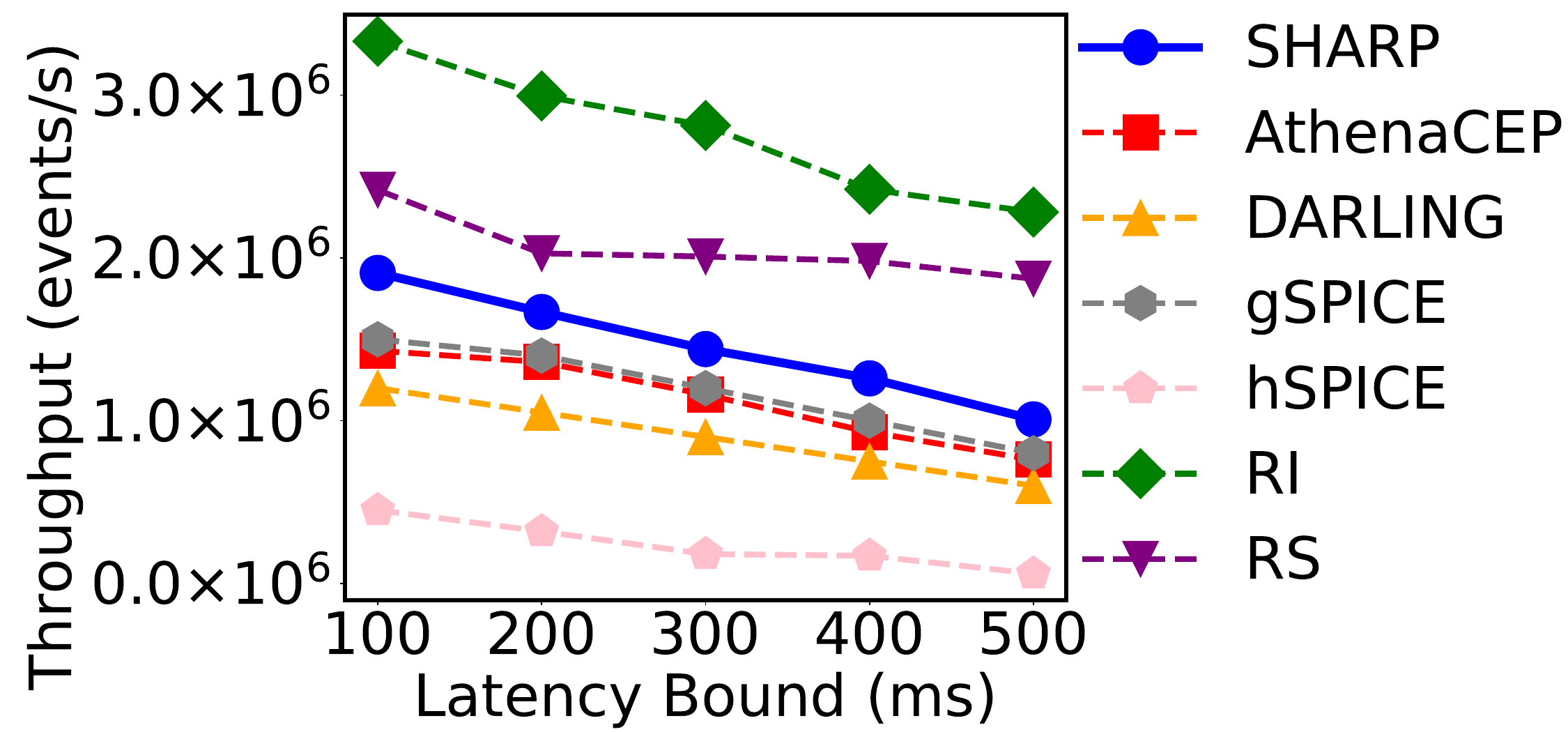}
      \caption{Throughput}
      \label{fig:citi_throughput}
    \end{subfigure}
    \caption{The overall performance of shared CEP patterns \code{P$_{3}$-P$_4$} over \code{Citi\_Bike}~\cite{bike}  at different latency bounds}
    \label{fig:citi_bike}
  \end{minipage}
\end{figure*}

\begin{figure*}[t]
	\centering
	\begin{minipage}[t]{0.4\linewidth}
		\centering
		\begin{subfigure}[t]{0.49\linewidth}
			\centering
			\includegraphics[width=\linewidth]{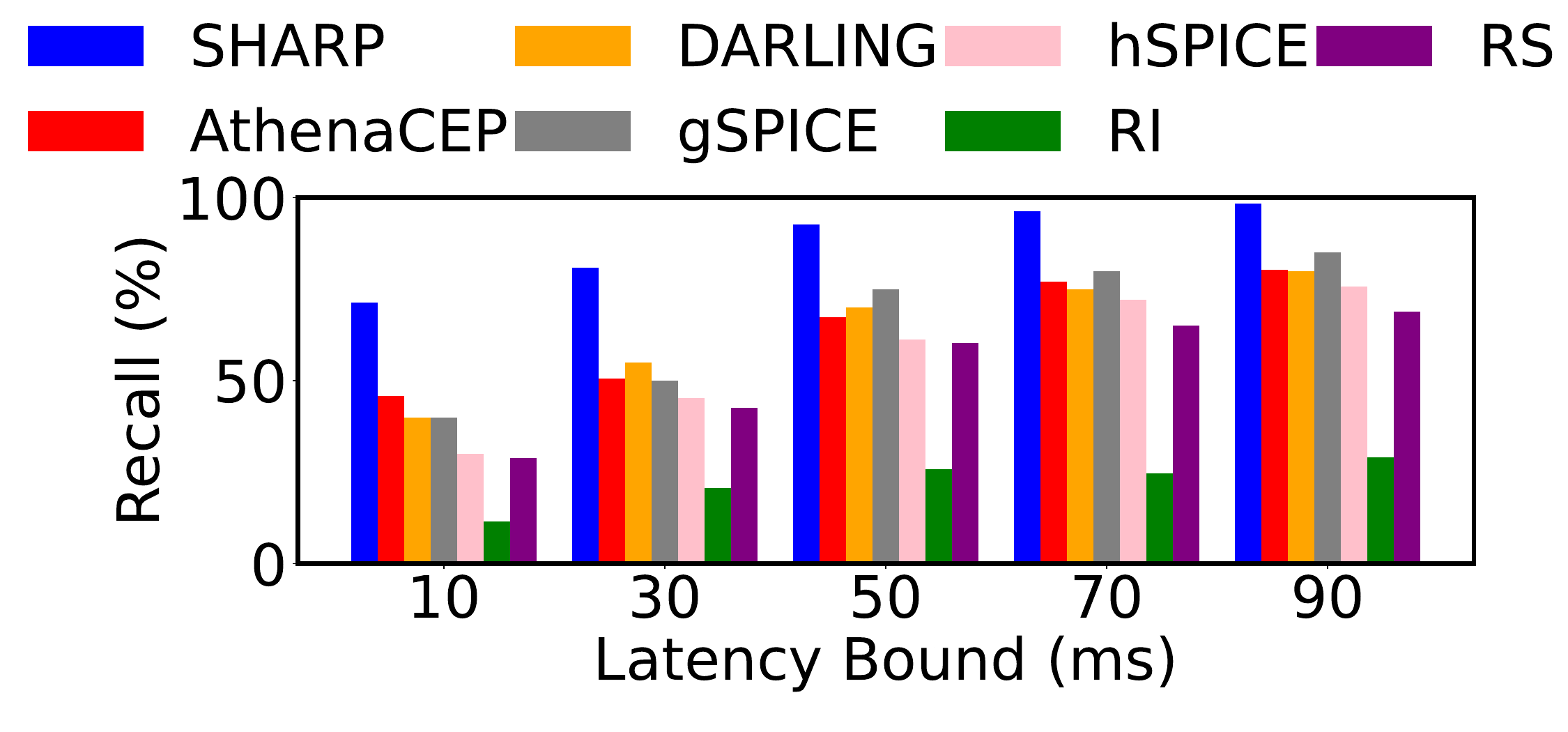}
			\caption{Recall}
			\label{fig:mr_recall}
		\end{subfigure}
		\hfill
		\begin{subfigure}[t]{0.49\linewidth}
			\centering
			\includegraphics[width=\linewidth]{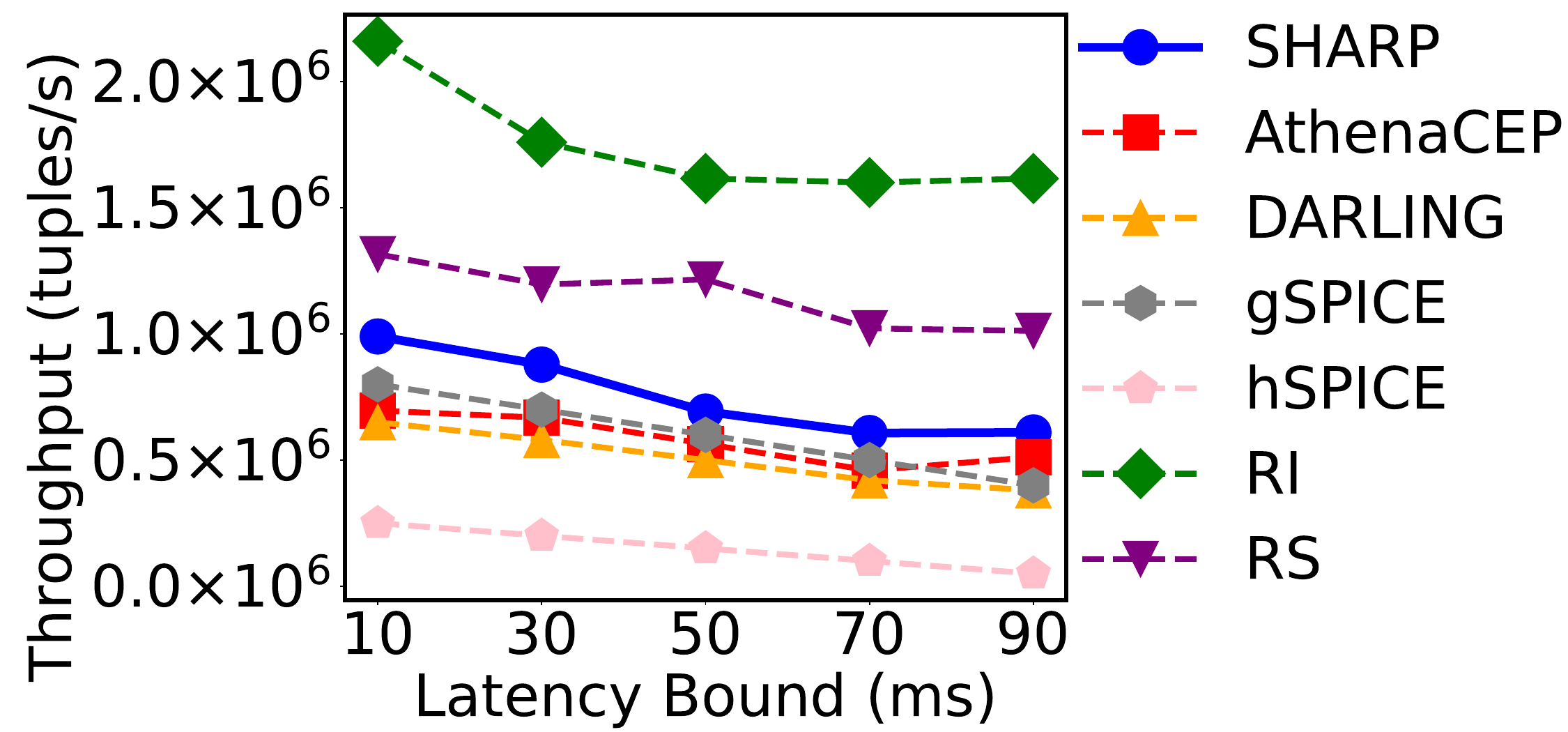}
			\caption{Throughput}
			\label{fig:mr_throughput}
		\end{subfigure}
		\caption{
			The overall performance of shared \texttt{\footnotesize MATCH\_RECOGNIZE} patterns \texttt{\footnotesize P$_{5}$-P$_6$} over \code{Crimes}~\cite{chicago_data} at different latency bounds}
		\label{fig:mreg}
	\end{minipage}
	\hfill
	\hspace{0.5em}
	\begin{minipage}[t]{0.58\linewidth}
		\centering
		\begin{subfigure}[t]{0.325\linewidth}
			\centering
			\includegraphics[width=\linewidth]{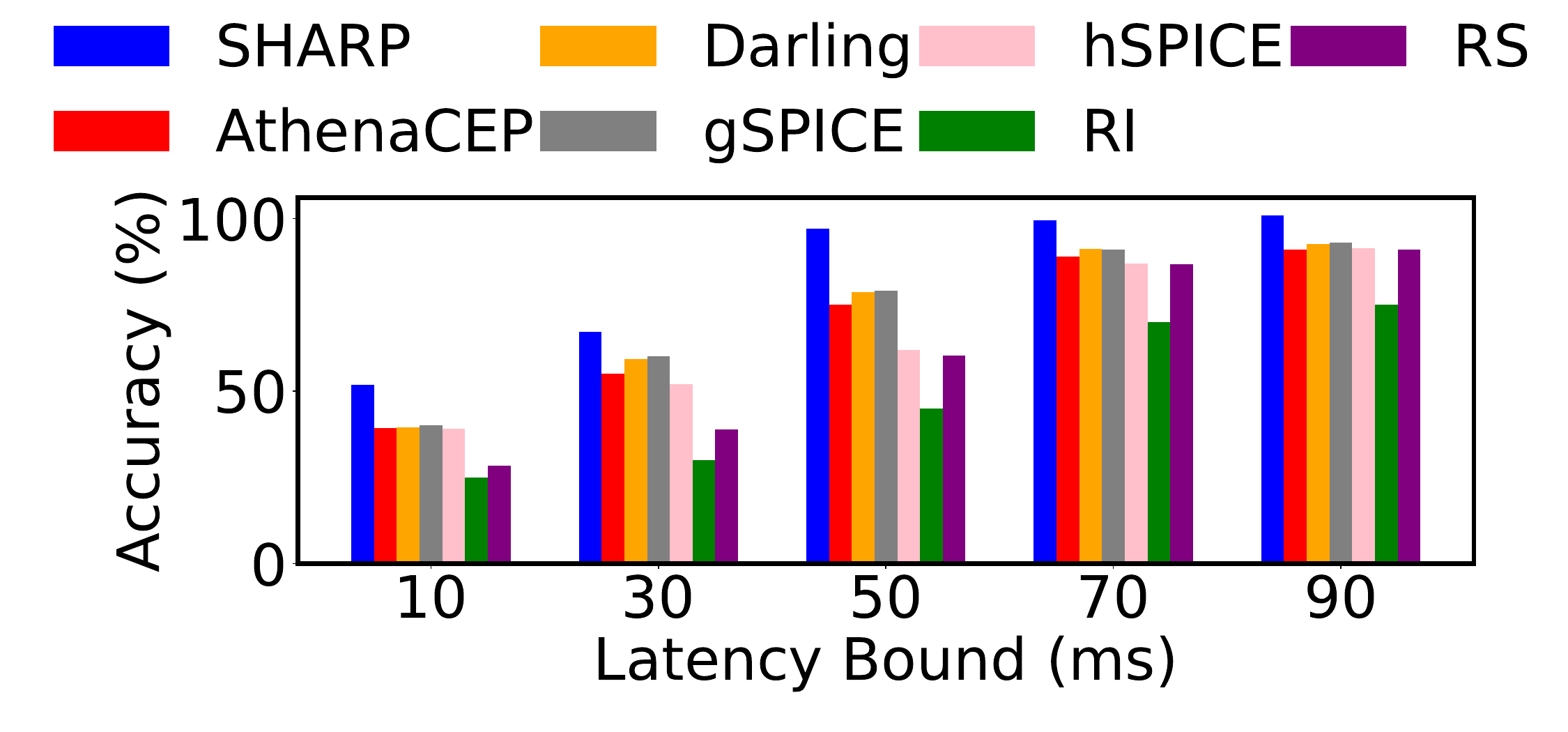}
			\caption{Accuracy}
			\label{fig:grag_acc}
		\end{subfigure}
		\hfill
		\begin{subfigure}[t]{0.325\linewidth}
			\centering
			\includegraphics[width=\linewidth]{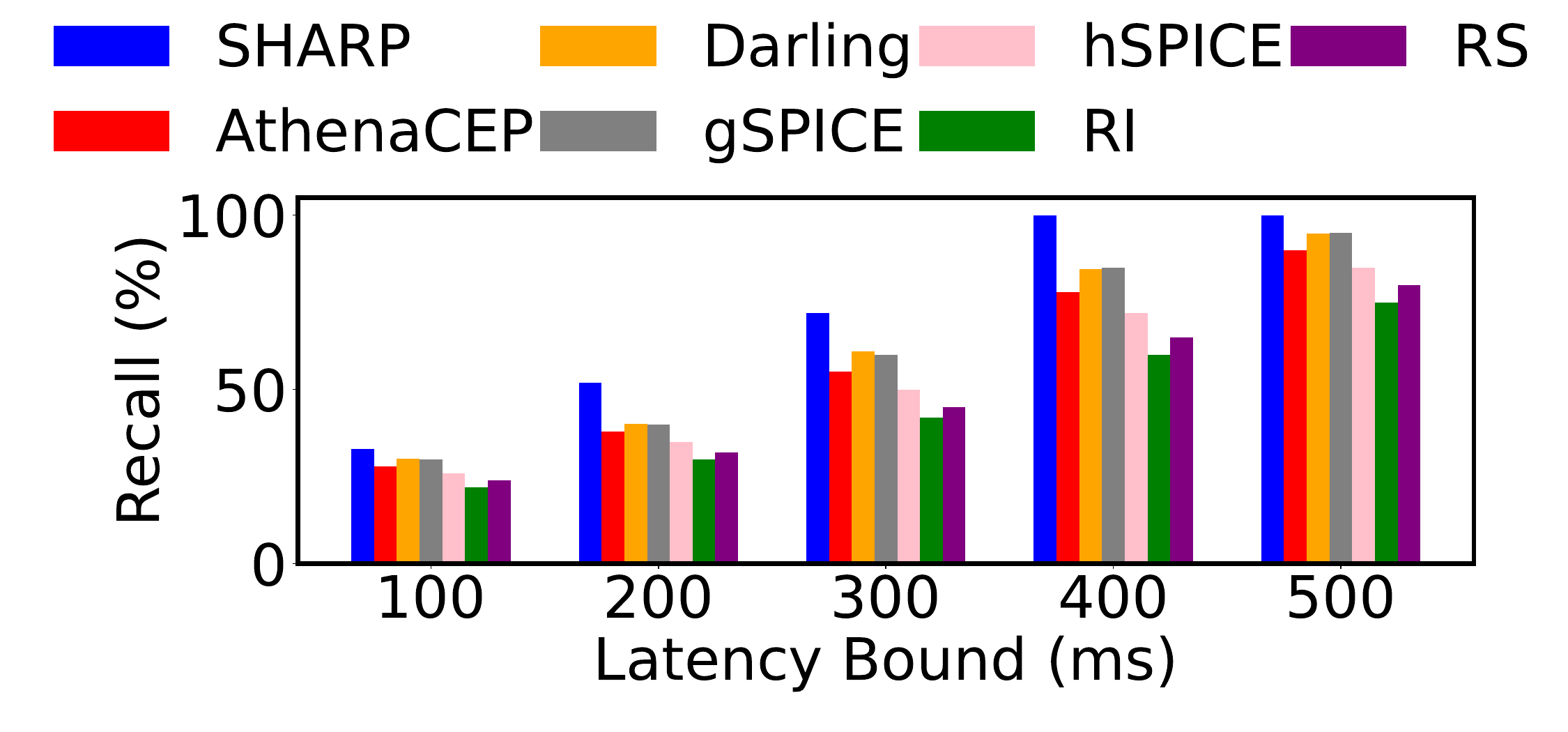}
			\caption{Recall}
			\label{fig:grag_recall}
		\end{subfigure}
		\hfill
		\begin{subfigure}[t]{0.325\linewidth}
			\centering
			\includegraphics[width=\linewidth]{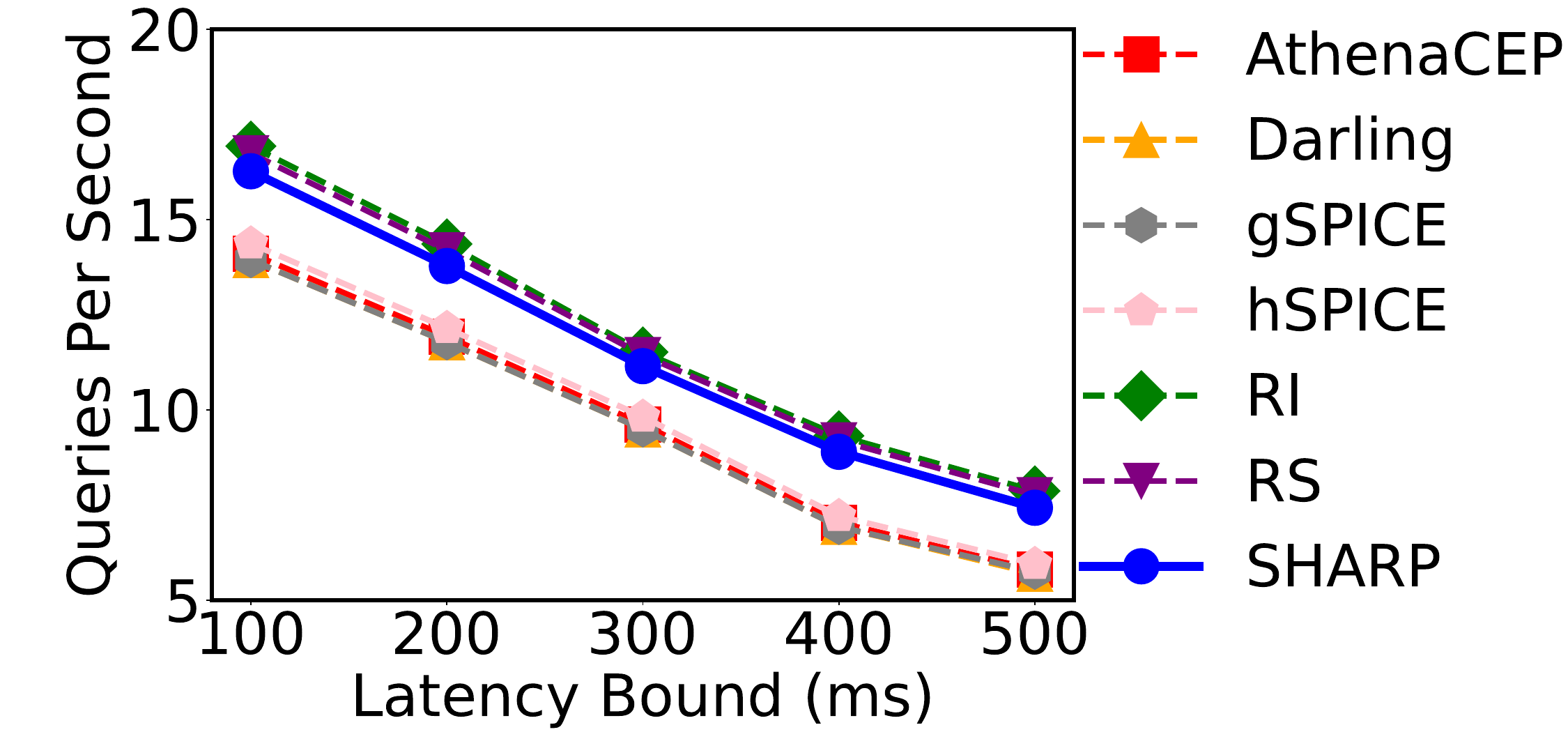}
			\caption{Throughput Improvement}
			\label{fig:grag_qps}
		\end{subfigure}
		\caption{The overall performance of GraphRAG on the \code{Meta-QA} benchmark~\cite{Zhang_Dai_Kozareva_Smola_Song_2018}  (\code{P$_{39}$-P$_{14,910}$}) at different latency bounds}
		\label{fig:grag}
	\end{minipage}
\end{figure*}

\tinyskip
\mypar{Baselines}
We compare \sys to six baselines:
\textbf{(1)} \code{Random input} (\code{RI}) selects input data randomly. 
\textbf{(2)}  \code{Random state} (\code{RS}) selects partial matches randomly. 
\textbf{(3)} \code{DARLING} \cite{chapnik2021darling} selects input data based on utility
and the queue buffer size. 
\textbf{(4)} \code{AthenaCEP} \cite{zhao2020load} selects the combination of
input data and partial matches to process based on its cost model. 
\textbf{(5)} \code{gSPICE} \cite{slo2023gspice} selects input data using a decision-tree-based black-box model trained on data properties.
\textbf{(6)} \code{hSPICE}~\cite{hSPICE} selects input data using a probabilistic model based on event type, position, and partial match state. \looseness = -2

\tinyskip
\mypar{Datasets} We have evaluated \sys and baseline approaches using two synthetic datasets and three real-world datasets:

\noindent
\textbf{(1)} \code{Synthetic Datasets} (\code{DS1} and \code{DS2}). 
\textbf{(i)} \code{DS1} contains tuples consisting of five uniformly-distributed attributes: a categorical type (\code{$\mathcal{U}(\{A, B,$}\code{$ C, D, E, F, G, H, I, J\})$}), a numeric \code{$ID$} (\code{$\mathcal{U}(1,10)$}), and numeric attributes \code{$X$}(\code{$\mathcal{U}(-90,90)$}), \code{$Y$}(\code{$\mathcal{U}(-180,180)$}), and \code{$V$} (\code{$\mathcal{U}(1, 3\times 10^6)$}).
\textbf{(ii)} \code{DS2} has similar settings, \ie a categorical type (\code{$\mathcal{U}(\{A,B,C,D,E,F\})$}), a numeric \code{$ID$} (\code{$\mathcal{U}(1,25)$}), and one numeric attribute \code{$X$} (\code{$\mathcal{U}(1,100)$}). \looseness=-1

\noindent
\textbf{(2)} \code{Citi\_Bike}~\cite{bike} is a publicly available dataset of bike trips in New York City. We use it for CEP patterns.

\noindent 
\textbf{(3)} \code{Crimes}~\cite{chicago_data} is a public crime record dataset
from the City of Chicago's Data Portal. We use it for \mreg patterns.

\noindent
\textbf{(4)} \code{KG-Meta-QA}~\cite{Zhang_Dai_Kozareva_Smola_Song_2018} is a knowledge graph that captures structured information of movies.
We use it for path query patterns in GraphRAG.

\tinyskip

\mypar{Patterns} We have evaluated 14,910 patterns, as shown in \T\ref{tab:queries}.
\textbf{\code{P$_1$-P$_6$}} are pattern \emph{templates} evaluated over synthetic and
real-world data by materializing the schema (\eg materializing \code{A}
in \code{P$_{3}$} with \code{bike\_trip}).
They cover three representative pattern-sharing schemes.
\code{P$_{1}$} and \code{P$_{2}$} share a Kleene closure sub-pattern
\code{SEQ(A,B$^+$)}.
\code{P$_{3}$} and \code{P$_{4}$} share the sub-pattern
\code{SEQ(A,B,C,D)} with computationally expensive predicates.
\code{P$_{5}$} and \code{P$_{6}$} share a negation pattern \code{SEQ(A,B,$\neg$C)}.
\textbf{\code{P$_7$–P$_{38}$}} evaluate \sys's scalability in terms of numbers of shared patterns, ranging from 2 to 32. These patterns are carefully designed to reflect a full spectrum of sharing schemes, as well as sequence overlap, Kleene closures, and negation operators.\looseness=-1
\textbf{\code{P$_{39}$-P$_{14,910}$}} evaluate the patterns in GraphRAG, taken from the 14,872 patterns in
Meta-QA benchmark~\cite{Zhang_Dai_Kozareva_Smola_Song_2018}.

\tinyskip

\mypar{Metrics} We measure the end-to-end performance of results quality and throughput for pattern matching under a range of strict latency bounds (wall-clock time).
We configure these latency bounds (in \S\ref{subsec: overall_effect}) based on the SLO of applications in real-world datasets and the latency without state reduction in synthetic datasets. 
The result quality is assessed in \textit{recall}--the ratio of complete matches obtained with state reduction to all complete matches derived without it. We omit the accuracy of CEP and \mreg patterns, because they always output accurate results of complete matches (accuracy=100\%). 
For GraphRAG, we measure the end-to-end performance of the entire pipeline in \F\ref{fig:sharp_flow} \myc{1}-\myc{6}, not only the pattern matching in the KG.
Its recall is the ratio of correct responses of LLM compared to the ground truth, while the accuracy is the ratio of correct answers to all LLM-generated responses.
For the \textit{throughput}, we report \emph{events or tuples
per second} for CEP and \mreg, and the \emph{LLM queries per second} (QPS) for GraphRAG. \looseness=-1

\begin{table}[t]
  \caption{14,910 patterns evaluated in the experiments}
  \resizebox{\columnwidth}{!}{
    \centering
    \begin{tabular}{ll}
        \toprule
        $\mathbf{P_1}$ & \texttt{SEQ(A a, B+ b[], C c, D d)} \\
        & \texttt{WHERE SAME [ID] AND} $\ \texttt{SUM}(\texttt{b[i].x}) < \texttt{c.x}$ \\

        $\mathbf{P_2}$ & \texttt{SEQ(A a, B+ b[], E e, F f)} \\
        & \texttt{WHERE SAME [ID] AND} $\ \texttt{a.x + SUM(b[i].x)} < \texttt{e.x + f.x}$ \\

        $\mathbf{P_3}$ & \texttt{SEQ(A a, B b, C c, D d, E e, F f, G g)} \\
        & \texttt{WHERE SAME [ID] AND} $\ a.v < b.v\ \texttt{AND}\ b.v + c.v < d.v$ \\
        & \texttt{AND} $\ 2r \cdot \texttt{arcsin}\left(\texttt{sin}^{2}\left(\frac{e.x - d.x}{2}\right) + \texttt{cos(d.x)cos(e.x)} \cdot \texttt{sin}^2\left(\frac{e.y - d.y}{2}\right)\right)^{1/2} \leq f.v$ \\

        $\mathbf{P_4}$ & \texttt{SEQ(A a, B b, C c, D d, H h, I i, J j)} \\
        & \texttt{WHERE SAME [ID] AND} $\ a.v < b.v\ \texttt{AND}\ b.v + c.v < d.v$ \\
        & \texttt{AND} $\ r \cdot \texttt{arccos}\left( \texttt{sin(d.x)sin(h.x) + cos(d.x)cos(h.x)cos(h.y - d.y)} \right) \leq i.v$ \\

        $\mathbf{P_5}$ & \texttt{SEQ(A a, B b, !C c, D d)} \\
        & \texttt{WHERE SAME [ID] AND} $\ a.x < b.x$ \\

        $\mathbf{P_6}$ & \texttt{SEQ(A a, B b, !C c, E e)} \\
        & \texttt{WHERE SAME [ID]} \\

        $\mathbf{P_{7}}$ & \texttt{SEQ(A, B, C)} \\
        $\mathbf{P_{8}}$ & \texttt{SEQ(A, B, E)} \\
        $\mathbf{P_{9}}$ & \texttt{SEQ(A, !E, C)} \\
        $\mathbf{P_{10}}$ & \texttt{SEQ(A, !E, D)} \\
        $\mathbf{P_{11}}$ & \texttt{SEQ(A, B+, C)} \\
        $\mathbf{P_{12}}$ & \texttt{SEQ(A, B+, D)} \\
        $\mathbf{P_{13}}$ & \texttt{SEQ(A, B, B, C)} \\
        $\mathbf{P_{14}}$ & \texttt{SEQ(A, C, D)} \\
        $\mathbf{P_{15}}$ & \texttt{SEQ(A, B, C, D)} \\
        $\mathbf{P_{16}}$ & \texttt{SEQ(A, B+, E)} \\
        $\mathbf{P_{17}}$ & \texttt{SEQ(A, !B, C)} \\
        $\mathbf{P_{18}}$ & \texttt{SEQ(A, !C, D)} \\
        $\mathbf{P_{19}}$ & \texttt{SEQ(A, B, D, E)} \\
        $\mathbf{P_{20}}$ & \texttt{SEQ(A, C, B, D)} \\
        $\mathbf{P_{21}}$ & \texttt{SEQ(A, !B, D, E)} \\
        $\mathbf{P_{22}}$ & \texttt{SEQ(A, B+, C, D)} \\

        $\mathbf{P_{23}}$ & \texttt{SEQ(A, G, H, I)} \\
        $\mathbf{P_{24}}$ & \texttt{SEQ(A, G, H+, J)} \\
        $\mathbf{P_{25}}$ & \texttt{SEQ(A, G, !I, J)} \\
        $\mathbf{P_{26}}$ & \texttt{SEQ(A, G, I, J, A)} \\
        $\mathbf{P_{27}}$ & \texttt{SEQ(A, G, J, H, B)} \\
        $\mathbf{P_{28}}$ & \texttt{SEQ(A, G, !H, J, C)} \\
        $\mathbf{P_{29}}$ & \texttt{SEQ(A, H, H, I)} \\
        $\mathbf{P_{30}}$ & \texttt{SEQ(A, G, H+, I, J)} \\
        $\mathbf{P_{31}}$ & \texttt{SEQ(A, G, A, B)} \\
        $\mathbf{P_{32}}$ & \texttt{SEQ(A, G, !J, C)} \\
        $\mathbf{P_{33}}$ & \texttt{SEQ(A, H, !J, D)} \\
        $\mathbf{P_{34}}$ & \texttt{SEQ(A, G, I, !H, E)} \\
        $\mathbf{P_{35}}$ & \texttt{SEQ(A, G, H, I, J, F)} \\
        $\mathbf{P_{36}}$ & \texttt{SEQ(A, J, G, I)} \\
        $\mathbf{P_{37}}$ & \texttt{SEQ(A, G, I+, A)} \\
        $\mathbf{P_{38}}$ & \texttt{SEQ(A, G, J, B+)} \\
        \addlinespace[0.5em]

        $\mathbf{P_{39}-P_{14,910}}$ & \text{\large \emph{14,872} queries from the Meta-QA benchmark~\cite{Zhang_Dai_Kozareva_Smola_Song_2018}.} \\
        \bottomrule
    \end{tabular}
  }
  \label{tab:queries}
\end{table}

\subsection{Overall Effectiveness and Efficiency}
\label{subsec: overall_effect}

We first investigate the overall performance of \sys in {\small CEP}, \code{MATCH\_}  \code{RECOGNIZE} and {\small GraphRAG} using synthetic and real-world datasets.

We execute the shared CEP patterns, \code{P$_3$} and \code{P$_4$},  over \code{DS1}.
\F\ref{fig:overall} demonstrates the results.
The latency without state reduction is 1035\unit{ms}. We set the latency bound ranging from 100\unit{ms} to 900\unit{ms}.
At all latency bounds, \sys achieves the highest recall value across all baselines ( \F\ref{fig:recall}), achieving over 95\% at 500--900\unit{ms}.
The margin becomes larger at tighter latency bounds.
At 100\unit{ms}, \sys achieves 70\% recall,  1.96$\times$, 4.10$\times$, 1.81$\times$, 4.3$\times$, 5.30$\times$, and 11.25$\times$ higher than \code{AthenaCEP}, \code{DARLING}, \code{gSPICE}, \code{hSPICE}, \code{RS} and \code{RI}. \looseness=-1

A similar trend is observed in the real-world \code{Citi\_Bike} dataset, where latency bounds (100–500\unit{ms}) align with SLOs in bike-sharing applications.~\cite{slo-example, pragmaticsre-slo-examples}.
As showed in \F\ref{fig:citi_recall}, \sys outperforms all baselines, improves the recall by $3.5\times$ (\code{Athena}\code{CEP}), $4.0\times$(\code{DARLING}), 2.4$\times($\code{gSPICE}), 2.8$\times($\code{hSPICE}), $2.8\times$(\code{RS}) and $7 \times$(\code{RI}).

We attribute \sys's high recall to the combination of 
\emph{pattern-sharing degree} (\S\ref{sec:clustering}) and the
cost model (\S\ref{sec:cost_model}) which capture both pattern-sharing
schemes and the cost of state, \ie partial matches.
\code{AthenaCEP} and \code{DARLING} do not consider shared patterns, resulting in lower recall. 
Although \code{hSPICE} and \code{gSPICE} consider multiple patterns, they ignore the interaction and interference among patterns in shared states, which again leads to lower recall. \looseness=-1

We then examine the throughput performance (\F\ref{fig:throughput} and
\F\ref{fig:citi_throughput}). \sys's throughput is higher than \code{AthenaCEP}
(1.8$\times$), \code{DARLING} (1.69$\times$), \code{gSPICE} (1.73$\times$), 
and \code{hSPICE} (2.1$\times$),
but lower than the random approaches. 
However, the high throughput of
\code{RS} and \code{RI} comes at the expense of poor recall (below 20\%). 
 Compared to all baselines, \sys strikes a better trade-off of recall and throughput.

\sys's superior performance stems from the PSD design (\S\ref{sec:clustering}), the cost model (\S\ref{sec:cost_model}) and efficient hierarchical state selection (\S\ref{sec:shedding}).  
PSD enables efficient separation of non-latency-violated state and the priority of pattern share degree in the shared patterns, which significantly reduces the search space compared to \code{AthenaCEP}, \code{DARLING}, \code{gSPICE}, and \code{hSPICE}. While the greedy selection in latency-violated state buffers further reduces and overlaps  \sys's overhead. \looseness=-1

Similar trends are observed in \mreg patterns and the GraphRAG application.
\F\ref{fig:mreg} shows the results of executing shared \mreg patterns, \code{P$_5$} and \code{P$_6$}, over  \texttt{Crimes}~\cite{chicago_data}, ranging the latency bounds from 10\unit{ms} to 90\unit{ms} (crime detection requires a sub-100\unit{ms} latency~\cite{sajjanar2025real,mishra2024abnormal}).
Here,  \sys outperforms all baselines in recall (\F\ref{fig:mr_recall}). At the 10\unit{ms} bound, \sys achieves the recall of 74\% that is $1.2\times$, $2.56\times$, 1.85$\times$, 1.84$\times$, $2.0\times$, and $3.7\times$ higher than \code{AthenaCEP}, \code{DARLING}, \code{gSPICE}, \code{hSPICE}, \code{RS} and \code{RI}. \sys's throughput again falls between that of the existing techniques (\code{AthenaCEP}, \code{DARLING}, \code{gSPICE}, and \code{hSPICE}) and the random approaches (\F\ref{fig:mr_throughput}).
  
For GraphRAG experiments, \sys executes the entire processing pipeline in \F\ref{fig:basic_flow} over \code{Meta-QA}~\cite{Zhang_Dai_Kozareva_Smola_Song_2018}. The average (14,872 LLM queries) end-to-end latency is 1,032\unit{ms} while the average KG pattern matching latency is 635\unit{ms}.
\F\ref{fig:grag} presents the average end-to-end results, varying the latency bound of KG pattern matching from 100\unit{ms} to 500\unit{ms} to align with the SLOs of information retrieval applications~\cite{google-slo-guide}.
Here, \sys outperforms the baselines in both accuracy and recall.
It keeps 100\% accuracy for latency bounds of 400\unit{ms} and 500\unit{ms}, and achieves 70\% accuracy at the tighter latency bound of 200\unit{ms} \ie 1.49$\times$, 1.29$\times$, 1.27$\times$, 1.56$\times$, 1.30$\times$, 1.78$\times$ and 2.1$\times$ higher than \code{AthenaCEP}, \code{DARLING}, \code{gSPICE}, \code{hSPICE}, \code{RS} and \code{RI}.
For the recall, \sys's performance margin becomes smaller at tighter latency bounds. At 100\unit{ms} bound, the recall values are comparable with baselines (see \F\ref{fig:grag_recall}).
Because the recall depends on the statistical efficiency of LLM-generated responses--a small set of responses will not cover the majority of ground truth. However, \sys's state selector ensures that over 80\% of the generated responses align the ground truth.
As for the throughput, \sys performs closely to \code{RS} and \code{RI}, and 1.41$\times$ higher than other baselines (\F\ref{fig:grag_qps}). This demonstrates \sys's low overhead even in complicated processing pipelines. 
\vspace{-1.2em}

 \begin{figure}[t]
 	\centering
 	\begin{subfigure}[b]{0.495\linewidth}
 		\centering
 		\includegraphics[width=\linewidth]{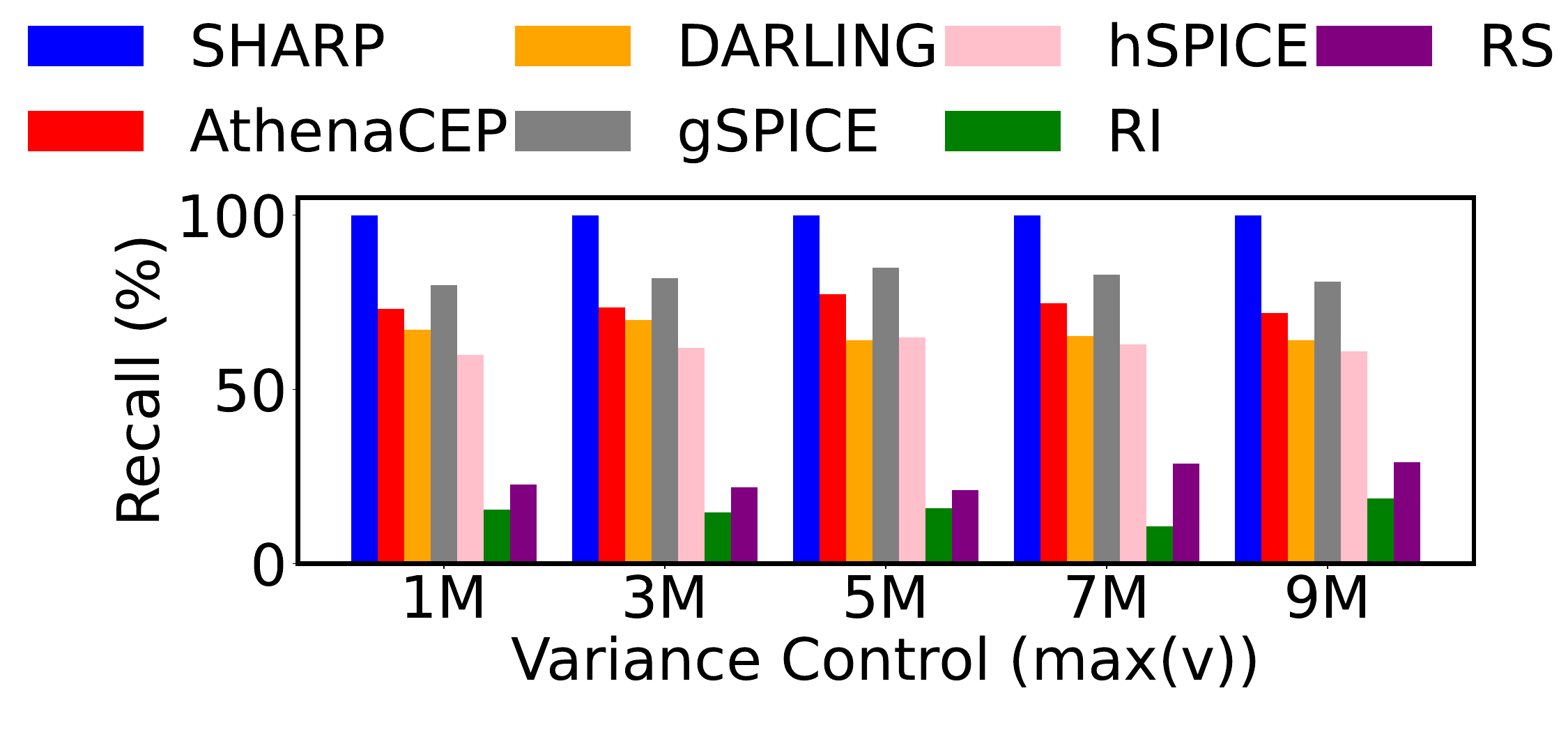}
 		\caption{CEP Recall}
 		\label{fig:query_sensitivity_recall}
 	\end{subfigure}
 	\hfill
 	\begin{subfigure}[b]{0.495\linewidth}
 		\centering
 		\includegraphics[width=\linewidth]{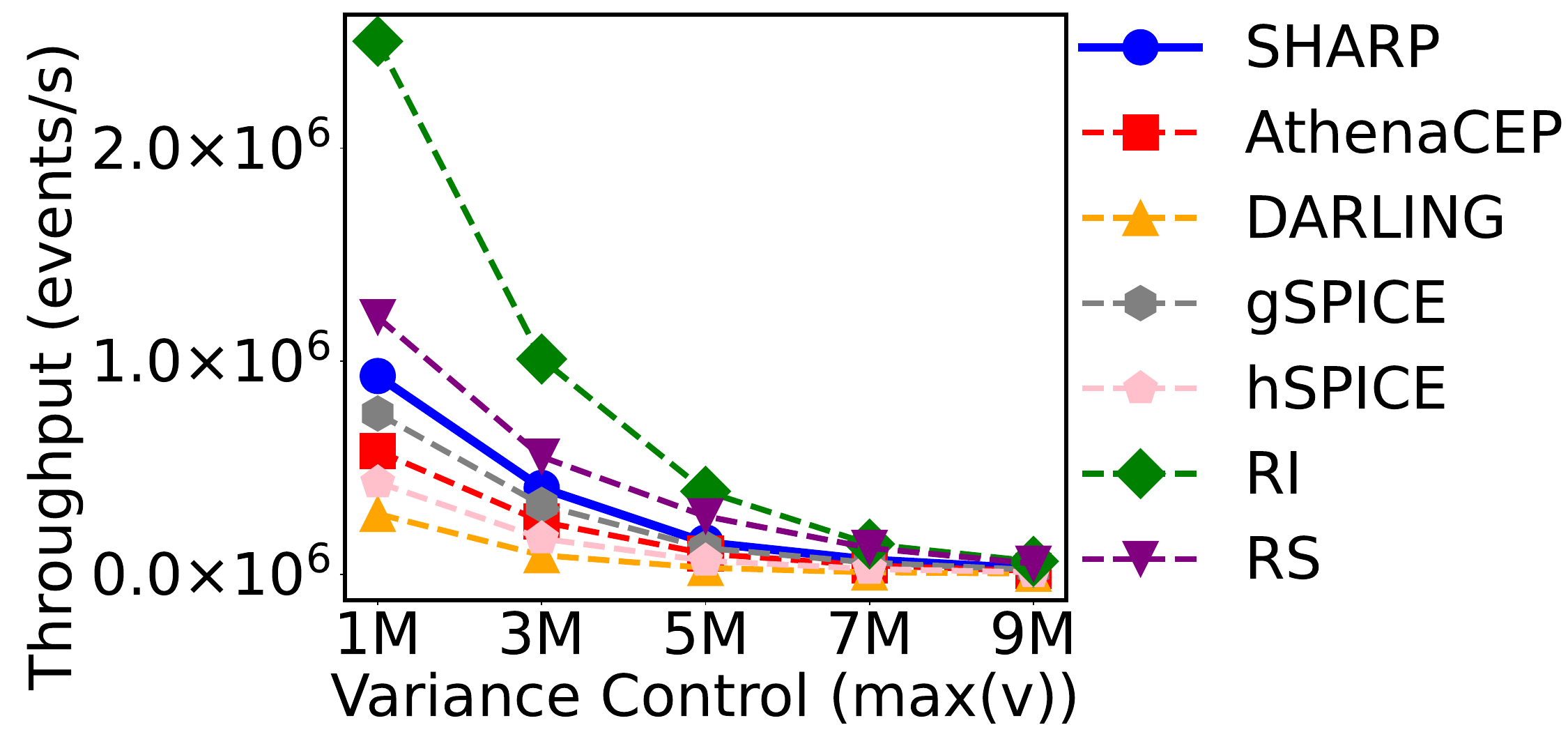}
 		\caption{CEP Throughput}
 		\label{fig:query_sensitivity_th}
 	\end{subfigure}

\medskip

 	\begin{subfigure}[b]{0.495\linewidth}
 		\centering
 		\includegraphics[width=\linewidth]{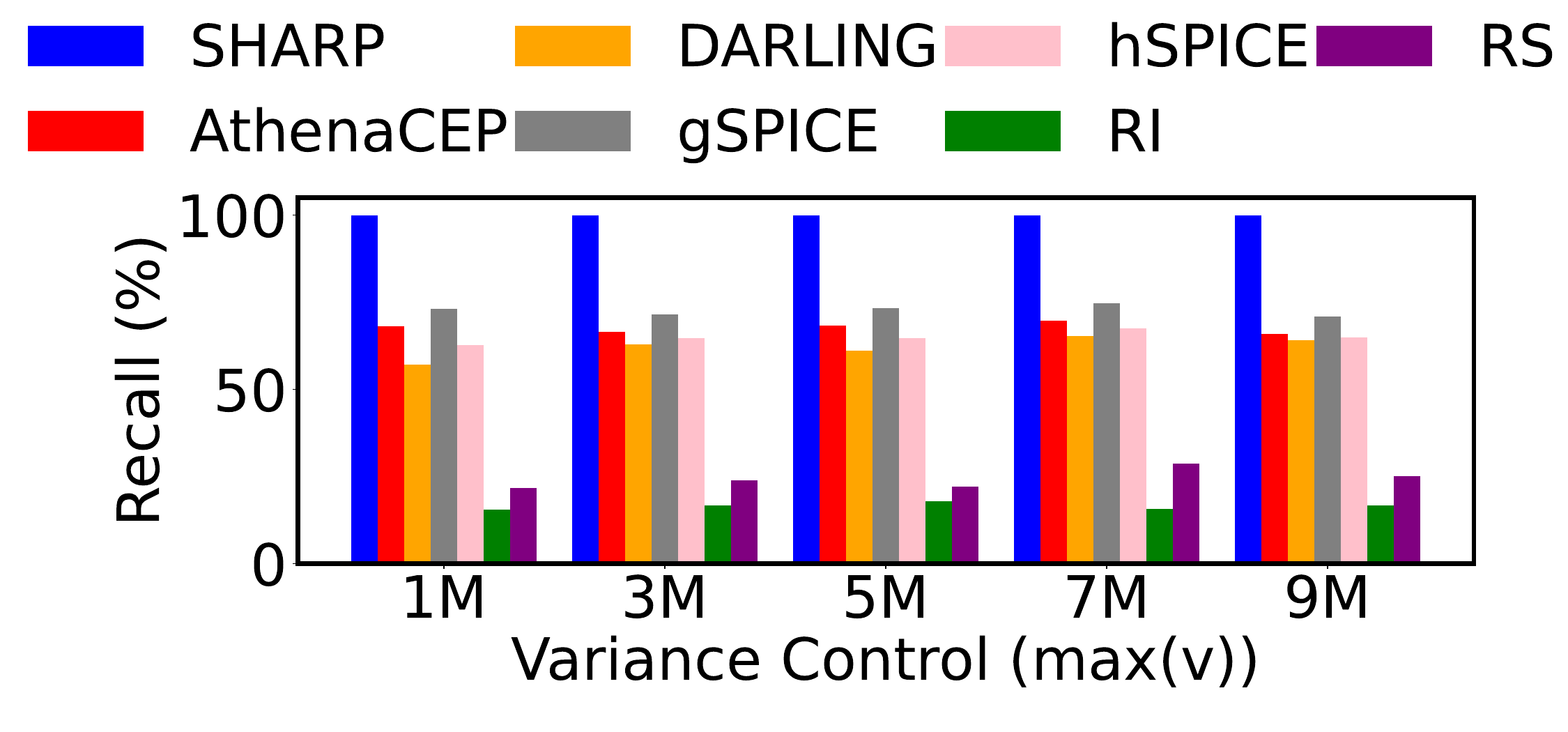}
 		\caption{\mreg Recall}
 		\label{fig:query_sensitivity_recall_mr}
 	\end{subfigure}
 	\hfill
 	\begin{subfigure}[b]{0.495\linewidth}
 		\centering
 		\includegraphics[width=\linewidth]{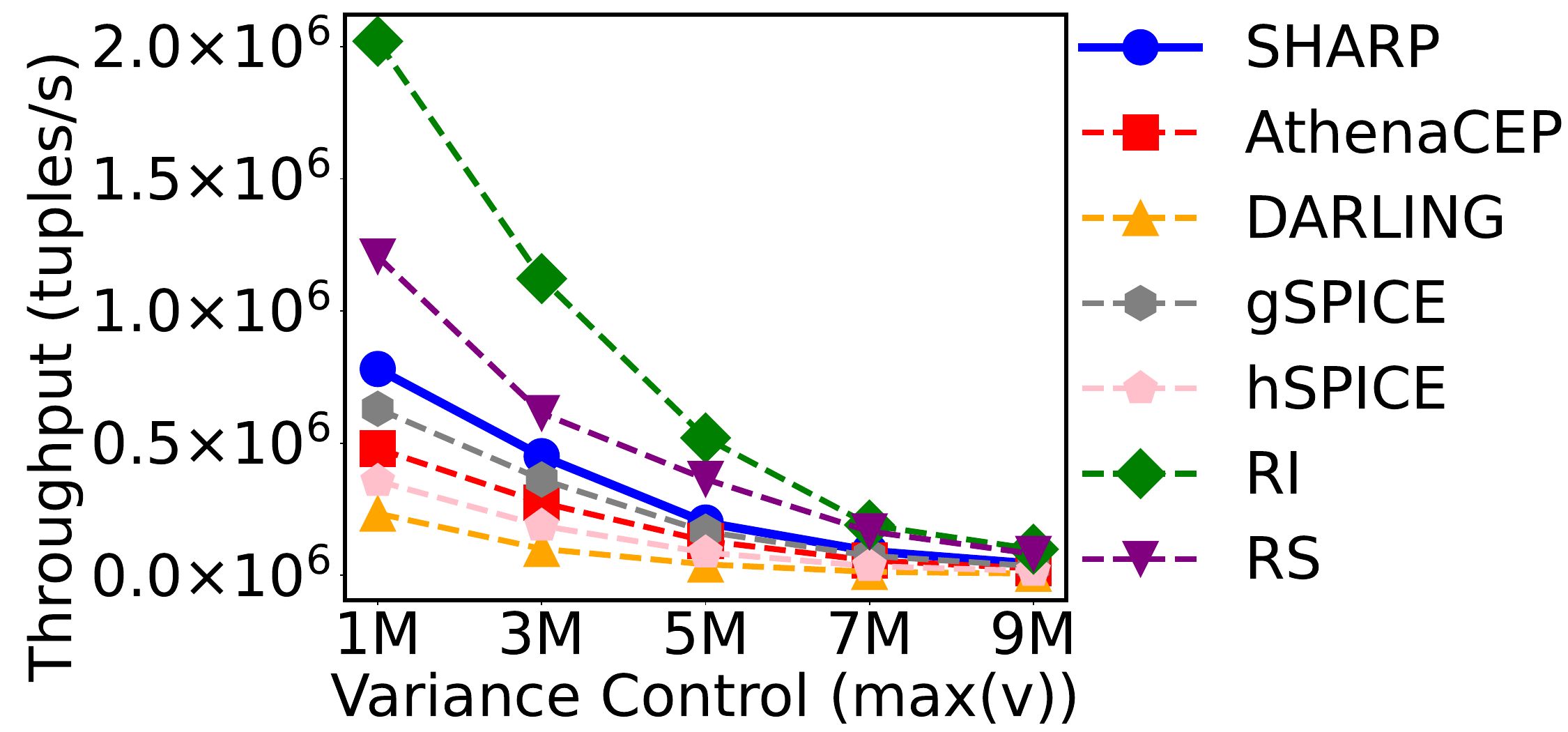}
 		\caption{\mreg Throughput}
 		\label{fig:query_sensitivity_recall_mr_th}
 	\end{subfigure}
 	\caption{
 		Impact of selectivity on shared patterns {\small \texttt P$_{3}$-P$_4$} over {\small \texttt DS1} \looseness=-1}
 	\label{fig:query_sensitivity}
 \end{figure}

\subsection{Sensitivity Analysis of Pattern Properties}
\label{sec:exe_sens}

Next, we examine \sys's sensitivity and robustness to various pattern properties,
considering selectivity, pattern length, and the time window size, 
because these properties affect the size of the state that changes in runtime.

\mypar{Selectivity} Pattern predicates select data and partial matches.
We control the selectivity by changing the value distribution of $V$ in
\code{DS1}. This affects \code{P$_3$} and \code{P$_4$} in
CEP and \code{P$_5$} and \code{P$_6$} in \mreg.
In particular, we change the distribution of $V$ from
$\mathcal{U}$(0, 1$\times$10$^6$) to $\mathcal{U}$(0, 1$\times$10$^9$),
increasing the selectivity for \code{P$_3$-P$_6$}.

\F\ref{fig:query_sensitivity_recall} shows that \sys keeps stable recall of 100\%
across all selectivity configurations, outperforming baselines.  In
contrast, the recall of baselines changes with the selectivity. For
instance, the recall of \code{AthenaCEP} drops by 10\% when \code{$V$}'s distribution
changes
from \code{$\mathcal{U}$}(0, 5$\times$10$^6$) to \code{$\mathcal{U}$}(0, 9$\times$10$^6$).
Specifically, the recall of \code{AthenaCEP} and \code{DARLING} drops by 10\% and 5\%.
The increased selectivity results in more partial matches, which lowers the throughput for all methods (\F\ref{fig:query_sensitivity_th}
The increasing selectivity generates more partial matches and therefore
results in lower throughput for \sys and all baselines
(\F\ref{fig:query_sensitivity_th}). We attribute \sys's robustness to its cost model, which efficiently adapts to changes in  selectivity.
\looseness=-1

\begin{figure}[t]
  \centering
  \begin{subfigure}[b]{0.495\linewidth}
    \centering
    \includegraphics[width=\linewidth]{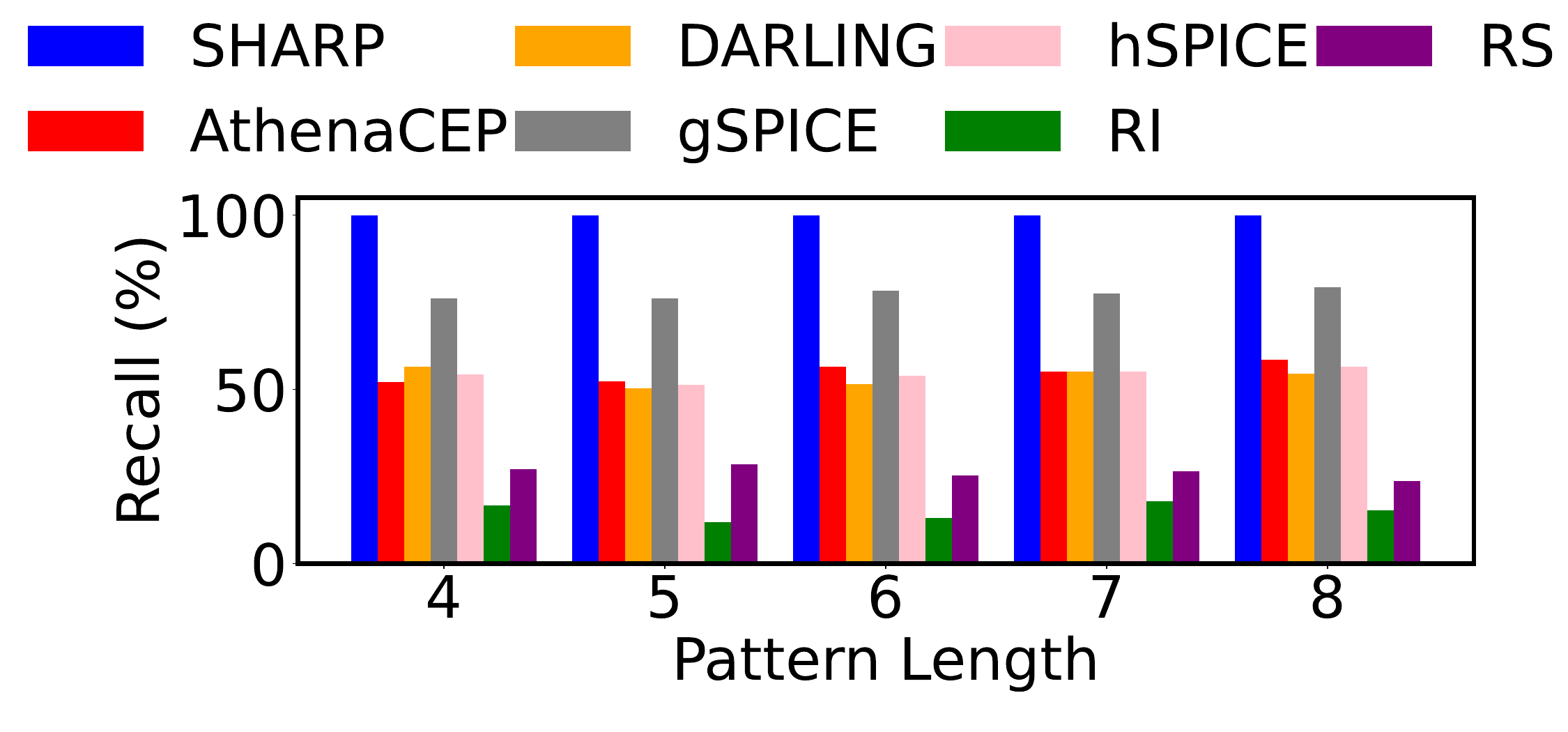}
    \caption{CEP Recall}
    \label{fig:patten_len_recall}
  \end{subfigure}
  \hfill
  \begin{subfigure}[b]{0.495\linewidth}
    \centering
    \includegraphics[width=\linewidth]{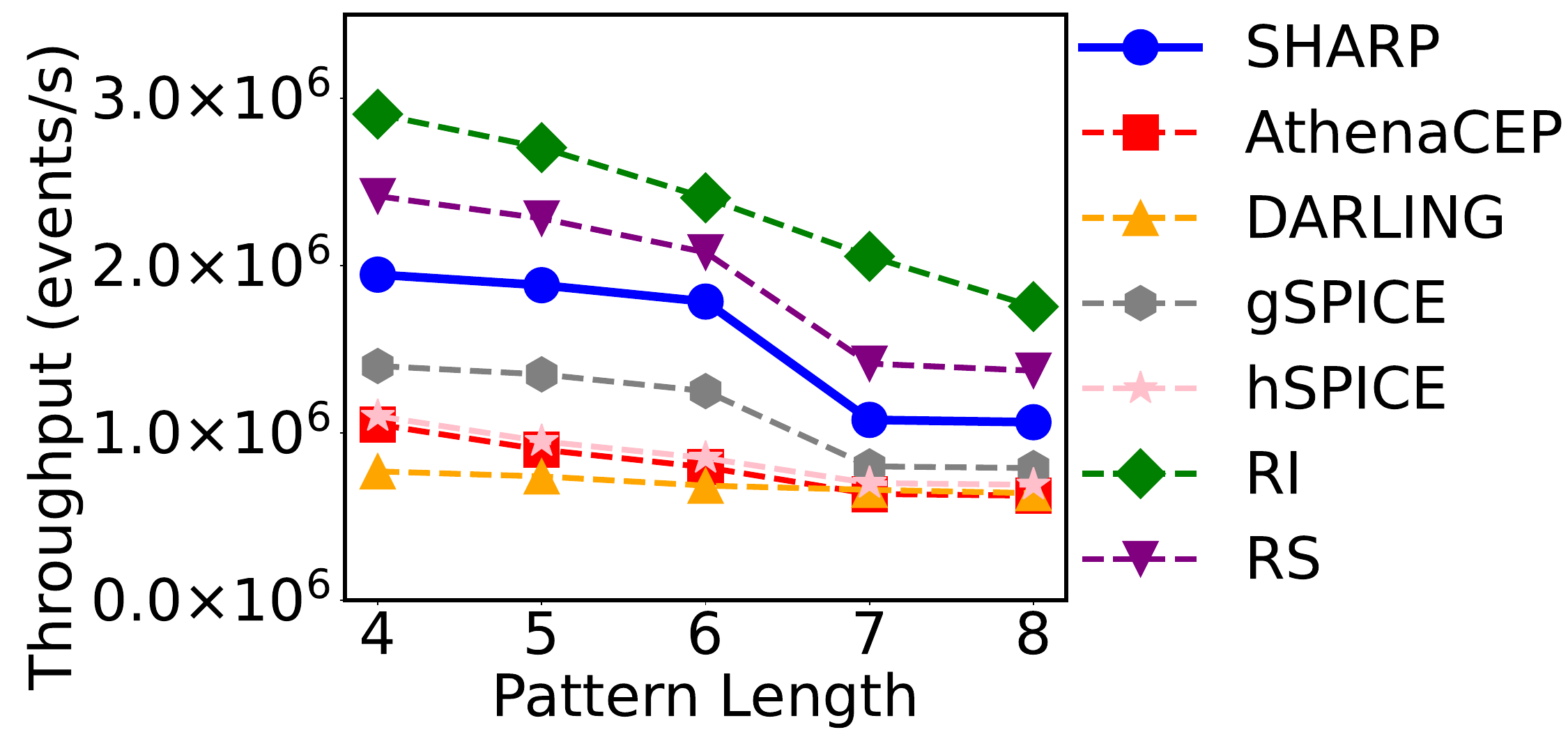}
    \caption{CEP Throughput}
    \label{fig:patten_len_th}
  \end{subfigure}

\medskip

  \begin{subfigure}[b]{0.495\linewidth}
    \centering
    \includegraphics[width=\linewidth]{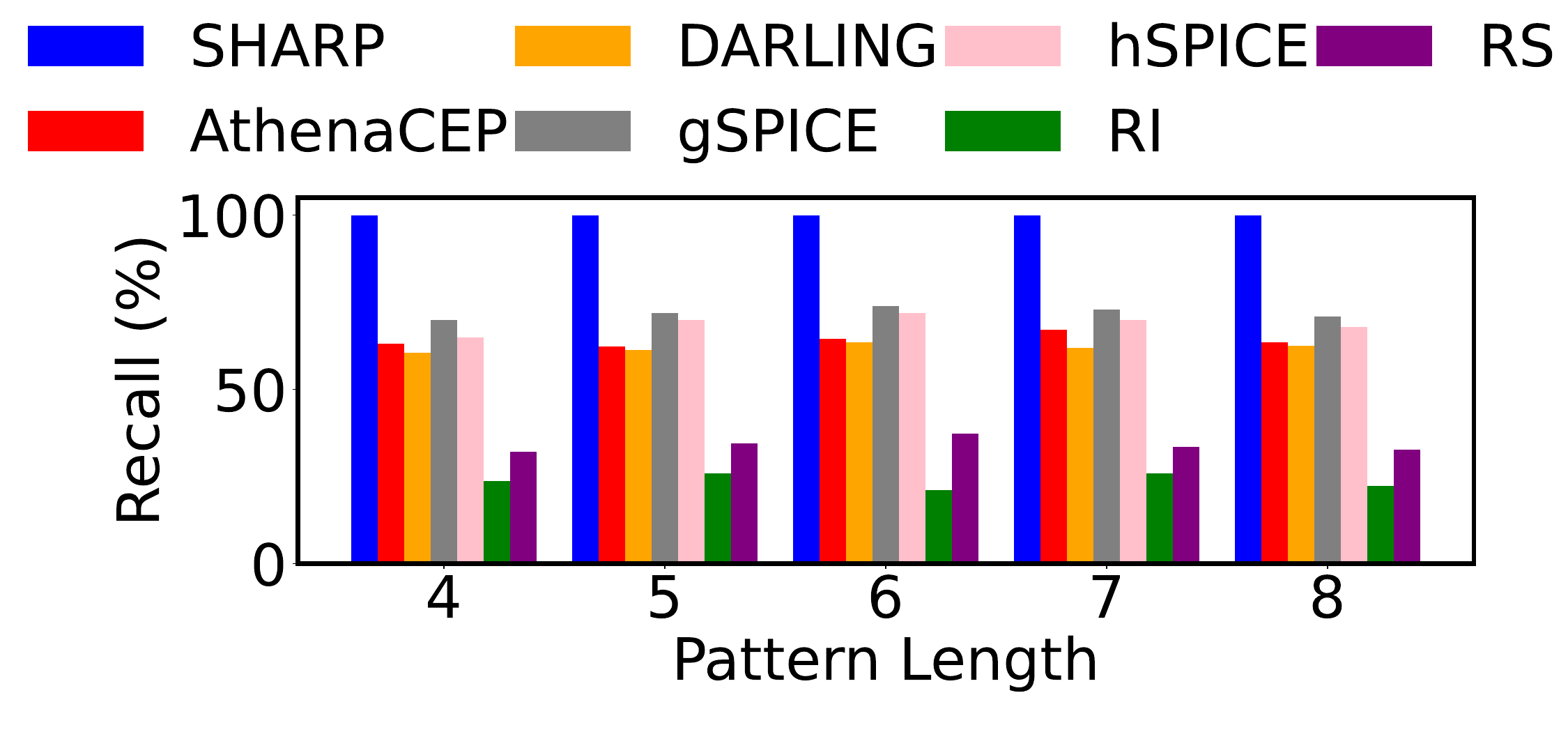}
    \caption{\mreg Recall}
    \label{fig:patten_len_recall_mr}
  \end{subfigure}
  \hfill
  \begin{subfigure}[b]{0.495\linewidth}
    \centering
    \includegraphics[width=\linewidth]{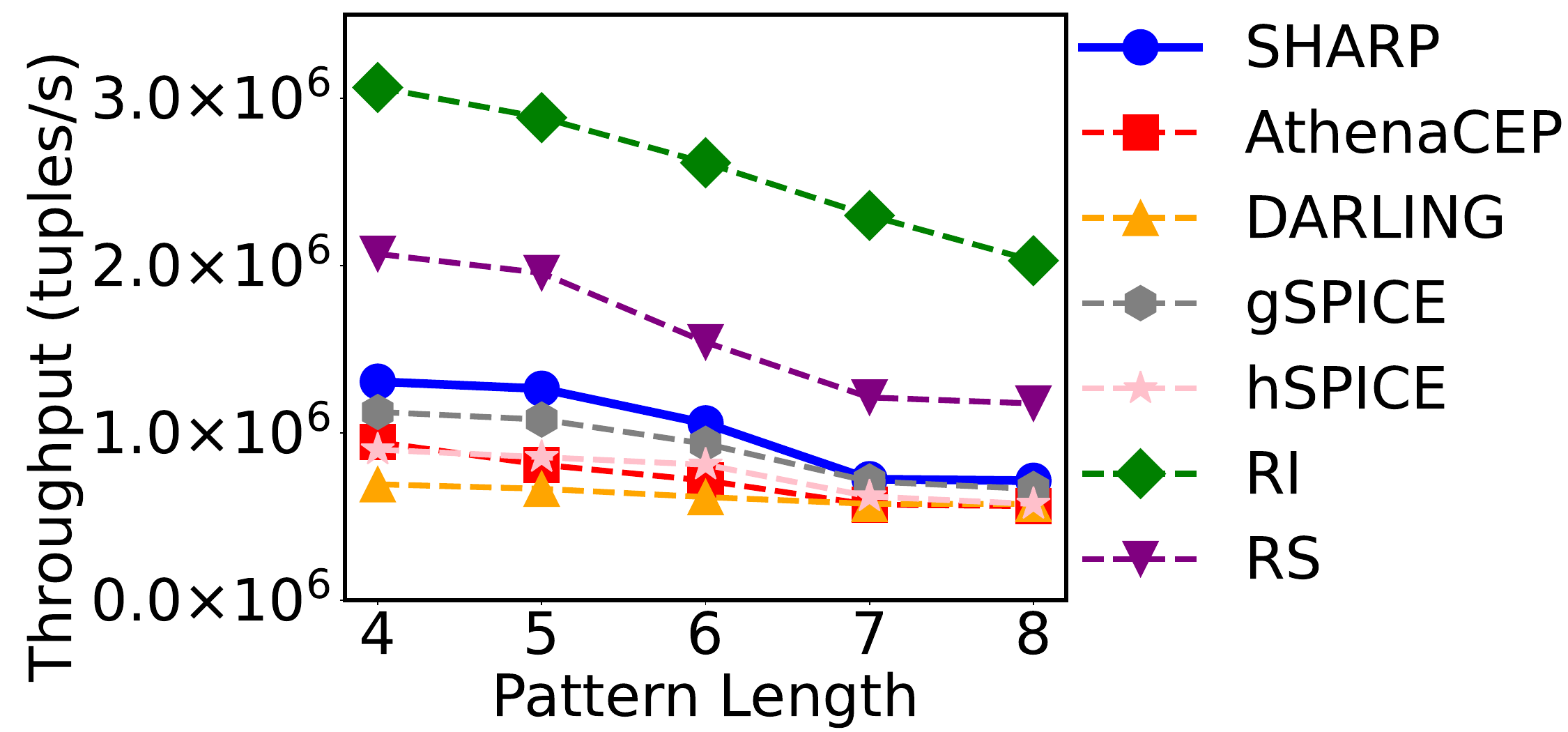}
    \caption{\mreg Throughput}
    \label{fig:patten_len_mr_th}
  \end{subfigure}
  \caption{
  	 Impact of pattern length 
on shared patterns \code{P$_{1}$-P$_2$} over \code{DS2}}
  \label{fig:pattern_len}
\end{figure}

\mypar{Pattern length} We control the pattern length of \code{P$_1$} and \code{P$_2$}, ranging from four to eight by changing the length of the Kleene closure, \code{B+}.
The patterns are executed for both CEP and \mreg over \code{DS1}.
\F\ref{fig:pattern_len} shows that the recall of \sys is not affected by pattern length (stable in 100\% recall), consistently outperforming baselines. In contrast, the recall fluctuates by 5.4\%, 4.7\%, 3.2\%, 7.8\%, 10.1\%, and 9.8\% for \code{AthenaCEP}, \code{DARLING}, \code{gSPICE}, \code{hSPICE}, \code{RI} and \code{RS}, respectively.
The increased pattern length leads to lower throughput due to more generated partial matches.
However,  \sys still outperforms all non-random baselines in throughput, \ie 25.9\%, 28.5\%, 23.2\%, and 25.2\% higher than \code{AthenaCEP}, \code{DARLING}, \code{gSPICE}, and \code{hSPICE}.
These results indicate that \sys is robust to changes in pattern length, and complex patterns can benefit more from \sys.

\begin{figure}[t]
	\centering
	\begin{subfigure}[b]{0.495\linewidth}
		\centering
		\includegraphics[width=\linewidth]{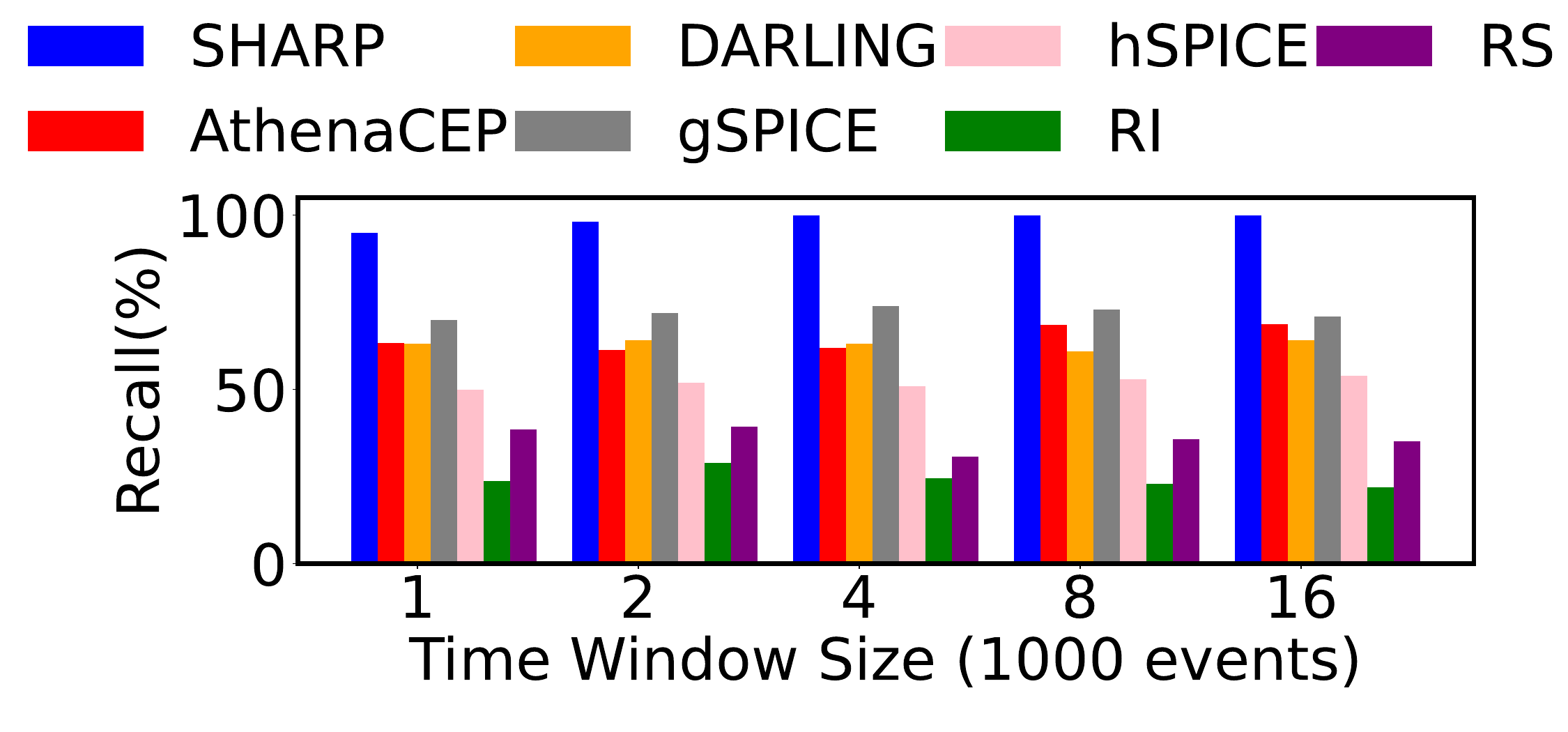}
		\caption{Recall}
		\label{fig:window_recall}
	\end{subfigure}
	\hfill
	\begin{subfigure}[b]{0.495\linewidth}
		\centering
		\includegraphics[width=\linewidth]{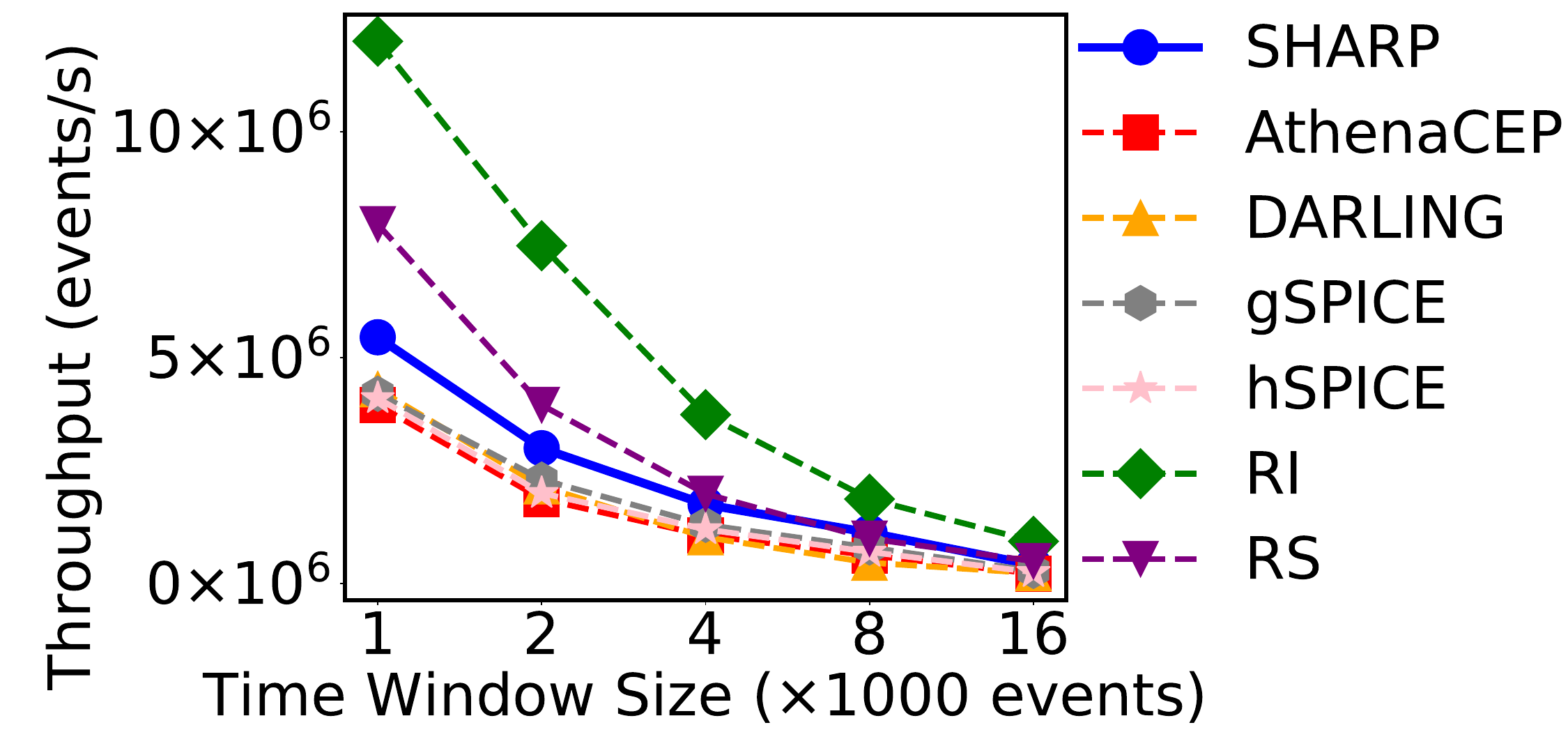}
		\caption{Throughput}
		\label{fig:window_throughput}
	\end{subfigure}
	\caption{
		Impact of time window size on patterns \code{P$_{3}$-P$_4$} over \code{DS1}\looseness=-1
		}
	\label{fig:window_size}
\end{figure}

\mypar{Time window size}
We change the size of the sliding time window of \code{P$_3$} and \code{P$_4$},
ranging from 1k to 16k events. The slide is one event. The patterns are
evaluated over the data stream from dataset \code{DS2}.
\F\ref{fig:window_recall} shows that \sys consistently yields the highest recall compared to baselines.
\sys's recall increases with increasing time window,  from 95\% to 100\%. This is because a larger time window provides more historical statistics for the cost model to learn, which allows \sys to select more promising states.
Regarding throughput, a larger time window increases the size of the maintained state, resulting in lower throughput for \sys and the baselines. However, on average, \sys's throughput remains 13.2\% higher than the best of the non-random baselines.\looseness=-1

\subsection{Impact of Pattern Sharing Schemes}
\label{sec:exe_sharingScheme}

\mypar{State materialization mechanism} We first analyse how state materialization approaches affect \sys's performance.
We consider (i) sharing by instance (\sys-Inst)---the shared states are immediately materialized~\cite{ma2022gloria, poppe2021share}, (ii) sharing by view (\sys-View)---the shared states are lazy-materialized until the complete matches are generated~\cite{ray2016scalable, DBLP:journals/pvldb/McSherryLSR20} and (iii) the extreme case that patterns maintain separate physical replicas of shared state (\sys-Sep).

\begin{figure}[t]
	\centering
	\begin{subfigure}[b]{0.495\linewidth}
		\centering
		\includegraphics[width=\linewidth]{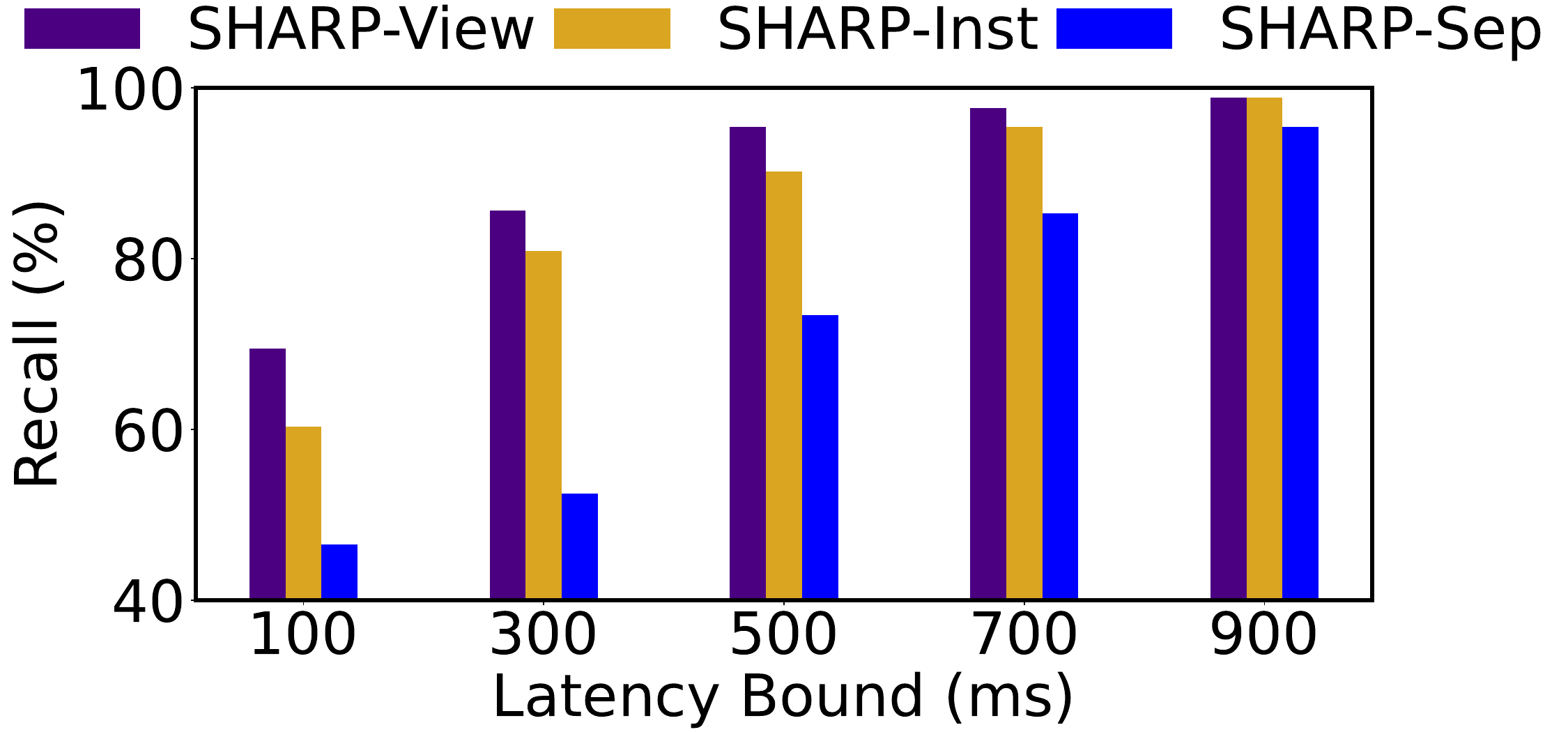}
		\caption{CEP Recall}
		\label{fig:memory_recall}
	\end{subfigure}
	\hfill
	\begin{subfigure}[b]{0.495\linewidth}
		\centering
		\includegraphics[width=\linewidth]{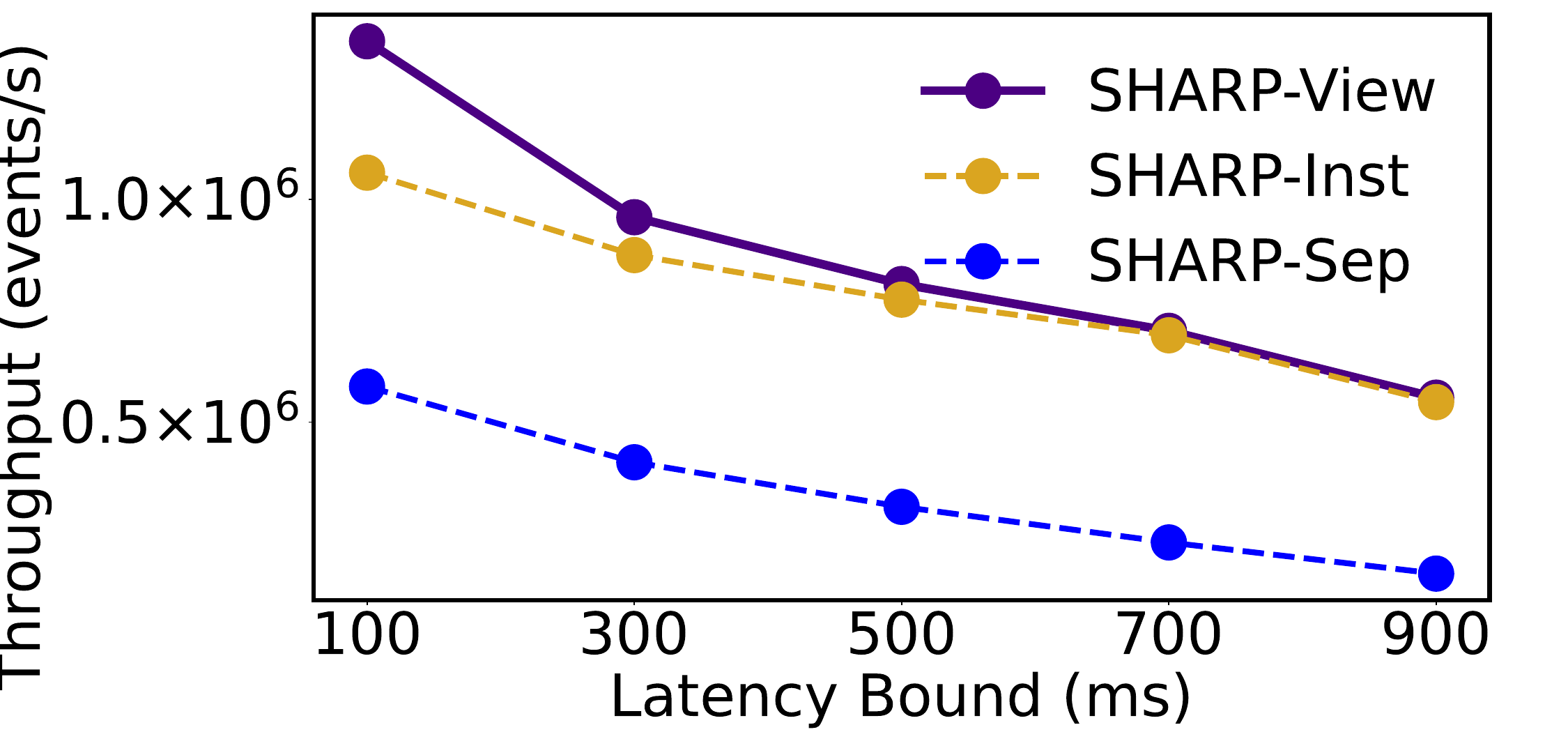}
		\caption{CEP Throughput}
		\label{fig:memory_throughput}
	\end{subfigure}

    \medskip
    
    \begin{subfigure}[b]{0.495\linewidth}
		\centering
		\includegraphics[width=\linewidth]{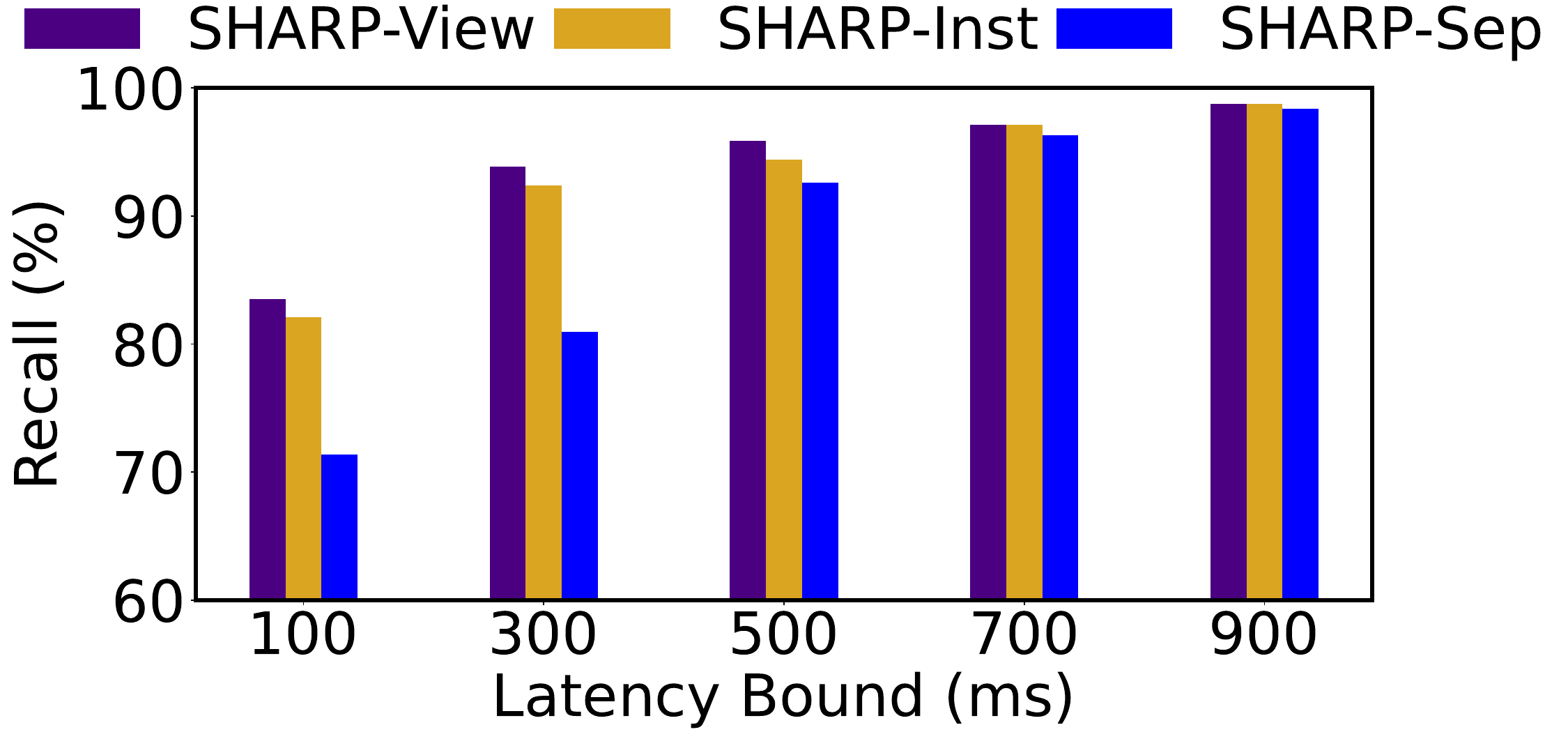}
		\caption{\mreg Recall}
		\label{fig:mr_recall_memory}
	\end{subfigure}
	\hfill
	\begin{subfigure}[b]{0.495\linewidth}
		\centering
		\includegraphics[width=\linewidth]{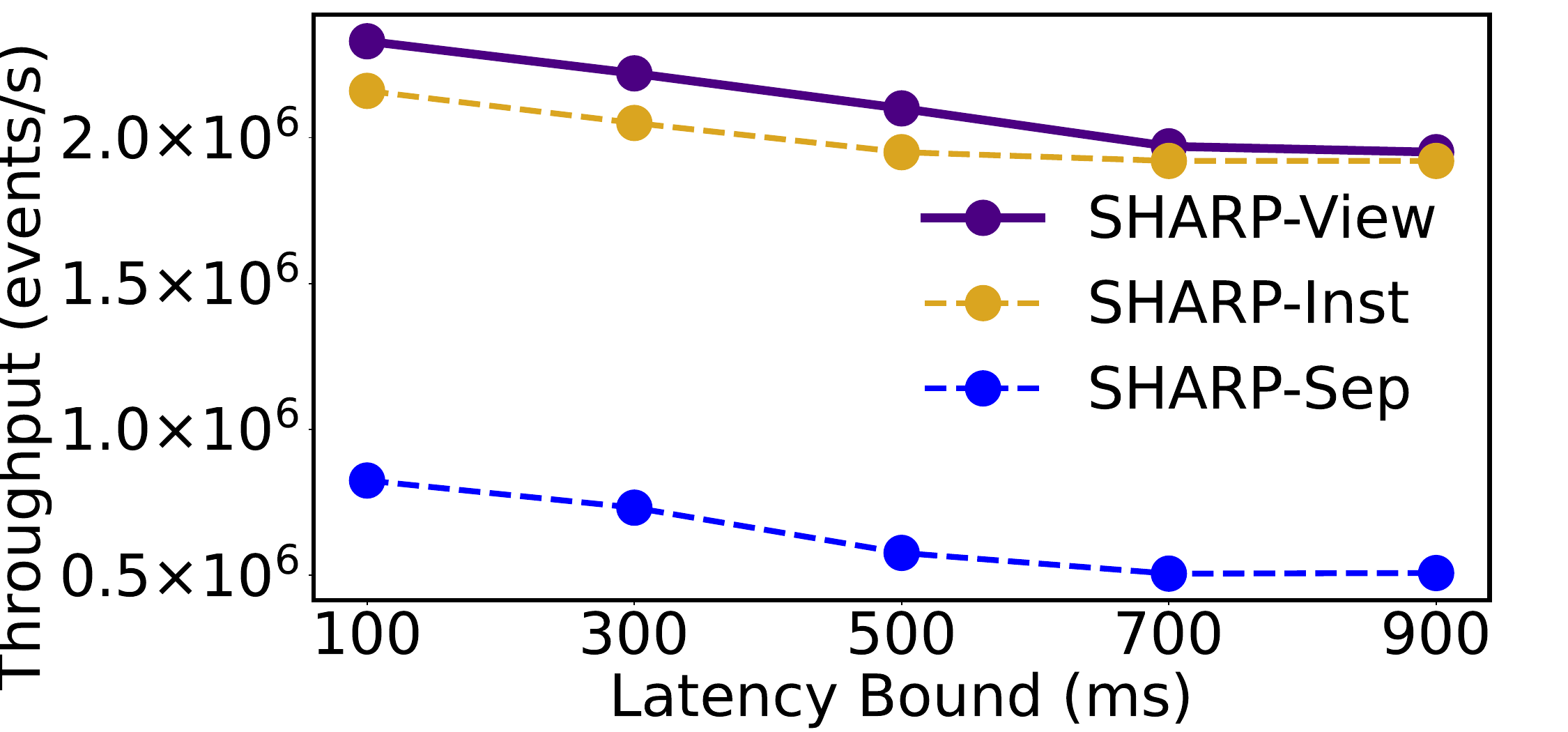}
		\caption{\mreg Throughput}
		\label{fig:mr_throughput_memory}
	\end{subfigure}
	\caption{Impact of state materialization  on patterns \code{P$_{3}$-P$_4$} over \code{DS1}}
	\label{fig:memory_CEP}
    \vspace{-1em}
\end{figure}

\F\ref{fig:memory_CEP}a-b (CEP) show the results of executing \code{P$_3$} and \code{P$_4$} over dataset \code{DS1}. \sys-View achieves the highest recall at all latency bounds (see \F\ref{fig:memory_recall}), 1.1$\times$ higher than \sys-Inst and 5$\times$ higher than \sys-Sep. Because \sys-View selects a single partial match for bespoke shared patterns at fine granularity by controlling the bitmap mask.
In contrast, \sys-Inst either selects a partial match for all shared patterns or neglects them, \ie a coarse-granular selection.
\sys-Sep is unaware of the shared state materialization and, therefore, cannot exploit the optimization opportunity of shared patterns.
For throughput (\F\ref{fig:memory_throughput}), \sys-View is higher than \sys-Inst due to its efficient in-memory reference count. \sys-Sep performs the worst because of its redundant state replicas. The similar trend is mirrored in \mreg (see \F\ref{fig:memory_CEP}c-d).
 \looseness=-1

\begin{figure}[t]
	\centering
	\begin{subfigure}[b]{0.495\linewidth}
		\centering
		\includegraphics[width=\linewidth]{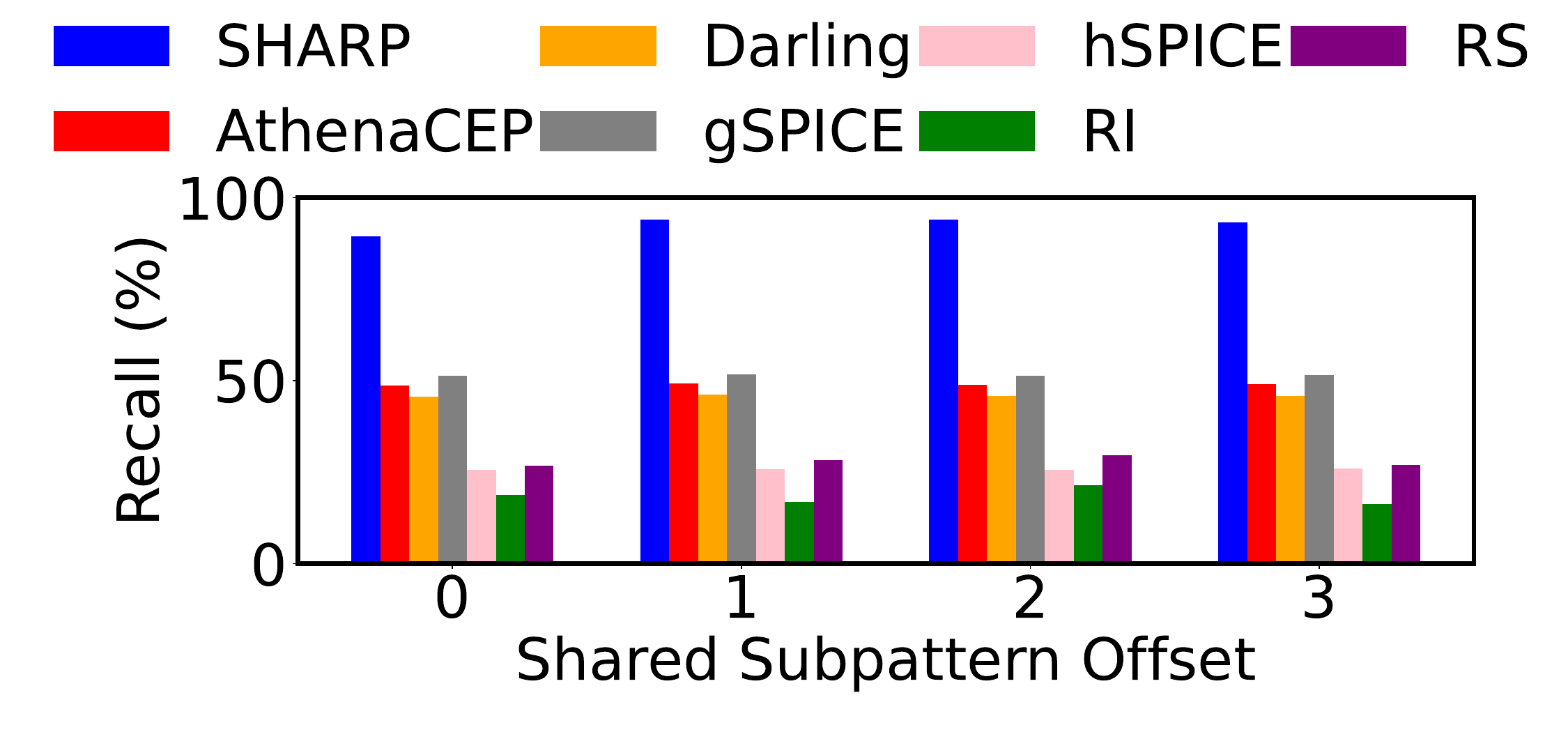}		\caption{CEP Recall}
		\label{fig:position_recall}
	\end{subfigure}
	\hfill
	\begin{subfigure}[b]{0.495\linewidth}
		\centering
		\includegraphics[width=\linewidth]{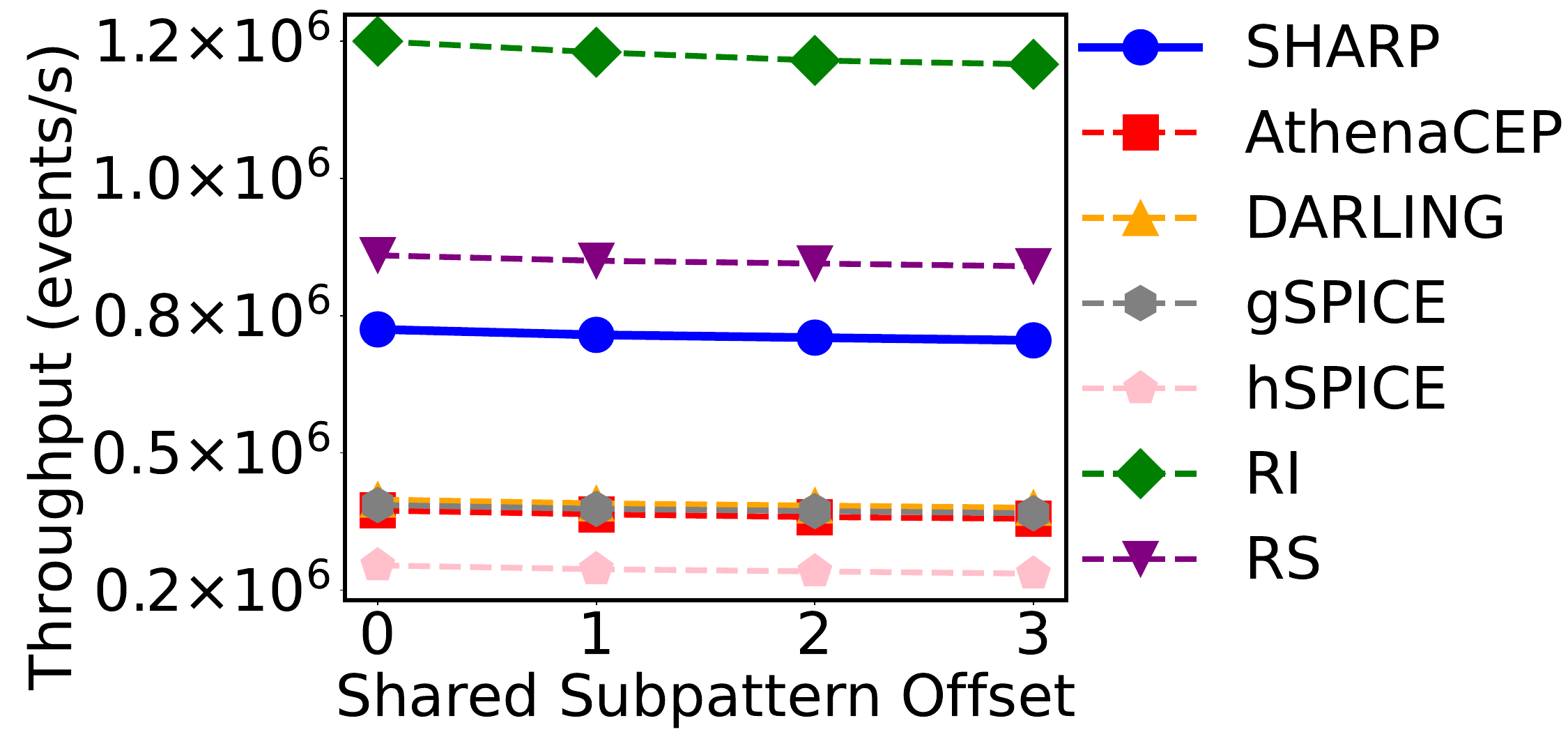}
		\caption{CEP Throughput}
		\label{fig:position_th}
	\end{subfigure}

    \medskip

    	\begin{subfigure}[b]{0.495\linewidth}
		\centering
		\includegraphics[width=\linewidth]{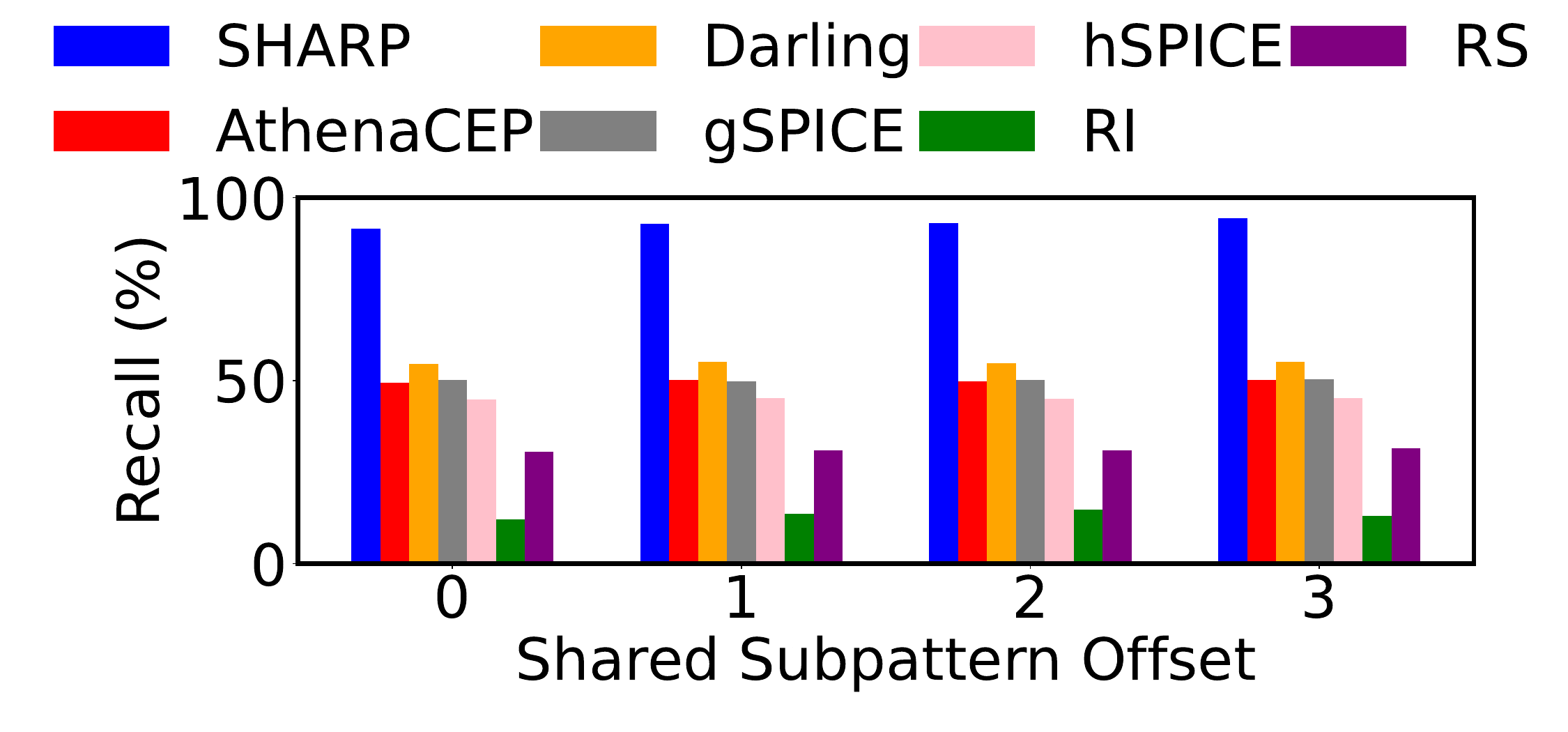}		\caption{\mreg Recall}
		\label{fig:position_recall_mr}
	\end{subfigure}
	\hfill
	\begin{subfigure}[b]{0.495\linewidth}
		\centering
		\includegraphics[width=\linewidth]{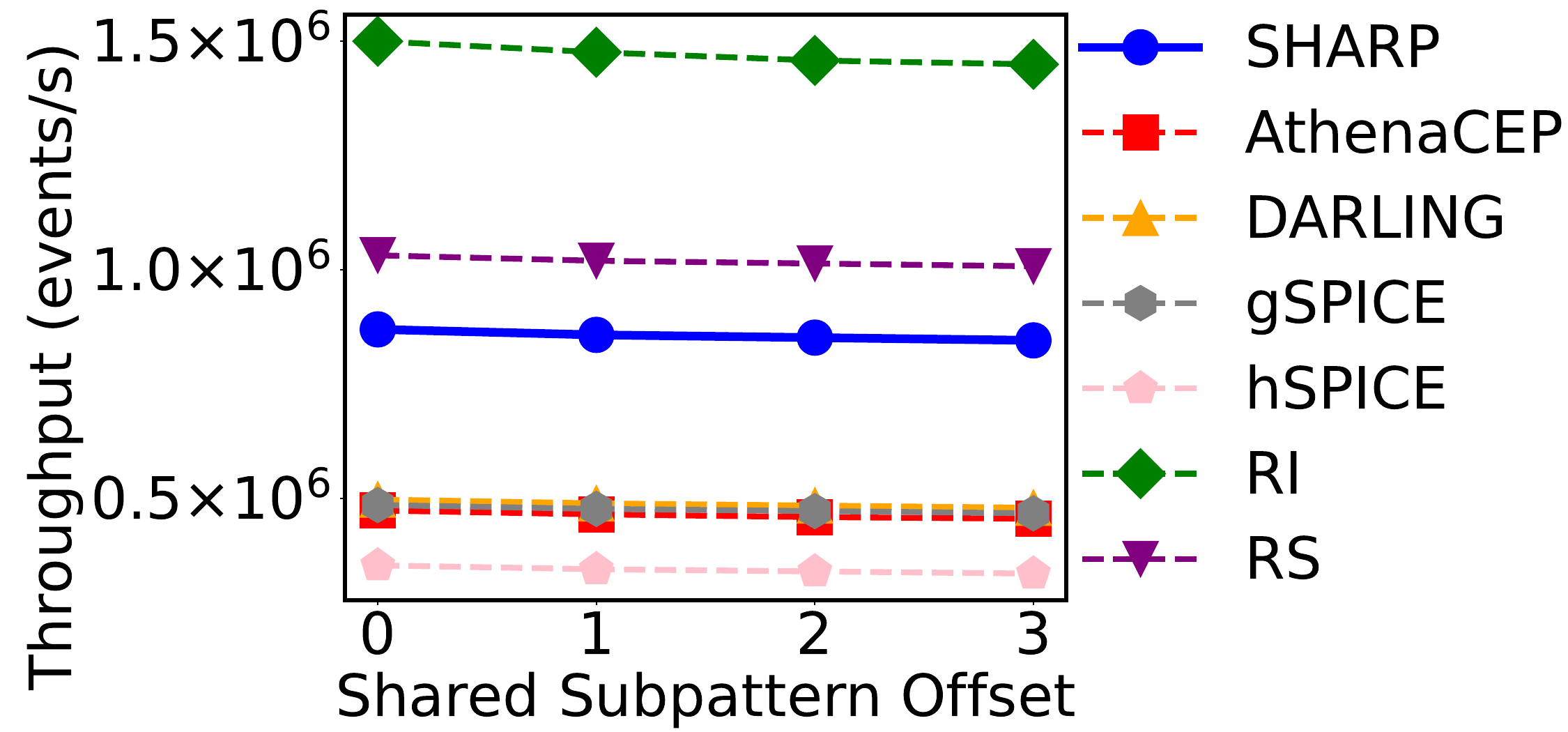}
		\caption{\mreg Throughput}
		\label{fig:position_th_mr}
	\end{subfigure}
	\caption{Impact of state sharing position on patterns \code{P$_{3}$-P$_4$} over \code{DS1}}
	\label{fig:position}
\end{figure}

\mypar{Sharing position in patterns} 
We investigate the impact of sharing position, using \code{P$_3$} and \code{P$_4$} on \code{DS1}.
To this end, we change the offset of the shared
sub-pattern from 0 to 3, and measure the recall.
\F\ref{fig:position} shows that the sharing position does not affect
\sys's performance, with higher recalls than baselines, up to 2.1$\times$. This is because \sys's PSD captures the sharing position and the state selector adapts to such a change.  In contrast, other baselines are unaware of the sharing position.

\subsection{Adaptivity to Concept Drifts}
\label{sec:exe_adaptivity}

\begin{figure}[t]
  \centering
  \begin{subfigure}[b]{0.495\linewidth}
    \centering
    \includegraphics[width=\linewidth]{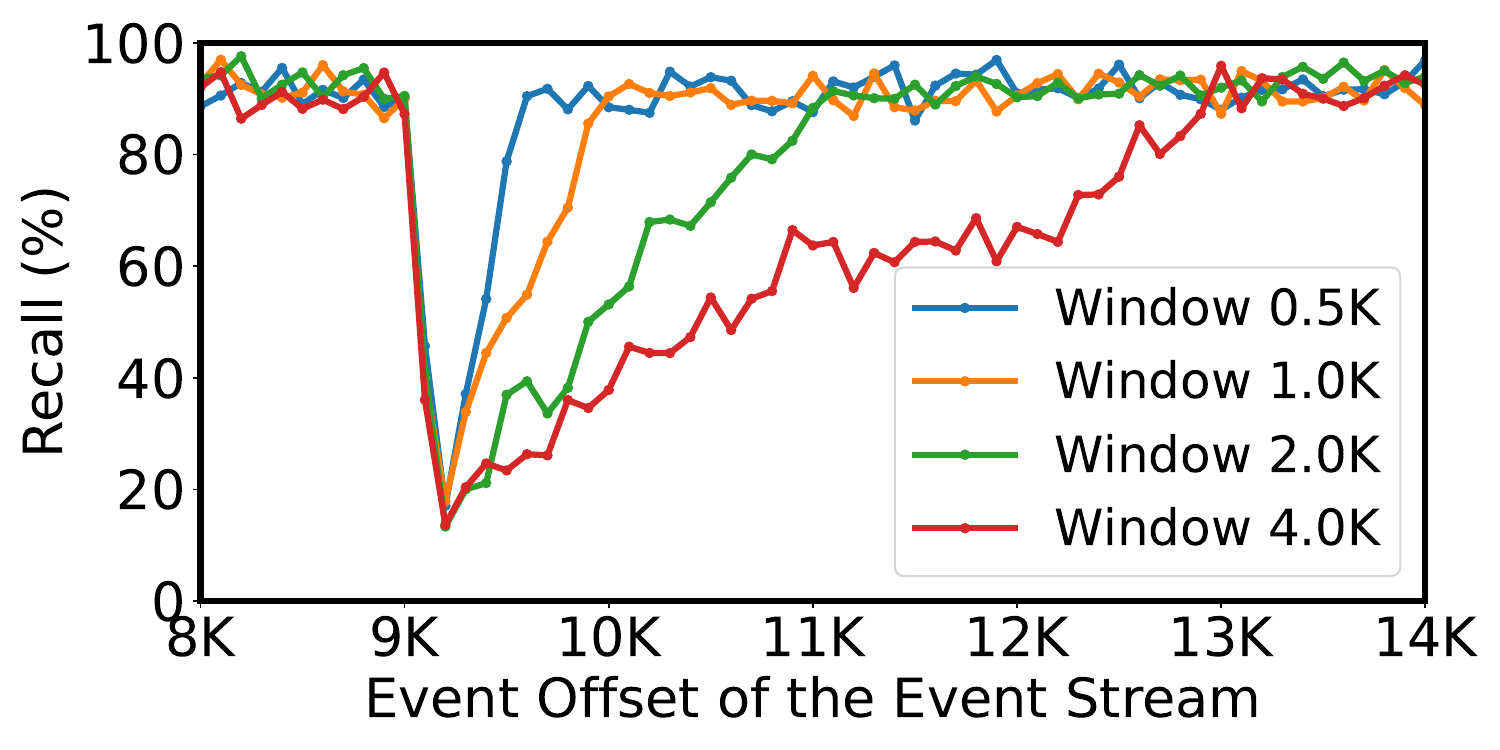}
    \caption{Recall of \code{P$_3$}}
    \label{fig:adaptivity_p1}
  \end{subfigure}
  \hfill
  \begin{subfigure}[b]{0.495\linewidth}
    \centering
    \includegraphics[width=\linewidth]{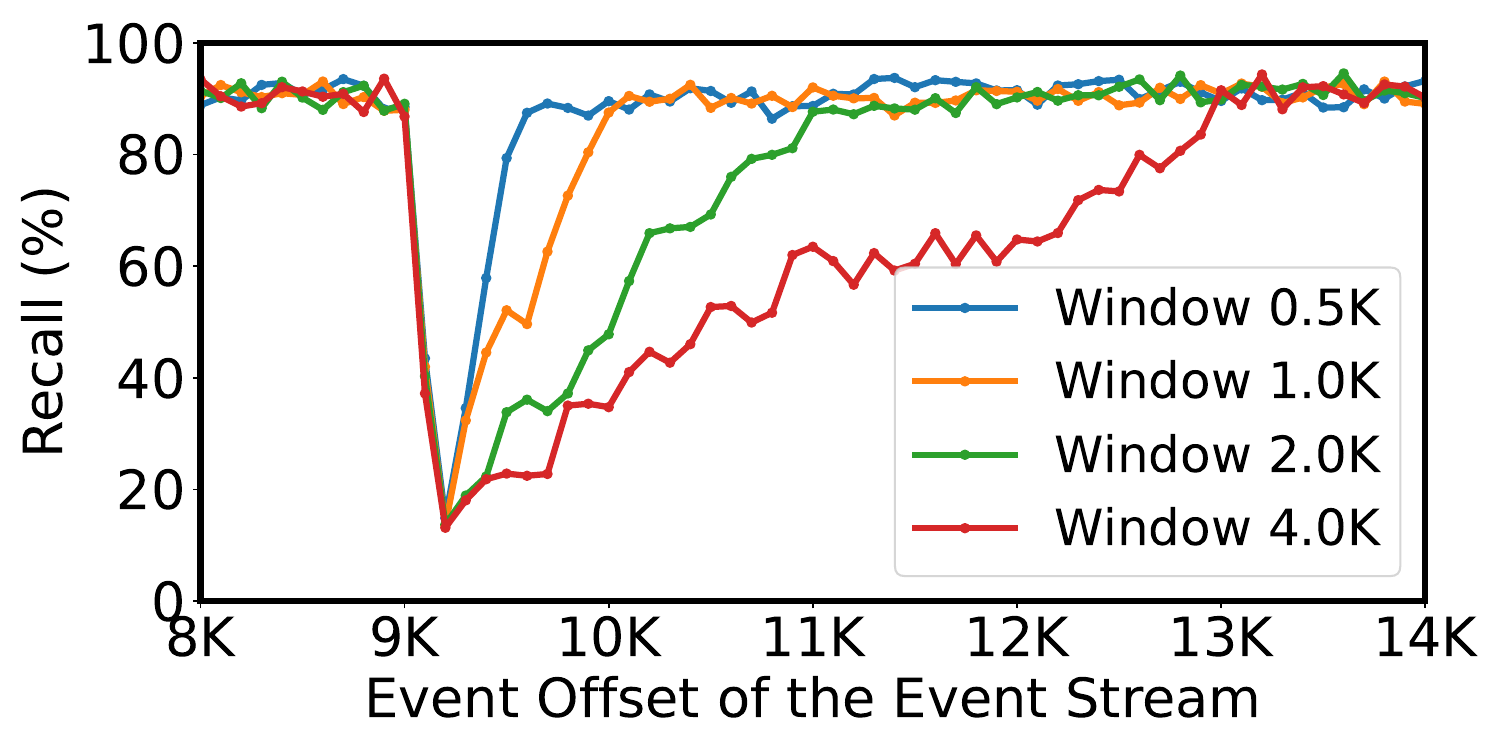}
    \caption{Recall of \code{P$_4$}}
    \label{fig:adaptivity_p2}
  \end{subfigure}
  \caption{ Adaptivity to concept drifts in shared patterns \code{P$_{3}$-P$_4$} over \code{DS1} \looseness=-1}
  \label{fig:Adaptivity}
\end{figure}

This section investigates \sys's adaptivity to concept drifts, using
 \code{P$_3$} and \code{P$_4$} over data streams derived from \code{DS1}.
To control the concept drift, we change the value distribution of \code{$D.V$} from \code{$\mathcal{U}$}(1$\times$10$^6$, 3.5$\times$ 10$^6$) to \code{$\mathcal{U}$}(1, 2$\times$10$^6$) at the offset of 9k in the event stream.
Fig.~\ref{fig:Adaptivity} shows an abrupt drop of recall (to 18\%) immediately after the concept drift.
This is because \sys's cost model is no longer accurate due to the flipped value distribution.
However, \sys is able to swiftly detect the drift and quickly updates its cost model to improve the recall.
After one time window, \sys improves the recall back to normal level and stabilizes between 95\% to 100\%.
The convergence is slower for larger time windows because of the longer lifespan of stale partial matches.

\begin{figure}
  \centering
  \begin{subfigure}{0.495\linewidth}
    \centering
    \includegraphics[width=\linewidth]{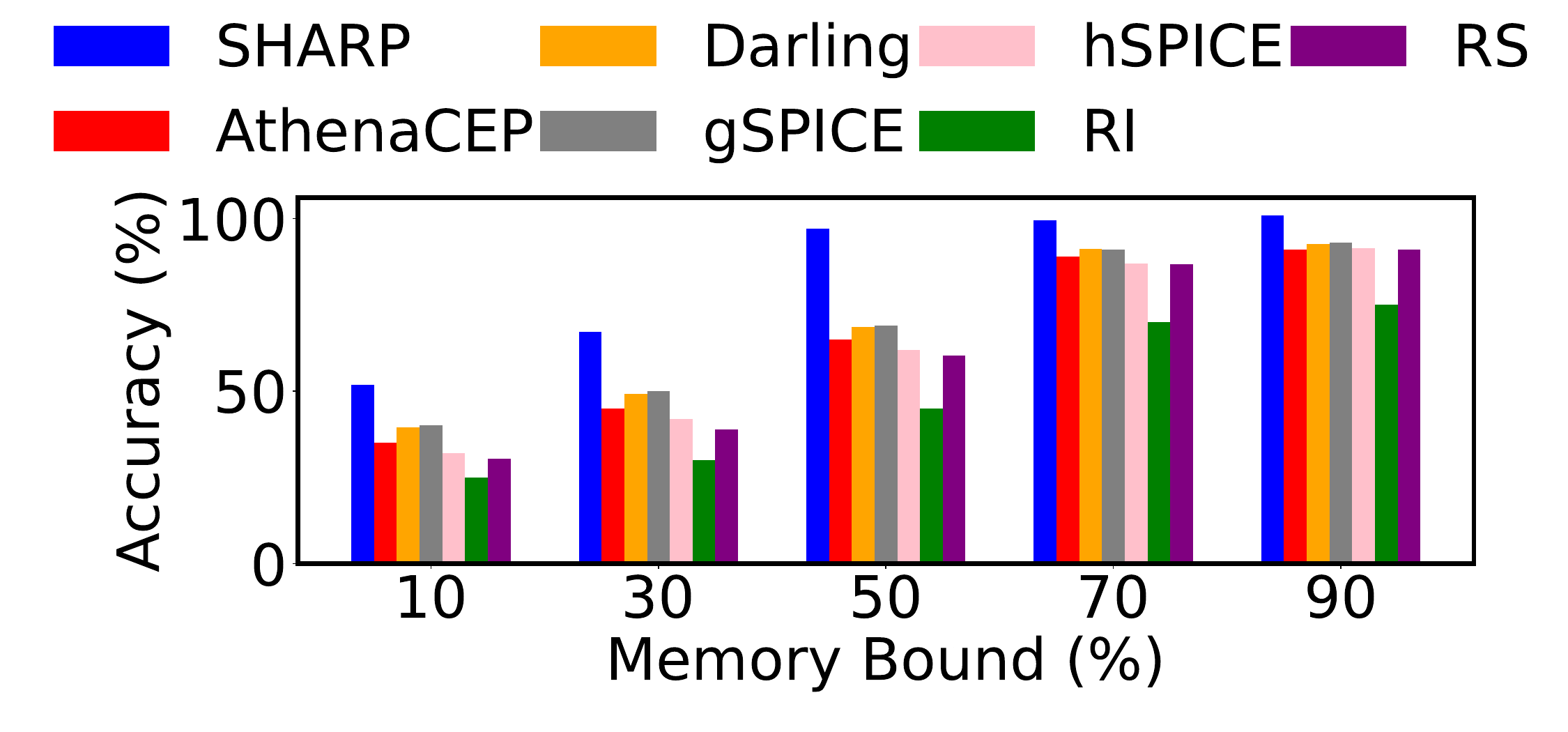}
    \caption{Accuracy}
    \label{fig:rag_hit}
\end{subfigure}
\hfill
\begin{subfigure}{0.495\linewidth}
    \centering
    \includegraphics[width=\linewidth]{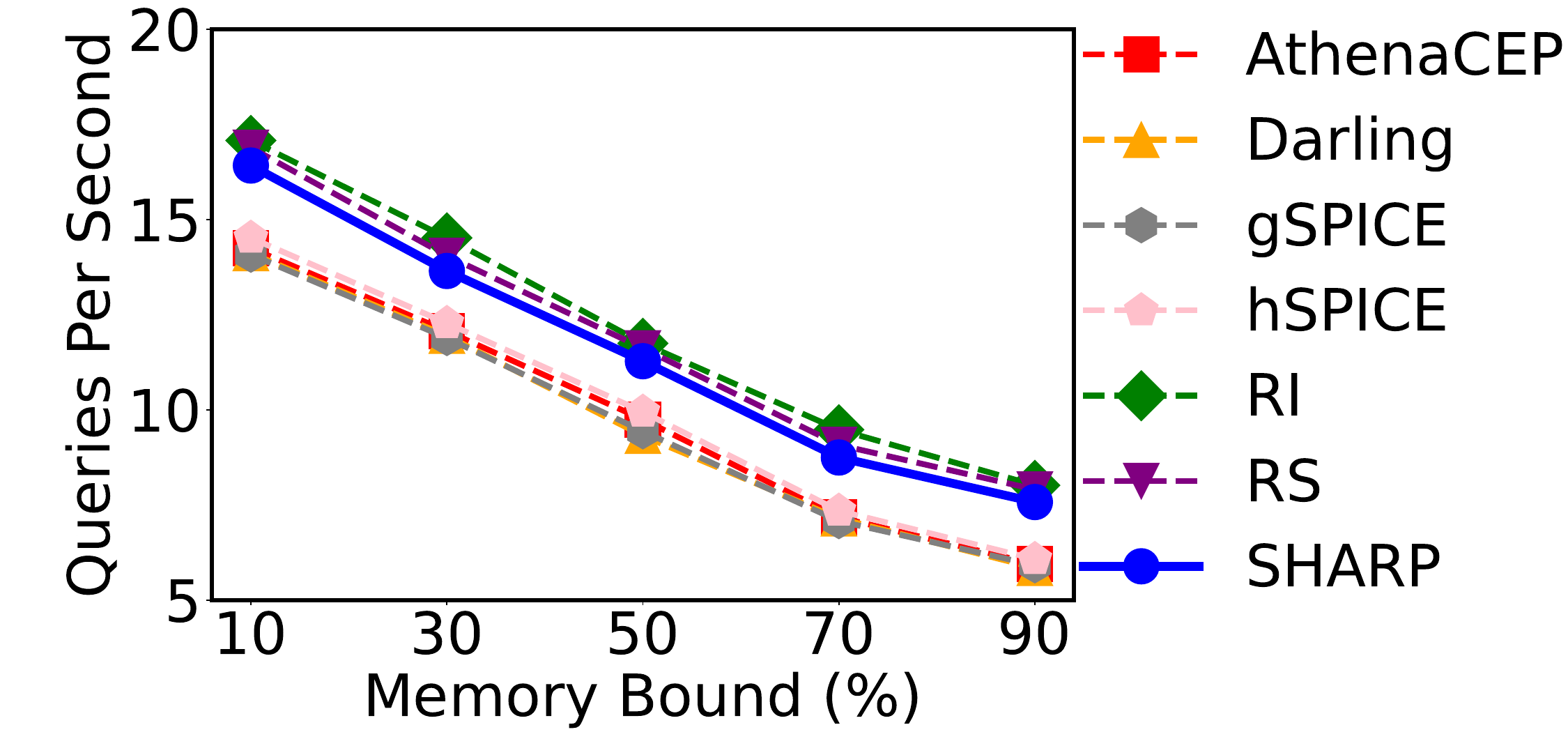}
    \caption{Throughput}
    \label{fig:rag_recall}
\end{subfigure}
\caption{Impact of memory constraint on GraphRAG over \code{Meta-QA}}
\label{fig:rag_memory}
\end{figure}

\subsection{Impact on Resource Constraints}
\label{sec:exe_resourceCon}

We examine how \sys handles limited memory capacity using GraphRAG with memory bounds from 90\% to 10\% of its original memory footprint. 
 We measure the accuracy and recall. 
As shown in \F\ref{fig:rag_memory},
\sys maintains the highest accuracy across all memory bounds, achieving 95\% accuracy with only 50\%  memory, that is, 
1.52$\times$, 1.31$\times$, 1.28$\times$, 1.60$\times$, 1.33$\times$, 1.83$\times$, and 2.05$\times$ higher than \code{AthenaCEP}, \code{DARLING}, \code{gSPICE}, \code{hSPICE}, \code{RS}, and \code{RI}, respectively.
Regarding recall, \sys outperforms baselines in all memory bounds by 1.5$\times$ (\code{AthenaCEP}), 1.4$\times$ (\code{DARLING}), 1.3$\times$ (\code{gSPICE}), 1.6$\times$ (\code{hSPICE}), 1.8$\times$ (\code{RS}) and 2.1$\times$ (\code{RI}). But the recall drops to 40\% at 10\% memory bound, because it depends on the statistical efficiency of LLM-generated responses--a small set 
cannot cover the majority of ground truth. \looseness=-1

\subsection{Adaptivity to Complex Pattern Interactions}
\label{sec:diff_bounds}

We investigate \sys's adaptivity to complex pattern interactions.
We (i) examine how changing one pattern's latency bound affects other shared patterns,  
and (ii) the impact of varying pattern latency bounds at opposite directions : increase one and decrease the other. \looseness=-1 

\begin{figure}[t]
  \centering
  \begin{subfigure}[b]{0.515\linewidth}
    \centering
    \includegraphics[width=\linewidth]{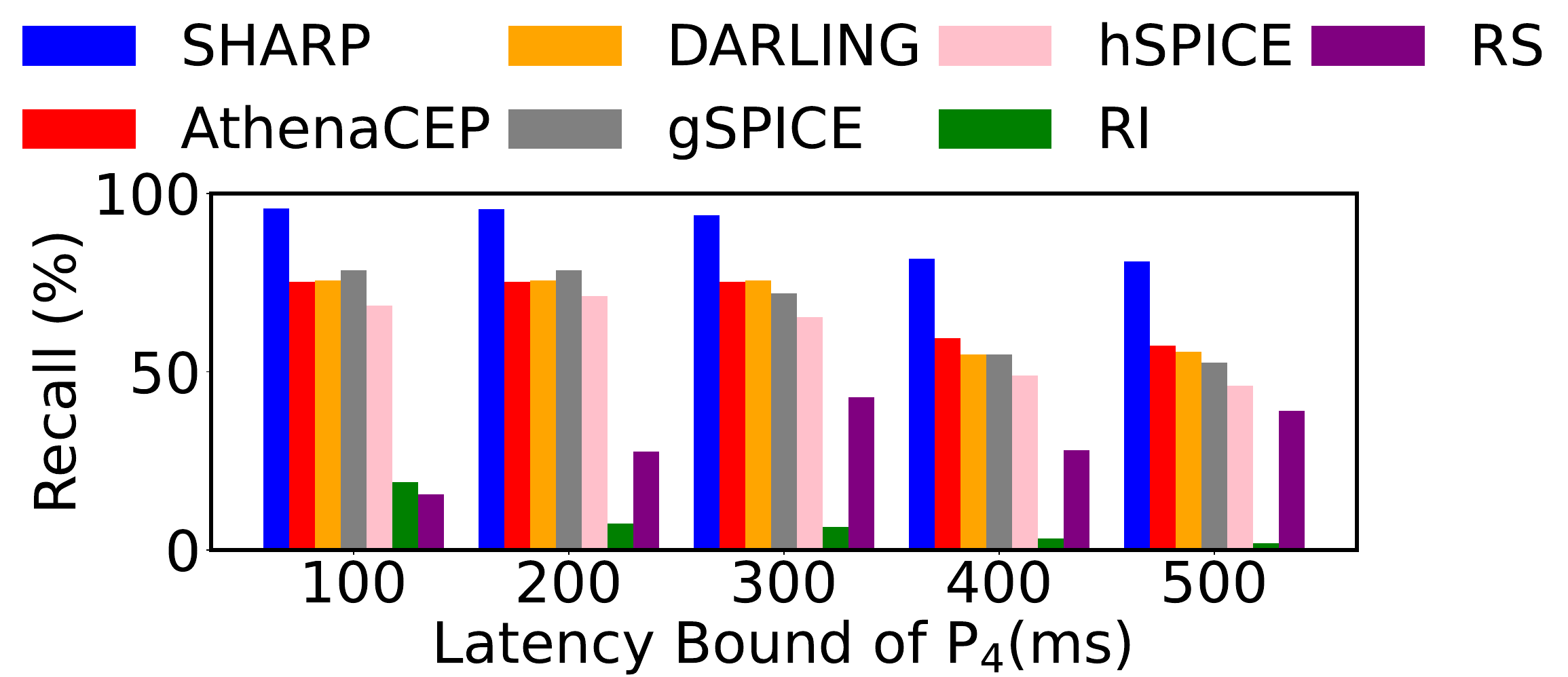}
    \caption{Recall of \code{P$_3$} under \underline{fixed} bound}
    \label{fig:single_query_0}
  \end{subfigure}
  \hfill
  \begin{subfigure}[b]{0.475\linewidth}
    \centering
    \includegraphics[width=\linewidth]{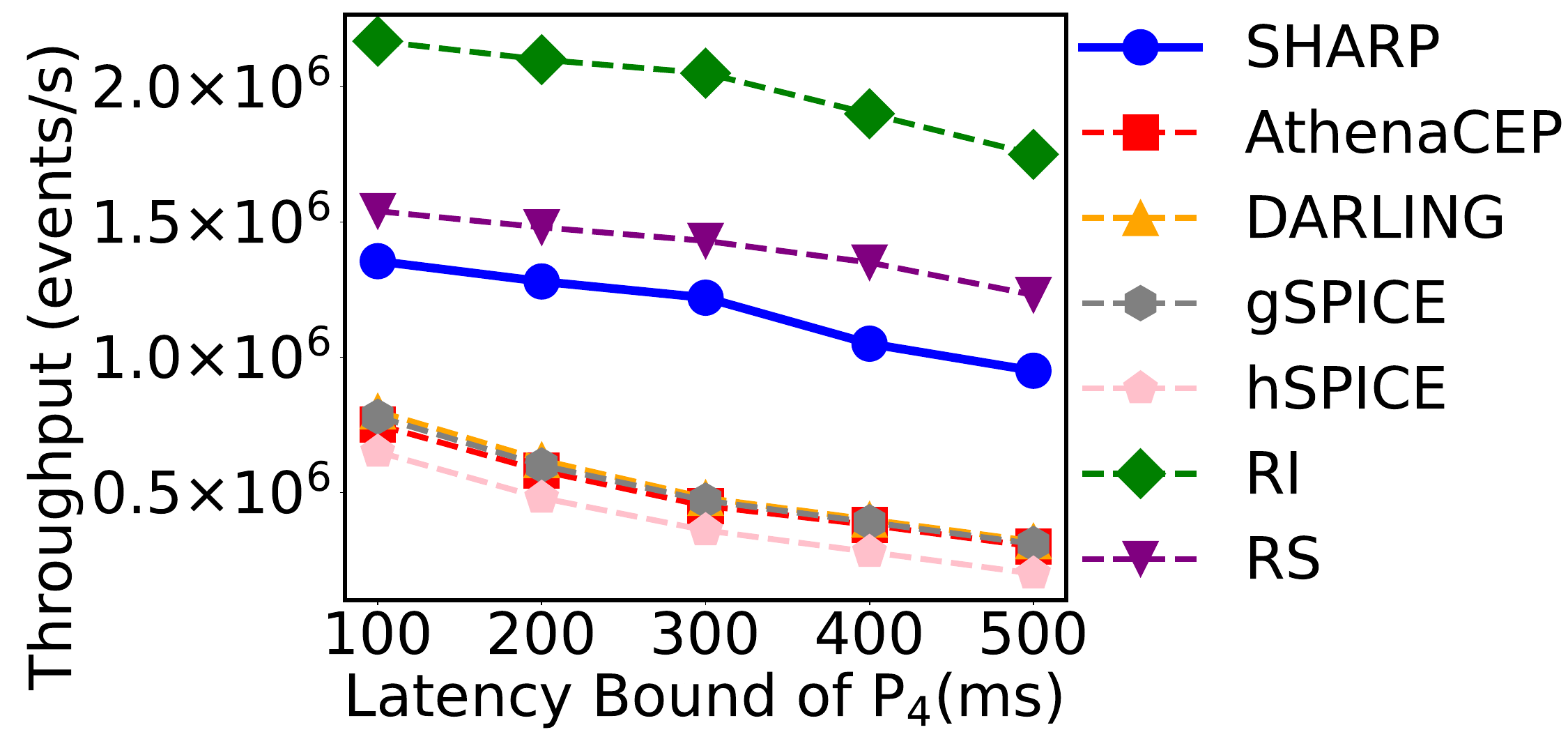}
    \caption{Throughput of \code{P$_3$} under \underline{fixed} bound}
    \label{fig:single_query_0_th}
  \end{subfigure}

  \medskip
  
    \begin{subfigure}[b]{0.515\linewidth}
    \centering
    \includegraphics[width=\linewidth]{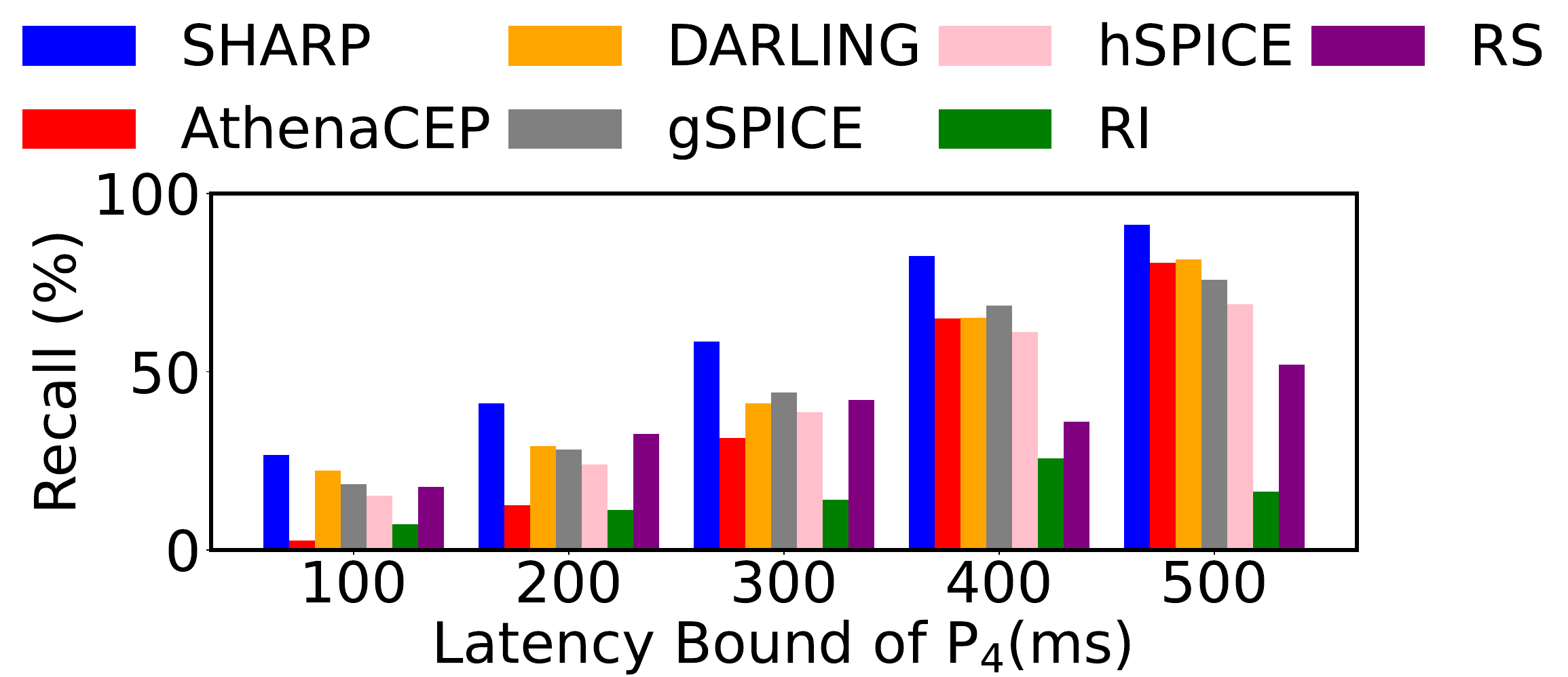}
    \caption{Recall of \code{P$_4$} under \underline{varying} bounds}
    \label{fig:single_query_1}
  \end{subfigure}
  \hfill
  \begin{subfigure}[b]{0.475\linewidth}
    \centering
    \includegraphics[width=\linewidth]{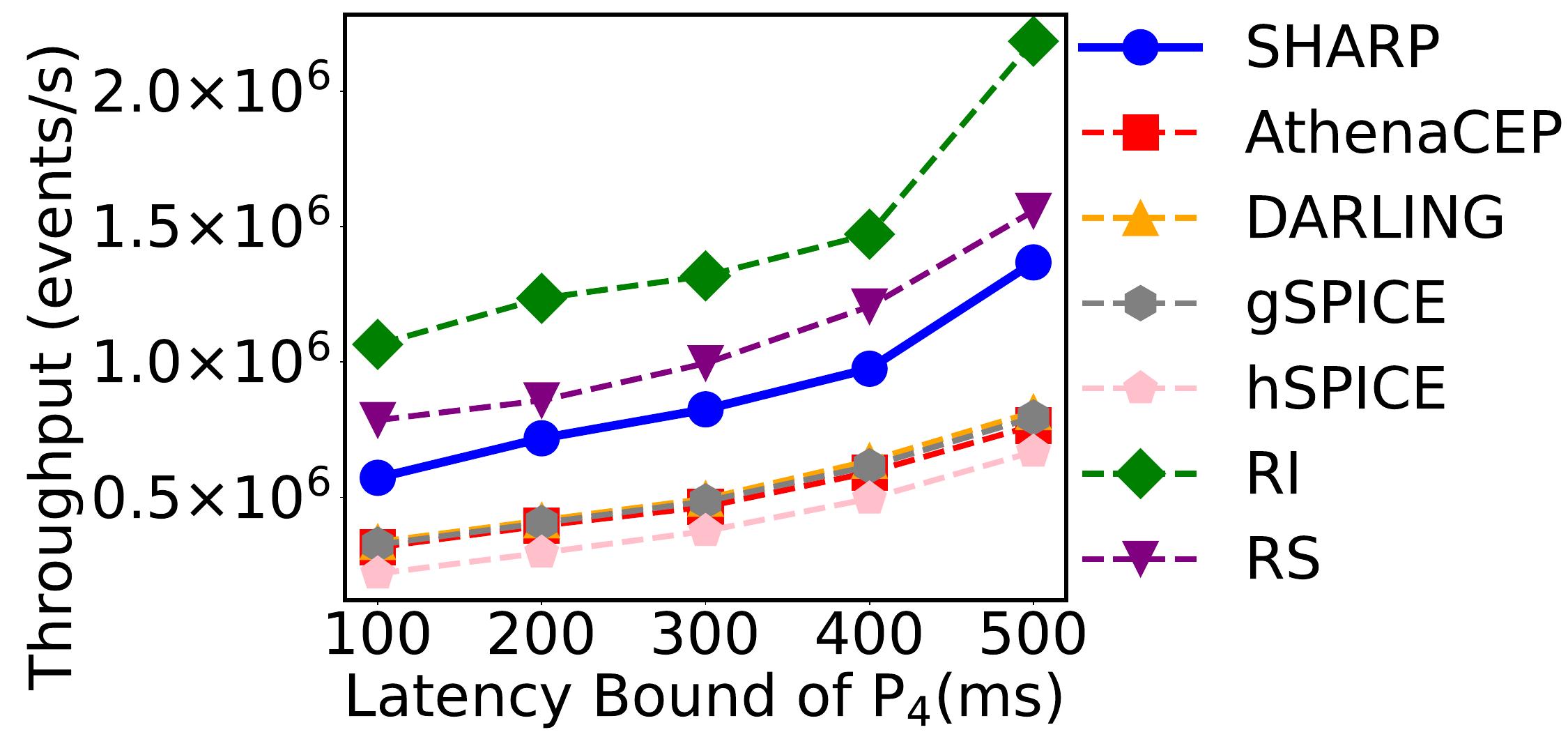}
    \caption{Throughput of \code{P$_4$} under \underline{varying} bounds}
    \label{fig:single_query_1_th}
  \end{subfigure}
  
  \caption{Impact of pattern interactions by varying P$_4$'s latency bound from 100\unit{ms} to 500\unit{ms}, while fixing P$_3$'s latency bound at 500\unit{ms}}
  \label{fig:different_latency_ab}
\end{figure}

For (i), we fix \code{P$_3$}’s latency bound of 500\unit{ms} 
while ranging \code{P$_4$}'s from 100\unit{ms} to 500\unit{ms}, using \code{Citi\_Bike} datasets.
 \F\ref{fig:different_latency_ab} illustrates the recall of \code{P$_3$} and \code{P$_4$}.
 Here,  relaxing \code{P$_4$}'s latency bound reduces \code{P$_3$}'s recall (\F\ref{fig:single_query_0}).
  The reason is two-fold. First, relaxed bounds allow \code{P$_4$} to consume more computational resources, leaving less for \code{P$_3$}. 
  Second, relaxed bounds on  \code{P$_4$}  also increases the number of  partial matches for \code{P$_3$}   (shared by \code{P$_3$} and \code{P$_4$}).
  \sys's PSD and cost model captures this, resulting in only 14\% drops in recall , compared to  \code{AthenaCEP}'s 19\%,  \code{DARLING}'s 20\%,  \code{hSPICE}'s 22\% and  \code{gSPICE}'s 26\%.     \looseness=-1
  
\begin{figure}[t]
  \centering
  \begin{subfigure}[b]{0.555\linewidth}
    \centering
    \includegraphics[width=\linewidth]{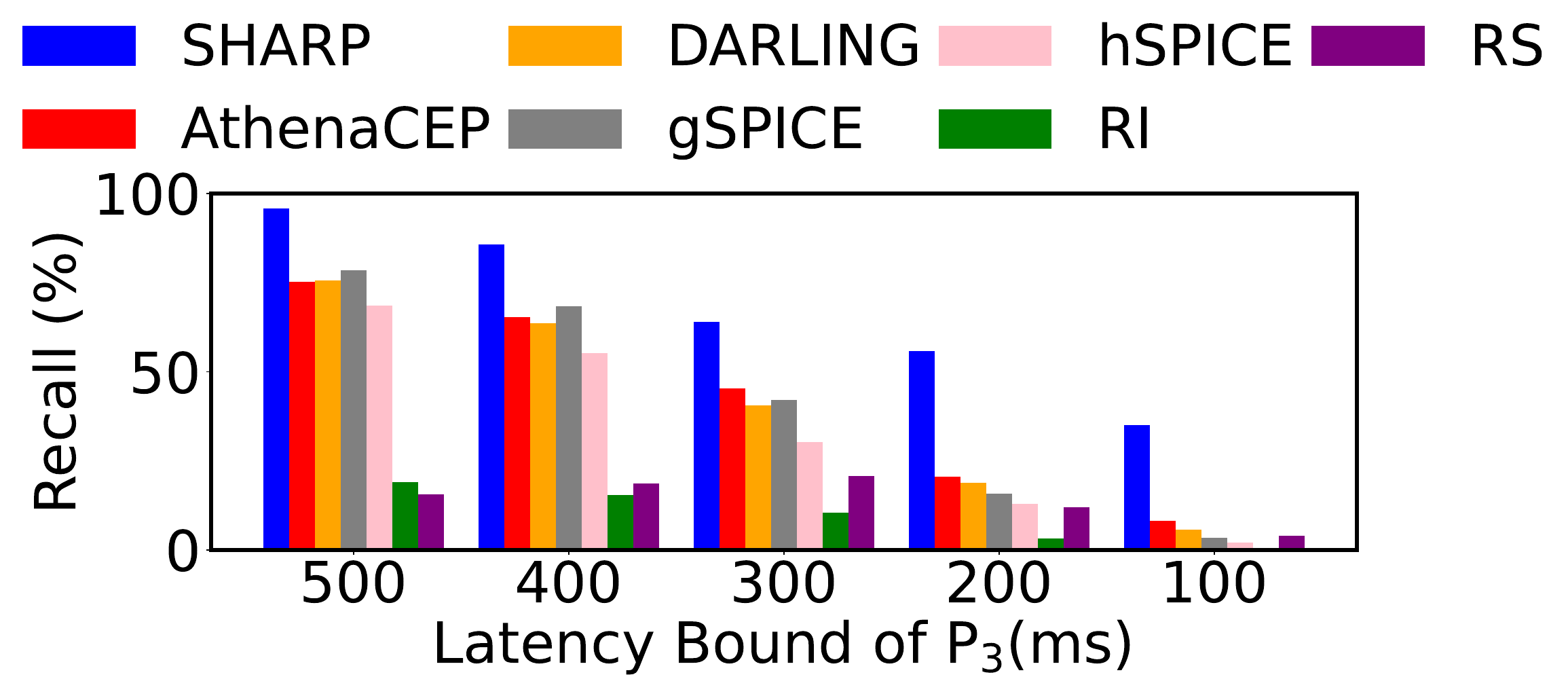}
    \caption{Recall of \code{P$_3$} under \underline{varying} bounds}
    \label{fig:single_query_c}
  \end{subfigure}
  \hfill
  \begin{subfigure}[b]{0.435\linewidth}
    \centering
    \includegraphics[width=\linewidth]{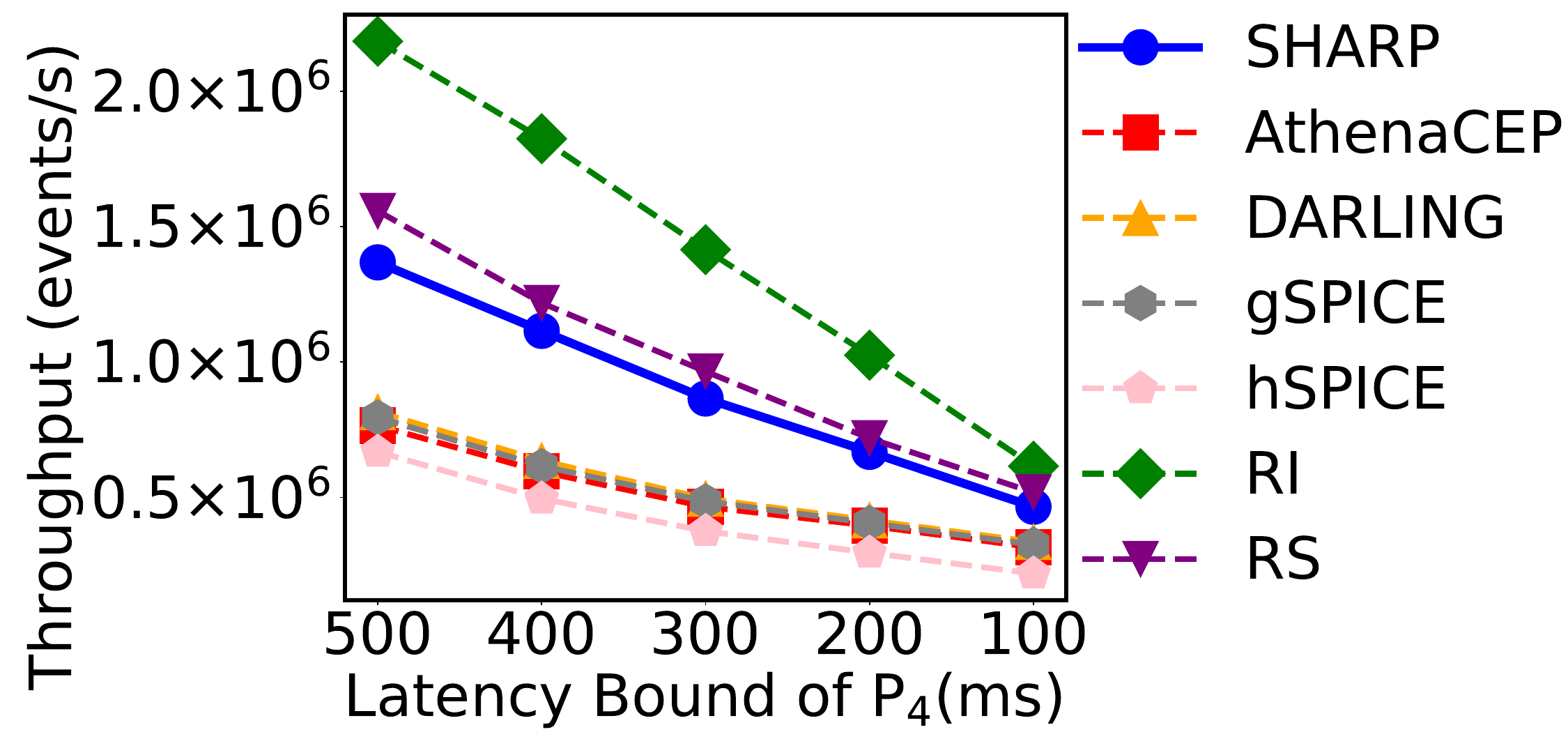}
    \caption{Throughput of \code{P$_3$} under \underline{varying} bounds}
    \label{fig:single_query_c_th}
  \end{subfigure}

    \medskip

  \begin{subfigure}[b]{0.555\linewidth}
    \centering
    \includegraphics[width=\linewidth]{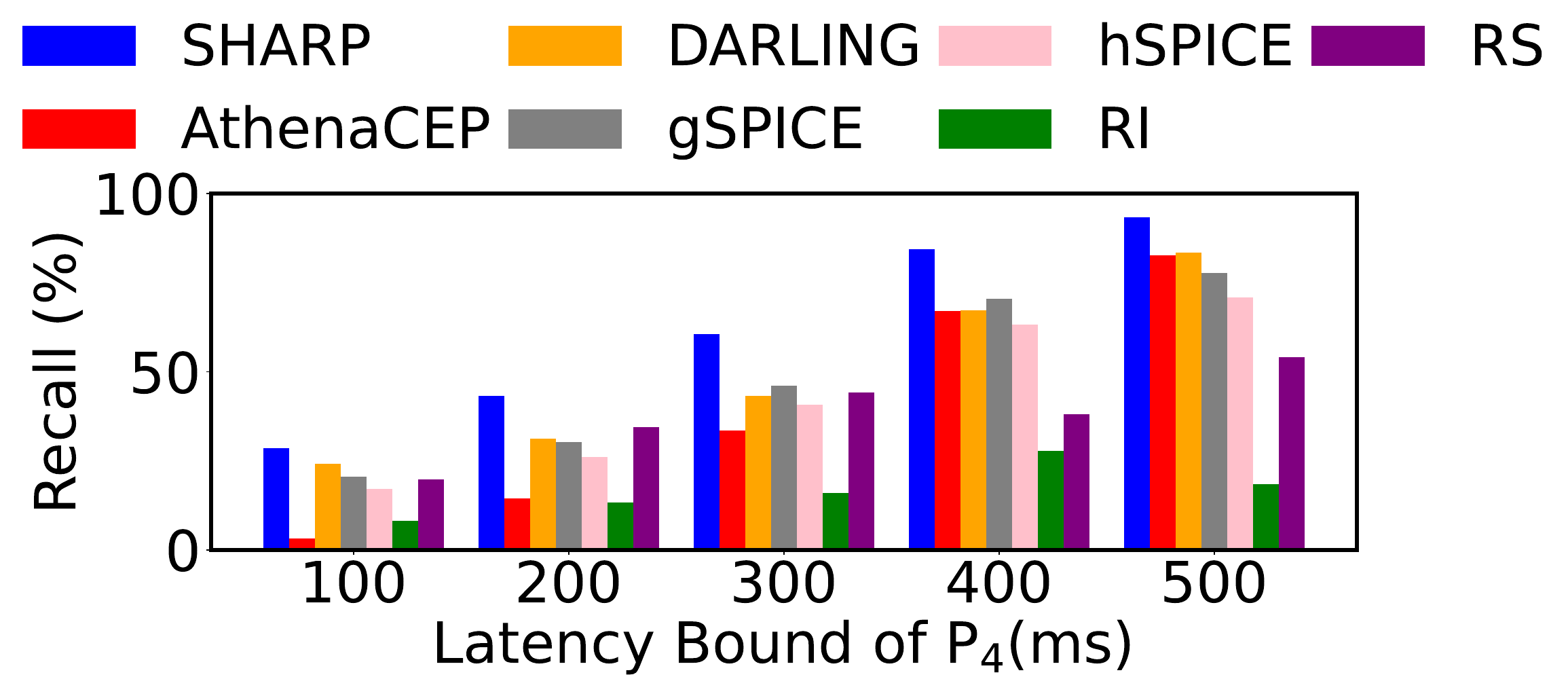}
    \caption{Recall of \code{P$_4$} under \underline{varying} bounds}
    \label{fig:single_query_d}
  \end{subfigure}
  \hfill
  \begin{subfigure}[b]{0.435\linewidth}
    \centering
    \includegraphics[width=\linewidth]{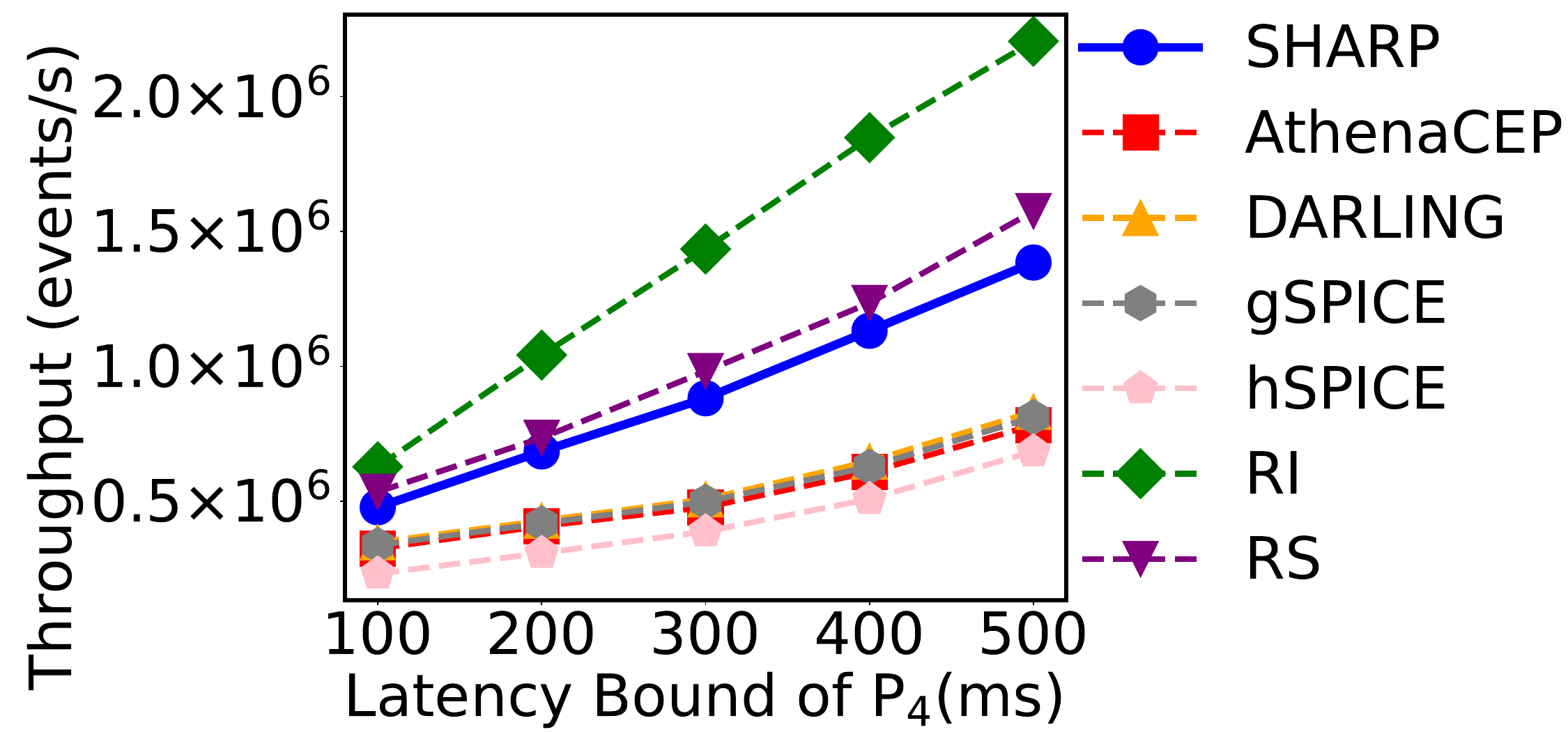}
    \caption{Throughput of \code{P$_4$} under \underline{varying} bounds}
    \label{fig:single_query_d_th}
  \end{subfigure}
  
  \caption{Impact of pattern interactions increasing P$_4$'s latency bound from 100\unit{ms} to 500\unit{ms}, while decreasing P$_3$'s from 500\unit{ms} to 100\unit{ms}}
  \label{fig:different_latency_cd}
\end{figure}

  For (ii), we decrease \code{P$_3$}'s latency bound from 500\unit{ms} to 100\unit{ms}, while increasing \code{P$_4$}'s from 100\unit{ms} to 500\unit{ms}.
   \F\ref{fig:different_latency_cd} shows decreasing recall in \code{P$_3$} and increasing recall in \code{P$_4$}. 
 Because \code{P$_3$}'s tighter latency bounds consumes less computational resources, making room for \code{P$_4$}'s increased consumption at relaxed  bounds. 
   Again, \sys's PSD and cost model  results in superior performance than \code{AthenaCEP} (4.2$\times$), \code{DARLING} (6.1$\times$), \code{hSPICE}  (12.5$\times$), and \code{gSPICE} (10.2$\times$). \looseness=-1
  
\subsection{Scalability Analysis}
\label{sec:scalability}
\begin{figure}[t]
  \centering
  \begin{subfigure}[b]{0.55\linewidth}
    \centering
    \includegraphics[width=\linewidth]{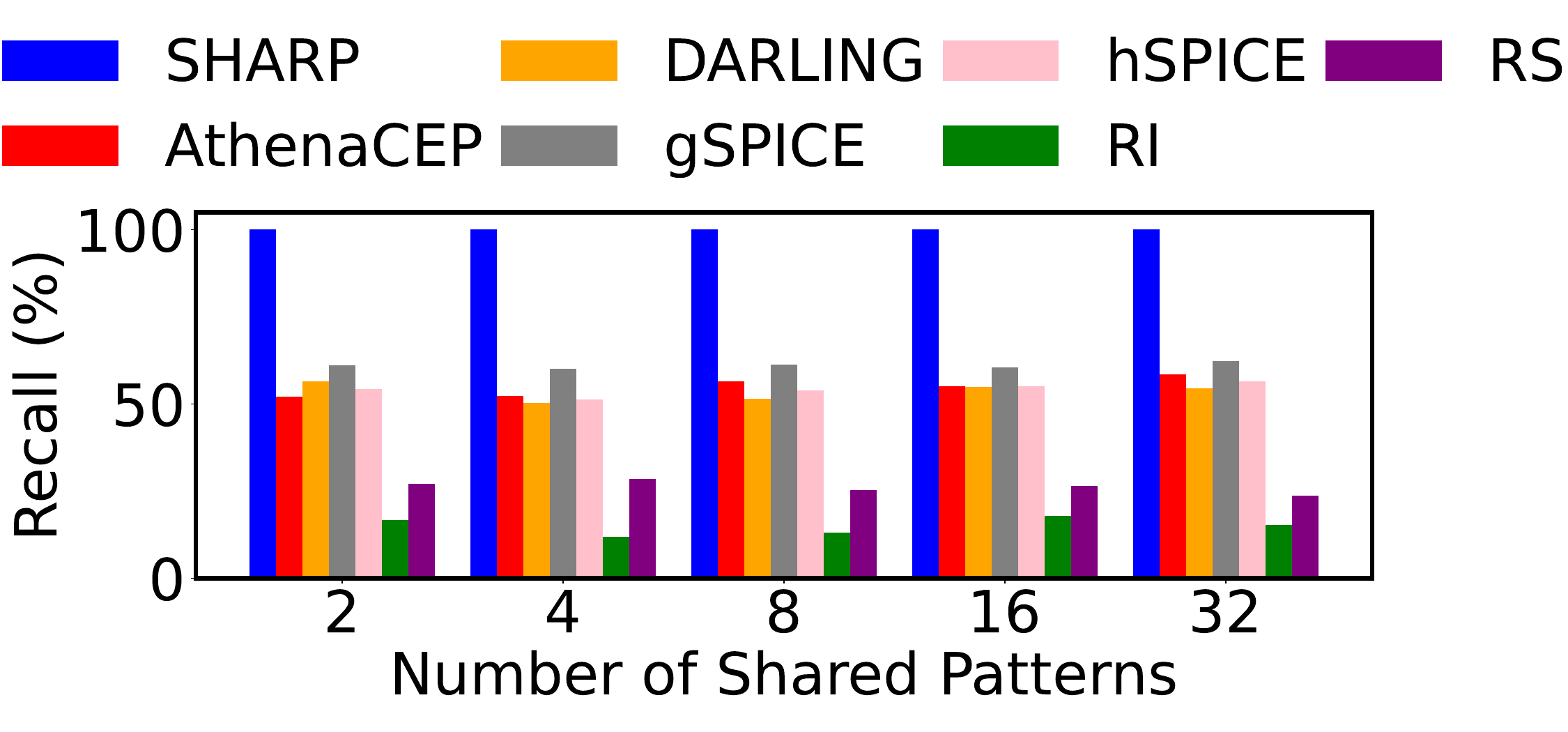}
    \caption{CEP Recall}
    \label{fig:scalability_cep_recall}
  \end{subfigure}
  \hfill
  \begin{subfigure}[b]{0.44\linewidth}
    \centering
    \includegraphics[width=\linewidth]{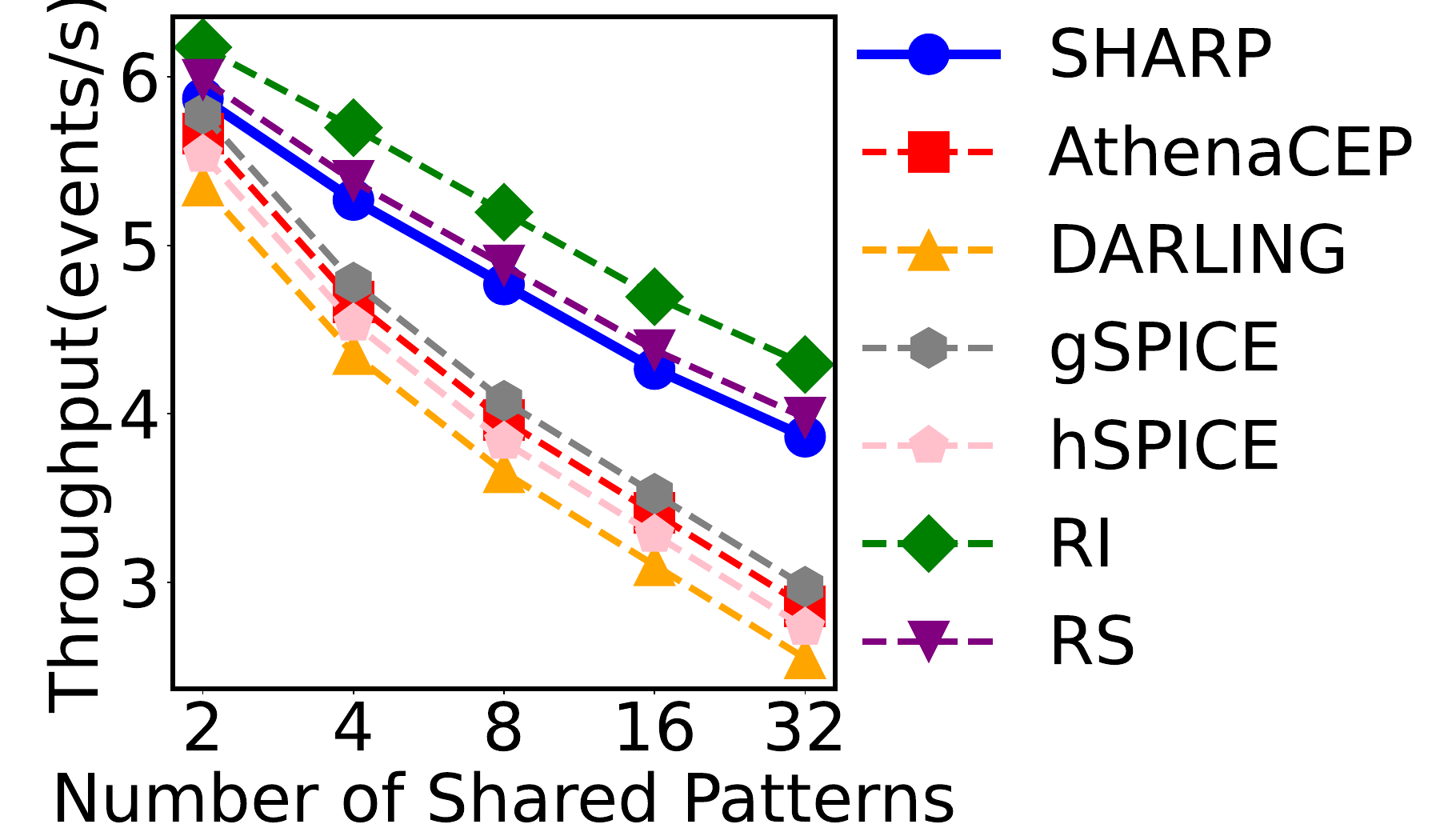}
    \caption{CEP Throughput (log scale)}
    \label{fig:scalability_cep_th}
  \end{subfigure}

\medskip

  \begin{subfigure}[b]{0.55\linewidth}
    \centering
    \includegraphics[width=\linewidth]{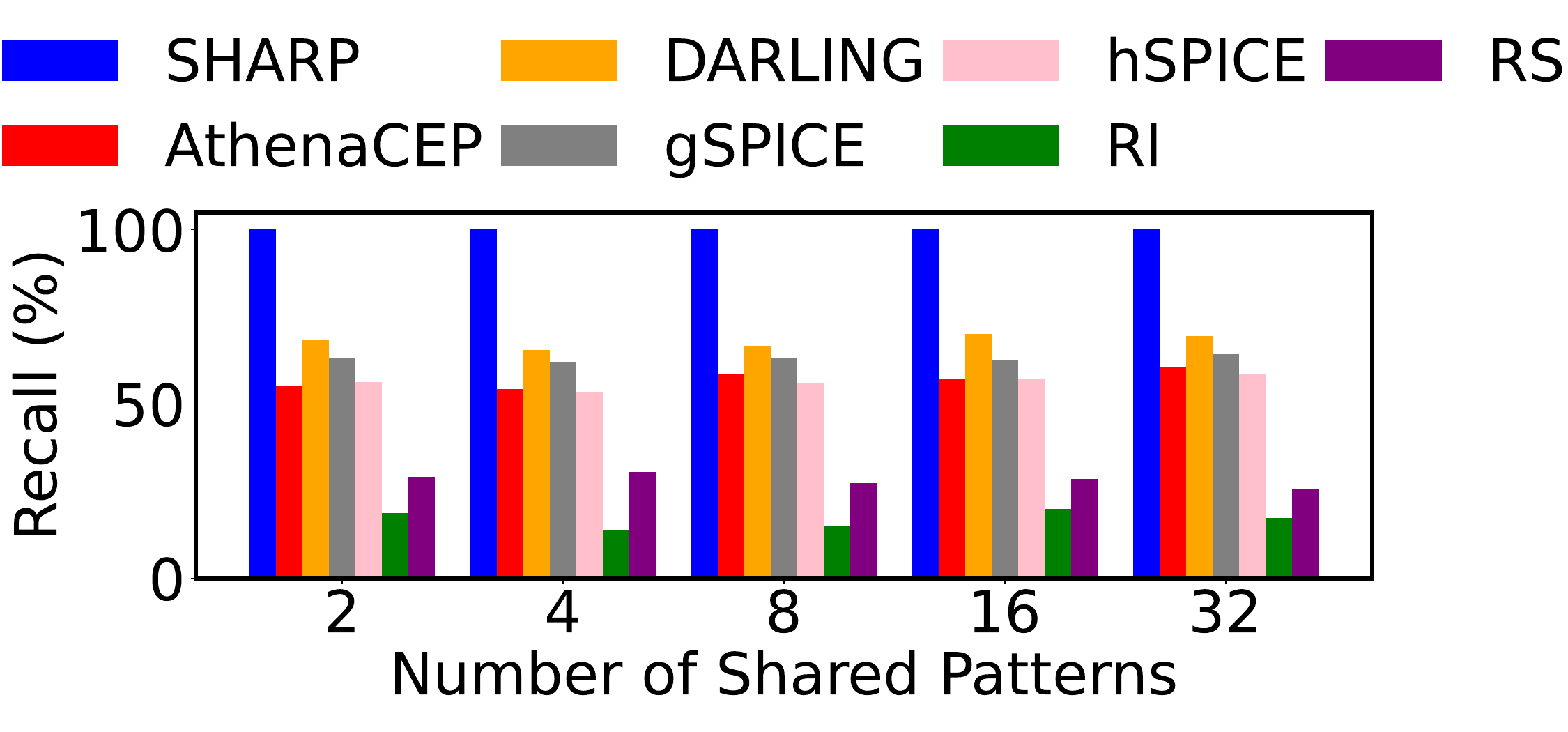}
    \caption{{\footnotesize \mreg} Recall}
    \label{fig:scalability_recall_mr}
  \end{subfigure}
  \hfill
  \begin{subfigure}[b]{0.44\linewidth}
    \centering
    \includegraphics[width=\linewidth]{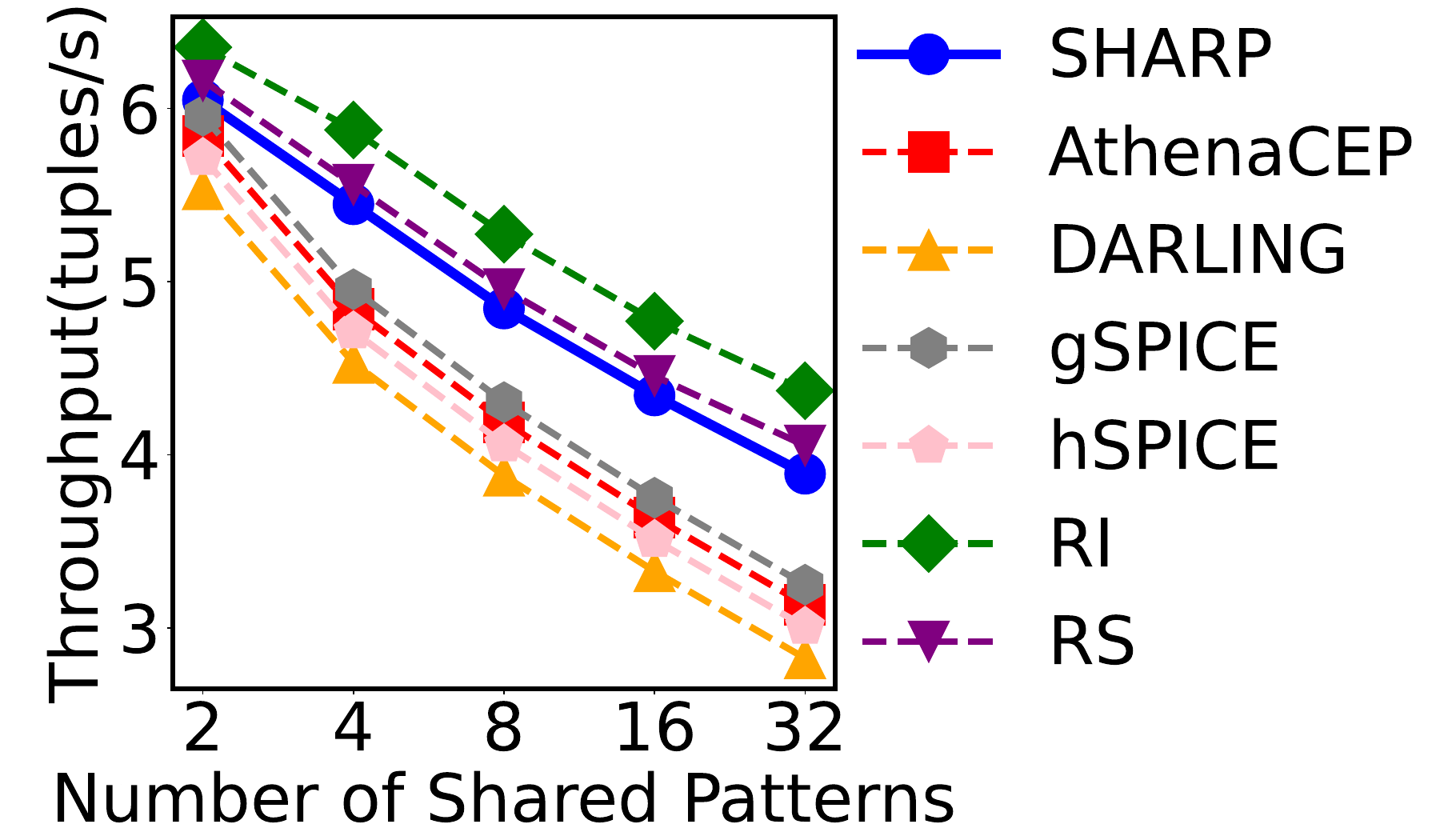}
    \caption{\mreg Throughput (log scale)}
    \label{fig:scalability_th_mr}
  \end{subfigure}

\medskip

  \begin{subfigure}[b]{0.55\linewidth}
    \centering
    \includegraphics[width=\linewidth]{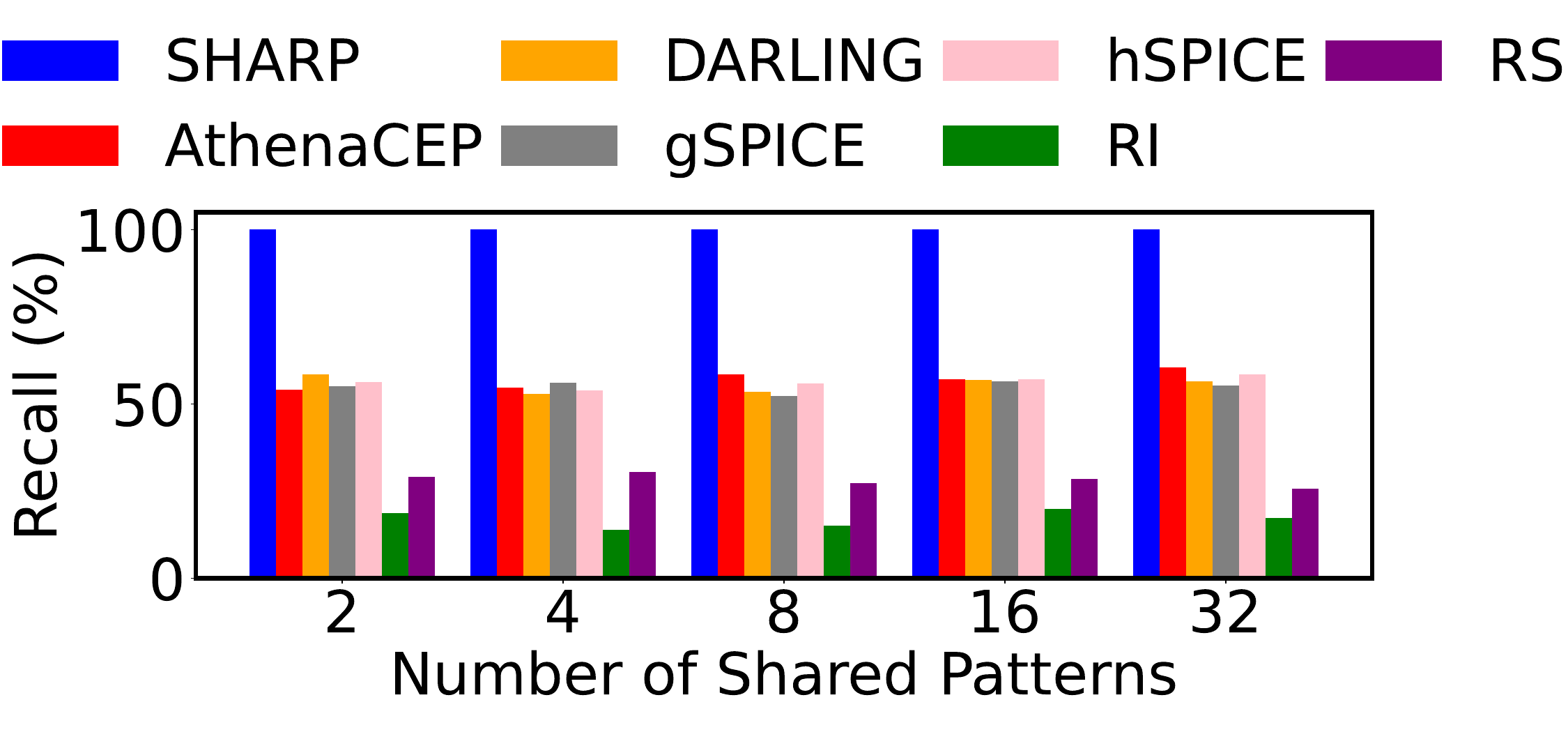}
    \caption{GraphRAG Recall}
    \label{fig:scalability_recall_rag}
  \end{subfigure}
  \hfill
  \begin{subfigure}[b]{0.44\linewidth}
    \centering
    \includegraphics[width=\linewidth]{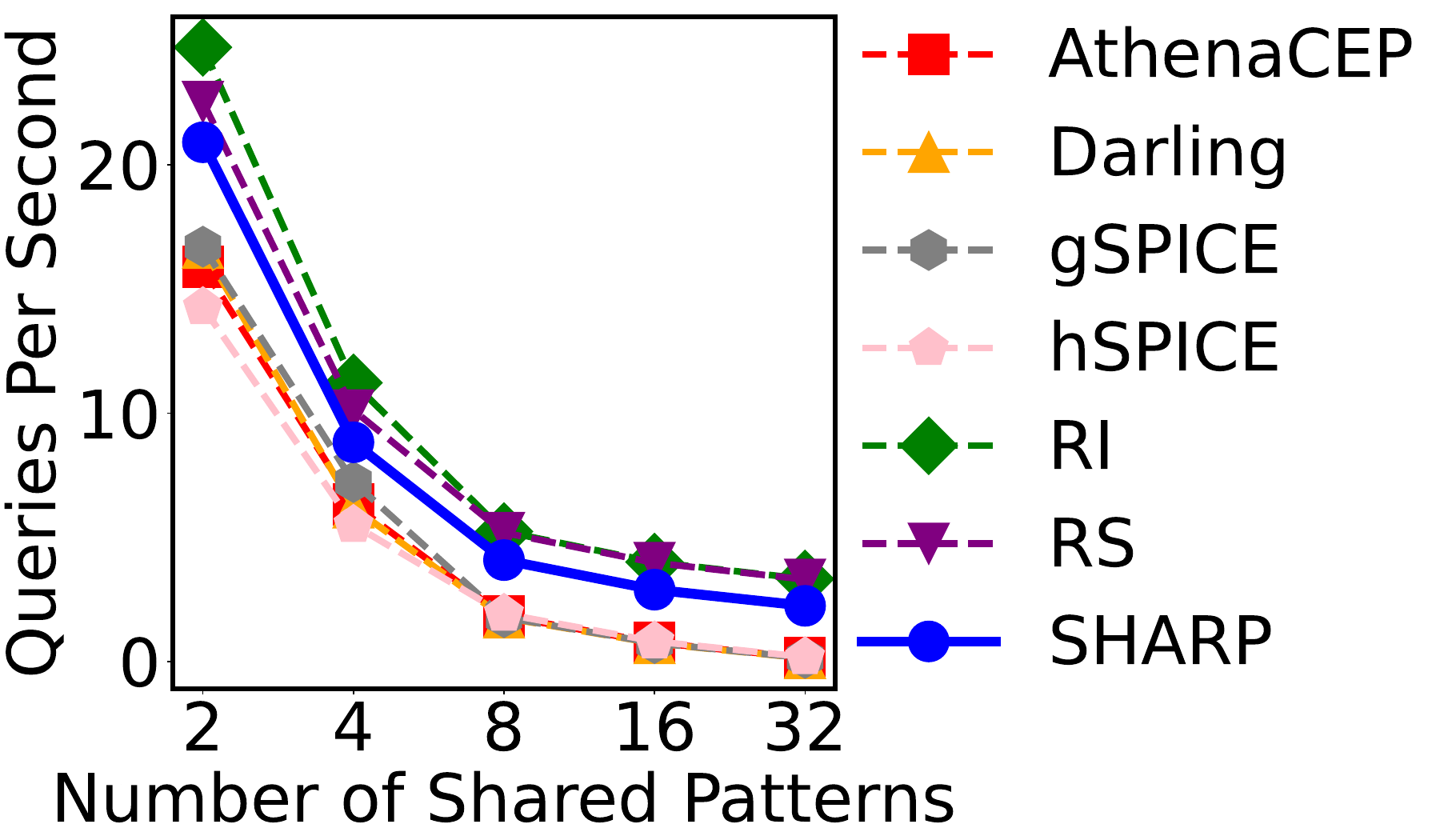}
    \caption{GraphRAG Throughput}
    \label{fig:scalability_th_rag}
  \end{subfigure}
  \caption{Scalability of \sys with shared CEP and \mreg patterns \code{P$_{7}$-P$_{38}$} and GraphRAG patterns from \code{Meta-QA}~\cite{Zhang_Dai_Kozareva_Smola_Song_2018}}
  \label{fig:scalability}
\end{figure}

\textcolor{black}{
We investigate \sys's scalability ranging the number of shared patterns (\code{P$_7$}–\code{P${_{38}}$}) from 2 to 32.
}. 
That is [\code{P$_7$}, \code{P$_8$}] for 2 sharing patterns, [\code{P$_7$}-\code{P$_{10}$}] for 4 sharing patterns, [\code{P$_7$}-\code{P$_{14}$}] for 8 sharing patterns, [\code{P$_7$}-\code{P$_{22}$}] for 16 sharing patterns and [\code{P$_7$}-\code{P$_{38}$}] for 32 sharing patterns.
For GraphRAG, we modify the query prompts of 32 patterns from \code{Meta-QA}~\cite{Zhang_Dai_Kozareva_Smola_Song_2018} to support pattern sharing.
\F\ref{fig:scalability} presents the recall and throughput results.
\sys consistently achieves the highest recall and throughput as the number of shared patterns increases. While its throughput is slightly below the random baselines, the poor recall of \code{RS} and \code{RI} renders them ineffective.
Note that increasing shared patterns significantly increases the number of partial matches by 2 to 16 orders of magnitudes, resulting in deteriorated throughput for all approaches. Nevertheless, \sys’s throughput degrades more gracefully than that of other baselines, demonstrating stronger scalability: it is 1.26$\times$ better than \code{gSPICE} and 1.25$\times$, 1.29$\times$, and 1.28$\times$ better than \code{hSPICE}, \code{AthenaCEP}, and \code{DARLING}, respectively.
This scalability is enabled by \sys’s PSD-based indexing and its cost model, which effectively capture state-sharing schemes.

\subsection{Impact of Selection / Consumption Policies} \label{sec:selection_consumption_policy}

\begin{figure}[t]
 	\centering
     	\begin{subfigure}[b]{0.495\linewidth}
 		\centering
 		\includegraphics[width=\linewidth]{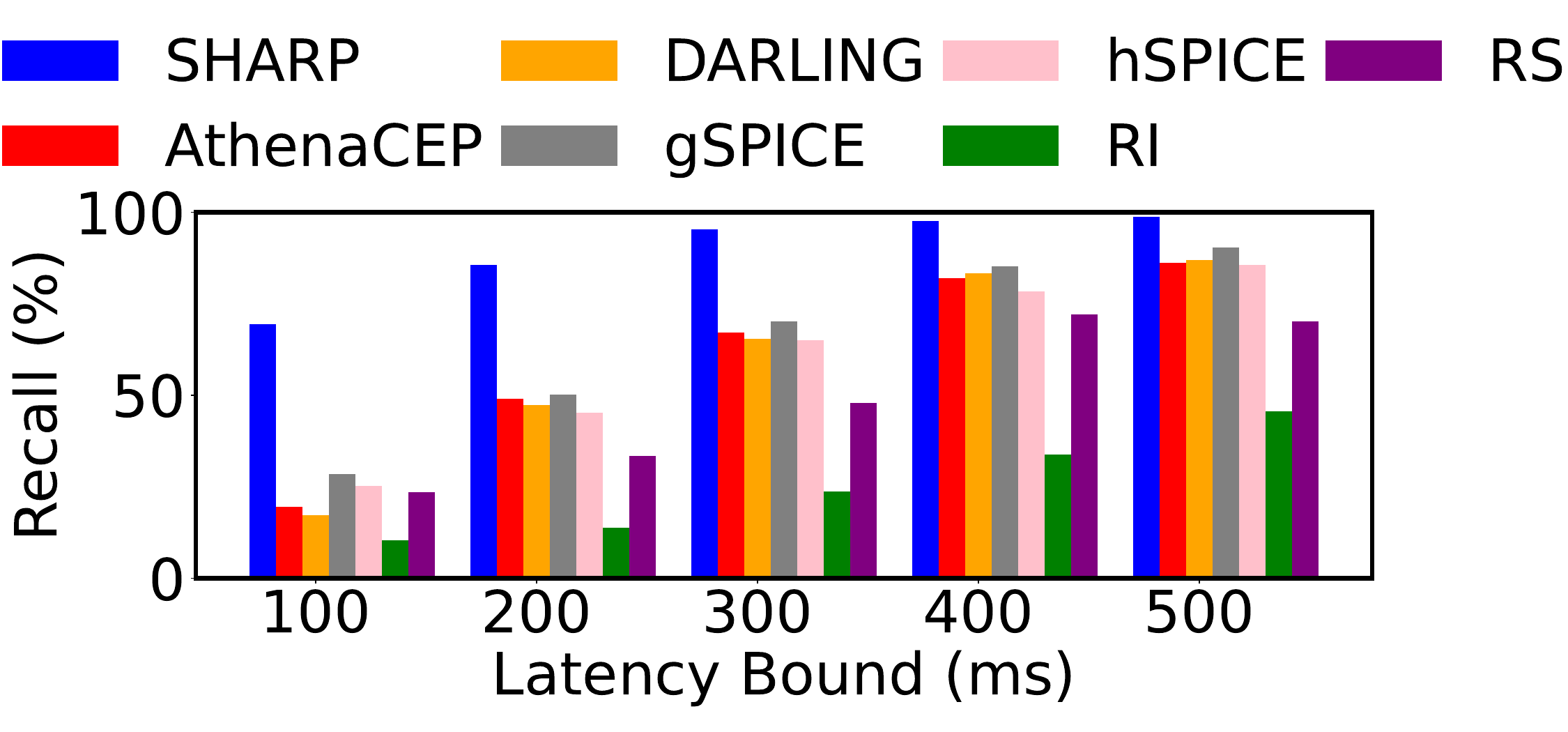}
 		\caption{Recall (Skip-till-any + Reuse)}
 		\label{fig:Skip-till-any-Reuse}
 	\end{subfigure}
 	\hfill
 	\begin{subfigure}[b]{0.495\linewidth}
 		\centering
 		\includegraphics[width=\linewidth]{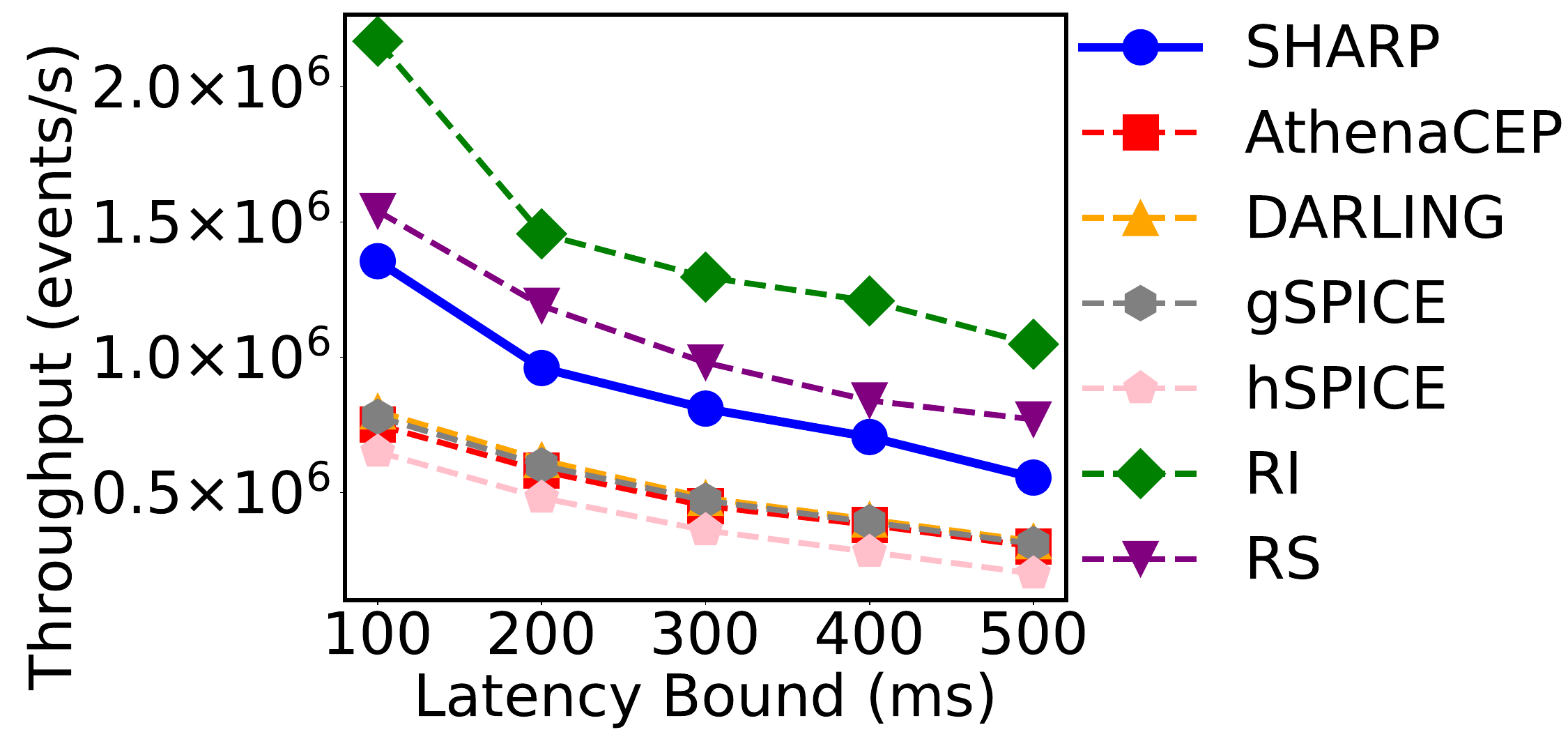}
 		\caption{Throughput (Skip-till-any + Reuse)}
 		\label{fig:Skip-till-any-Consume}
 	\end{subfigure}

    \medskip
    
 	\begin{subfigure}[b]{0.495\linewidth}
 		\centering
 		\includegraphics[width=\linewidth]{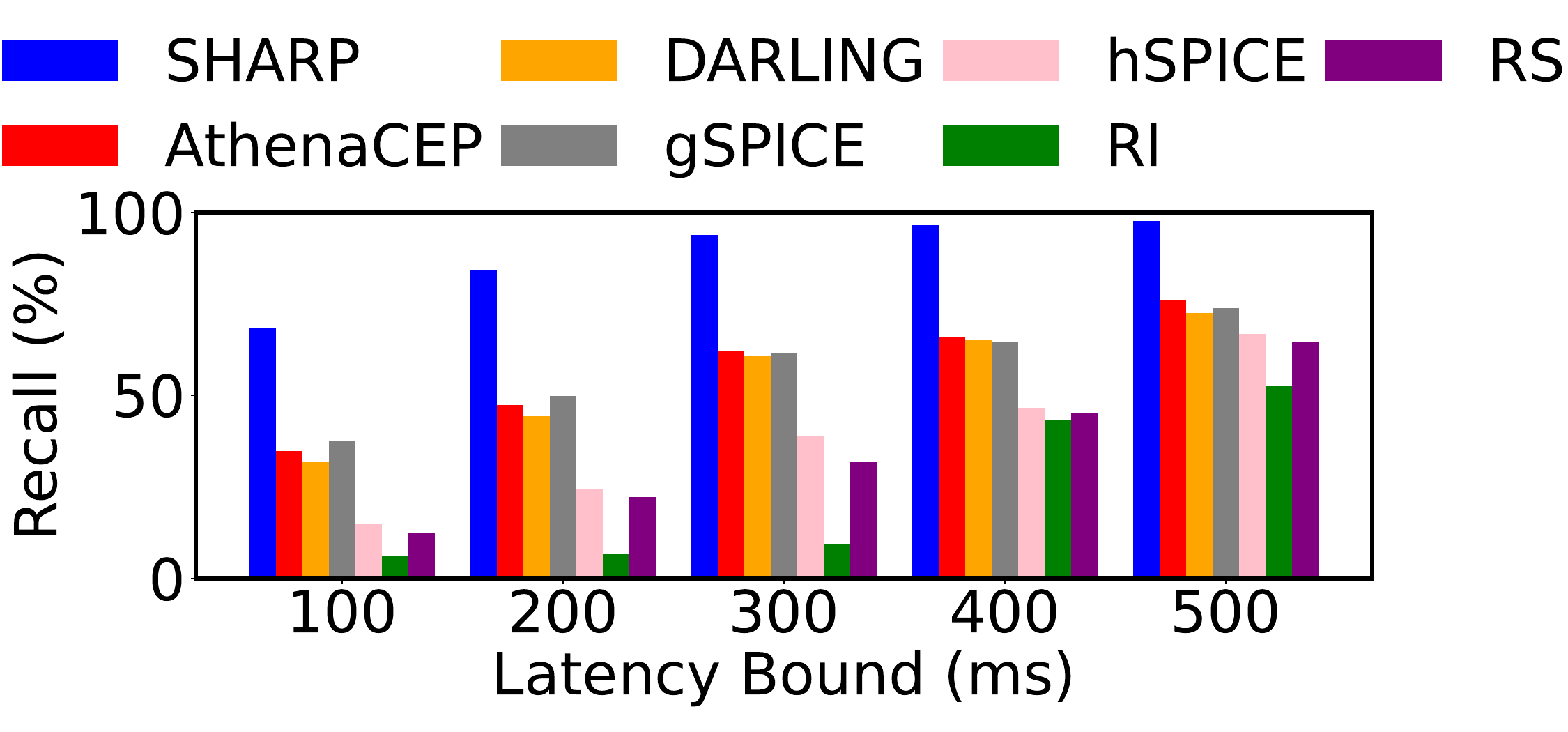}
 		\caption{Recall (Skip-till-any + Consume)}
 		\label{fig:Skip-till-next-Reuse}
 	\end{subfigure}
 	\hfill
 	\begin{subfigure}[b]{0.495\linewidth}
 		\centering
 		\includegraphics[width=\linewidth]{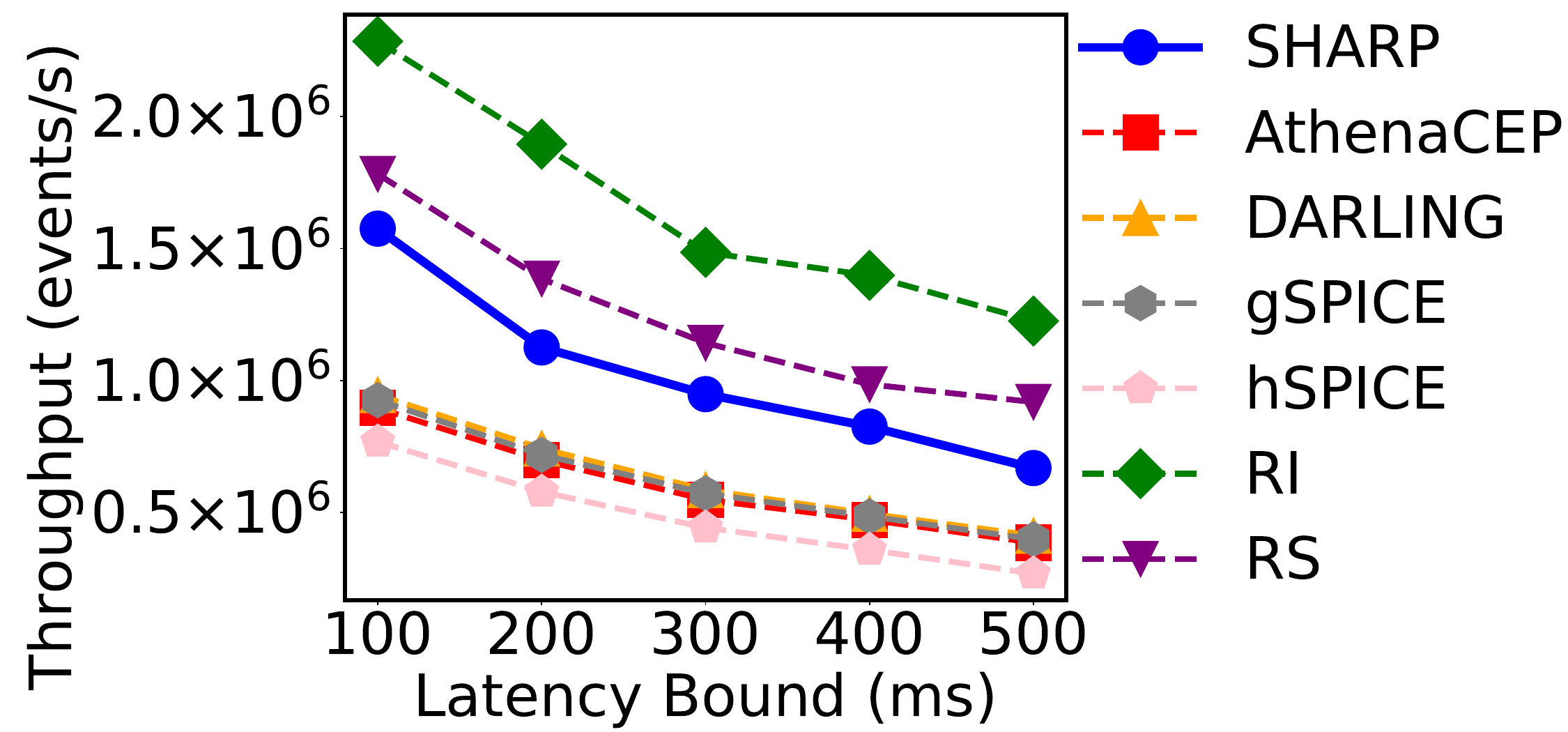}
 		\caption{Throughput (Skip-till-any + Consume)}
 		\label{fig:Strict-contiguity-Reuse}
 	\end{subfigure}

\medskip

 	\begin{subfigure}[b]{0.495\linewidth}
 		\centering
 		\includegraphics[width=\linewidth]{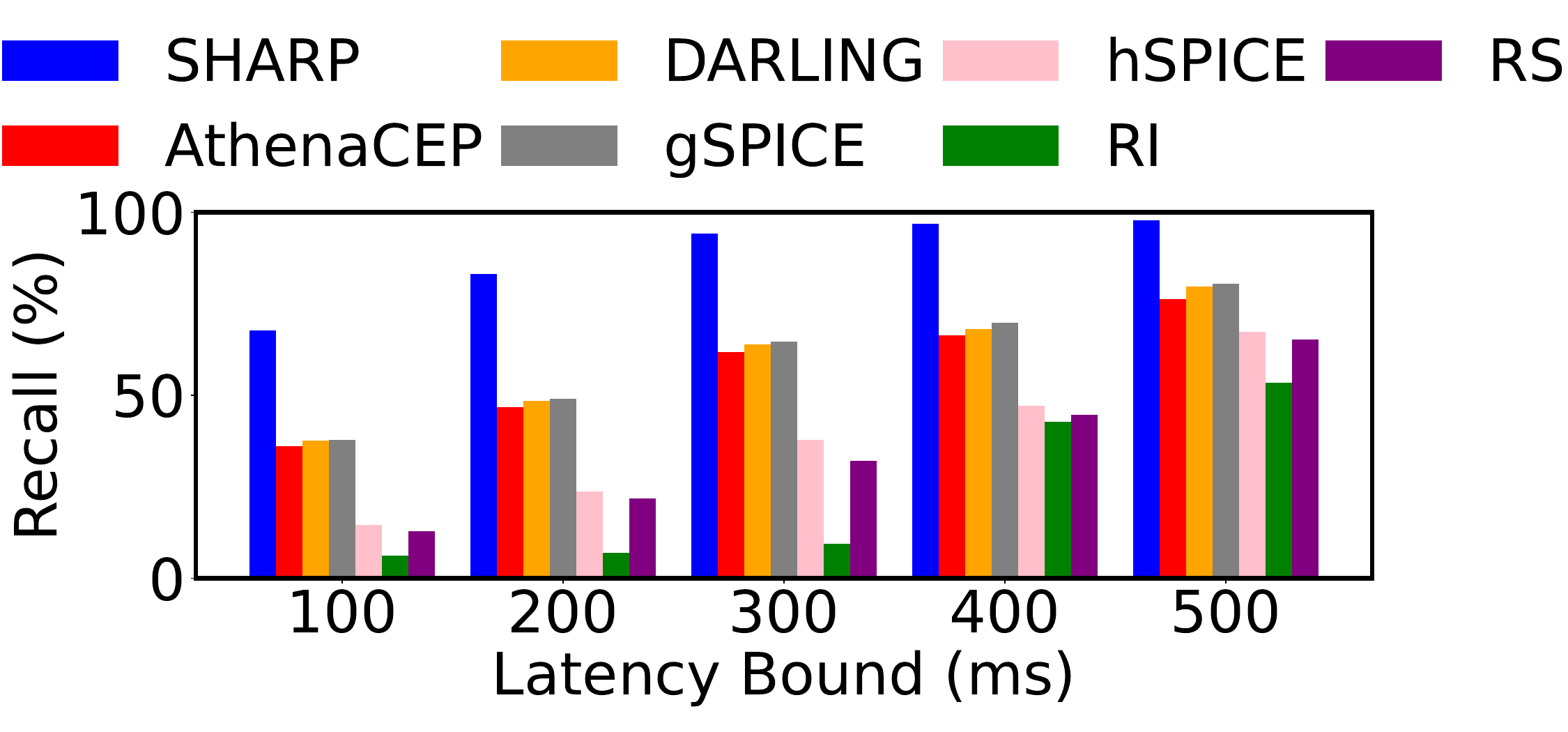}
 		\caption{Recall (Skip-till-next + Reuse)}
 		\label{fig:Skip-till-next-Consume}
 	\end{subfigure}
 	\hfill
 	\begin{subfigure}[b]{0.495\linewidth}
 		\centering
 		\includegraphics[width=\linewidth]{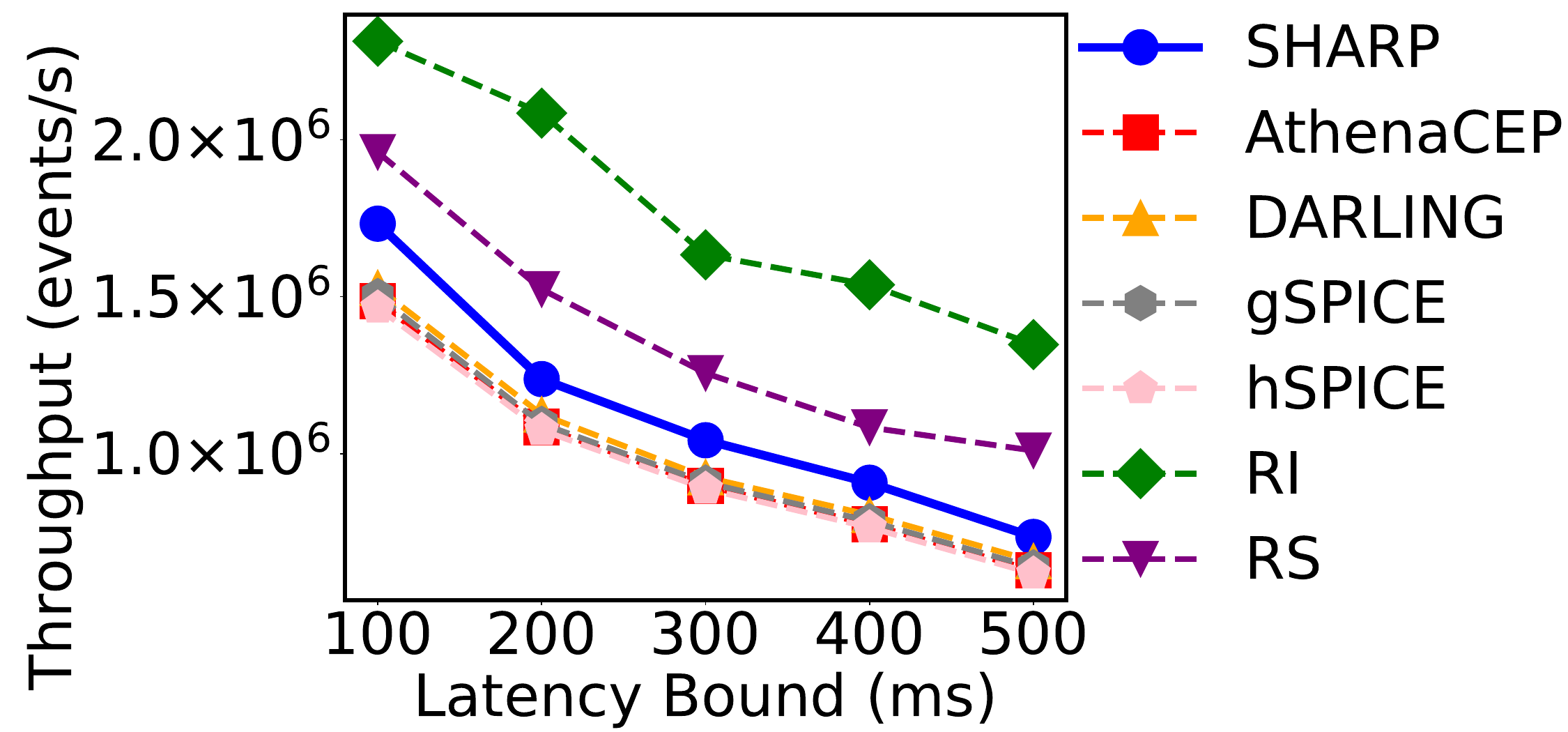}
 		\caption{Throughput (Skip-till-next + Reuse)}
 		\label{fig:Strict-contiguity-Consume}
 	\end{subfigure}

\medskip

     \begin{subfigure}[b]{0.495\linewidth}
 		\centering
 		\includegraphics[width=\linewidth]{figures/sensitivity/skip_till_any_reuse.pdf}
 		\caption{Recall (Skip-till-next + Consume)}
 		\label{fig:Skip-till-any-Reuse}
 	\end{subfigure}
 	\hfill
 	\begin{subfigure}[b]{0.495\linewidth}
 		\centering
 		\includegraphics[width=\linewidth]{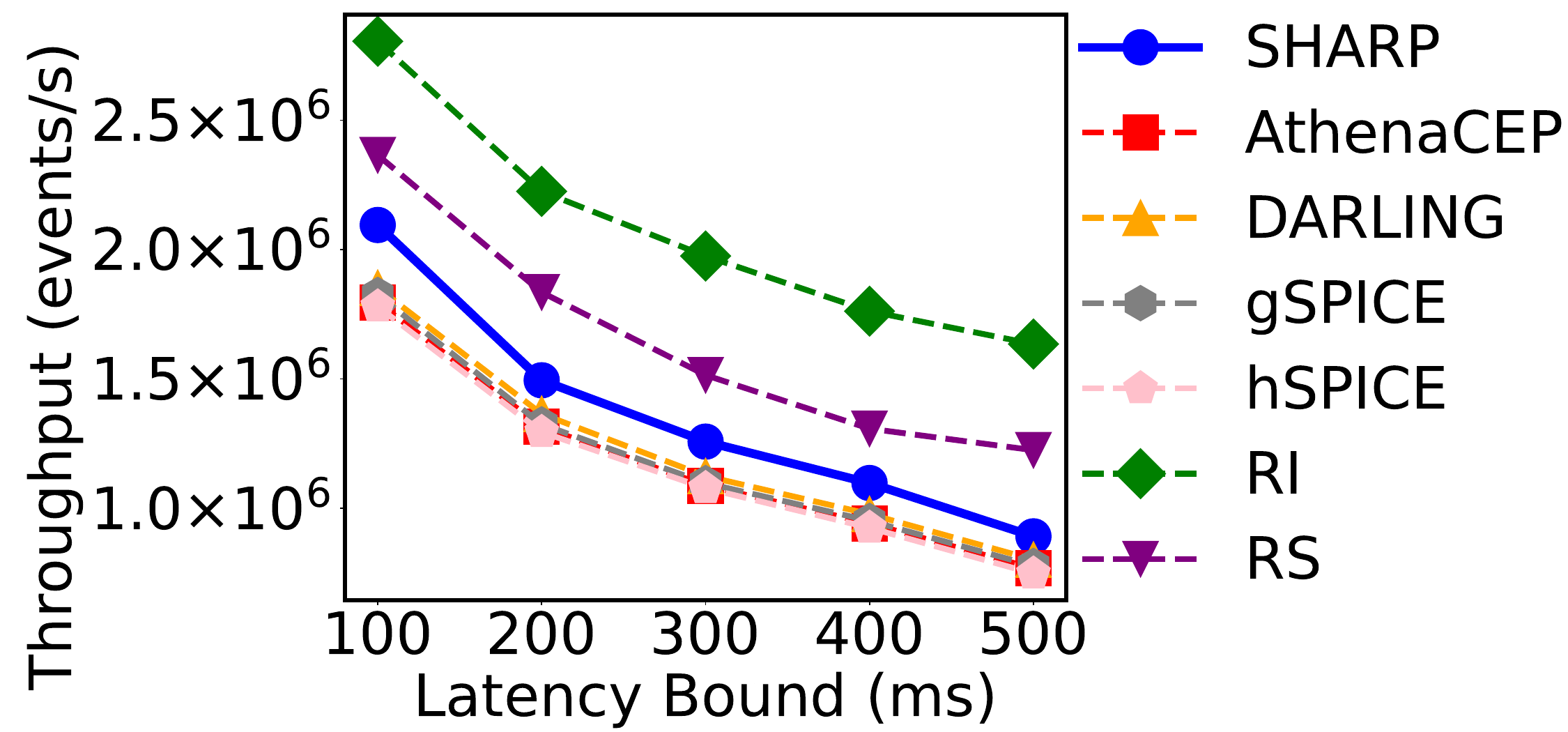}
 		\caption{Throughput(Skip-till-next + Consume)}
 		\label{fig:Skip-till-any-Consume}
 	\end{subfigure}

    \medskip
    
 	\begin{subfigure}[b]{0.495\linewidth}
 		\centering
 		\includegraphics[width=\linewidth]{figures/sensitivity/skip_till_next_reuse.pdf}
 		\caption{Recall (Strict-contiguity + Reuse)}
 		\label{fig:Skip-till-next-Reuse}
 	\end{subfigure}
 	\hfill
 	\begin{subfigure}[b]{0.495\linewidth}
 		\centering
 		\includegraphics[width=\linewidth]{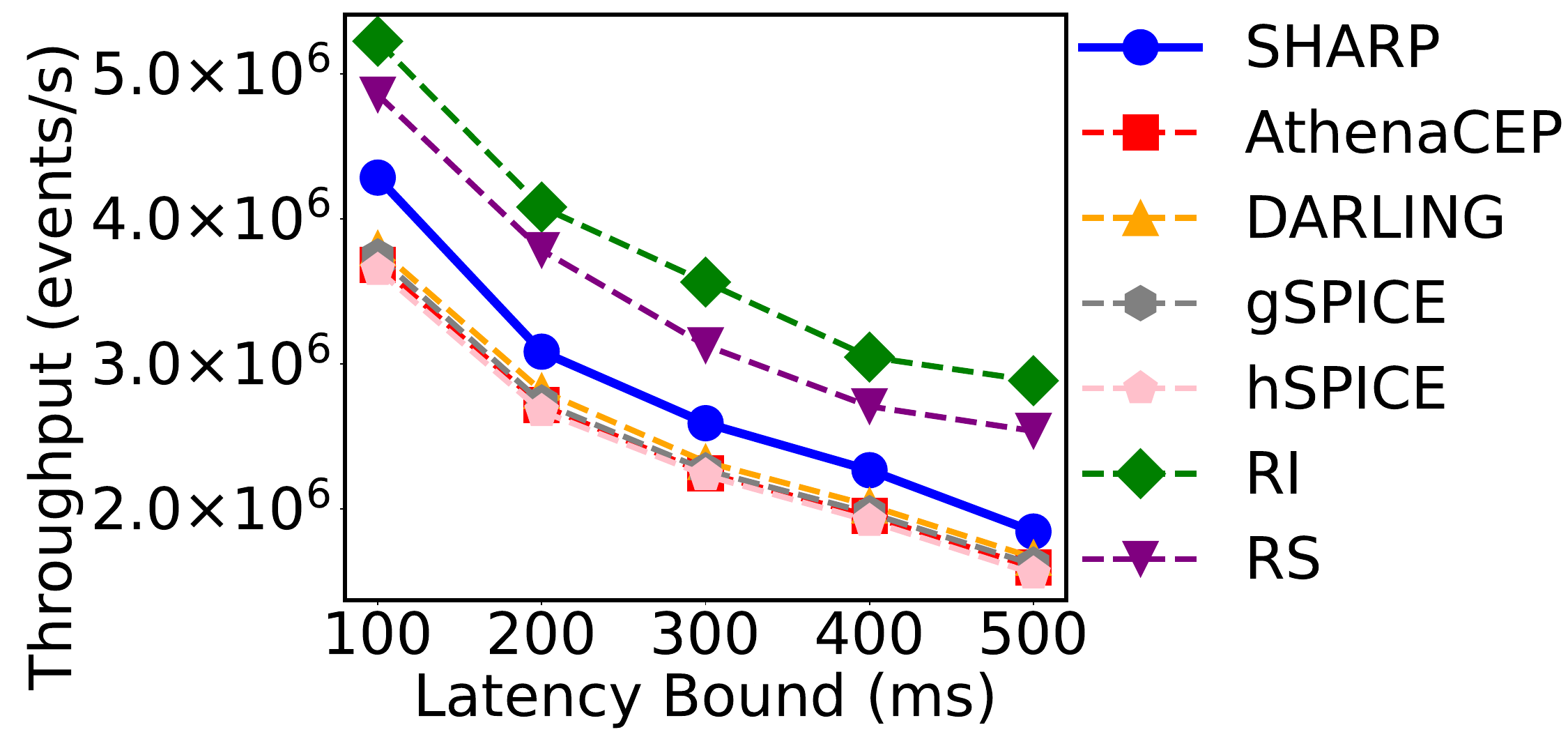}
 		\caption{Throughput (Strict-contiguity + Reuse)}
 		\label{fig:Strict-contiguity-Reuse}
 	\end{subfigure}

\medskip

 	\begin{subfigure}[b]{0.495\linewidth}
 		\centering
 		\includegraphics[width=\linewidth]{figures/sensitivity/strict_contiguity_reuse.pdf}
 		\caption{\footnotesize{Recall(Strict-contiguity+Consume)}}
 		\label{fig:Skip-till-next-Consume}
 	\end{subfigure}
 	\hfill
 	\begin{subfigure}[b]{0.495\linewidth}
 		\centering
 		\includegraphics[width=\linewidth]{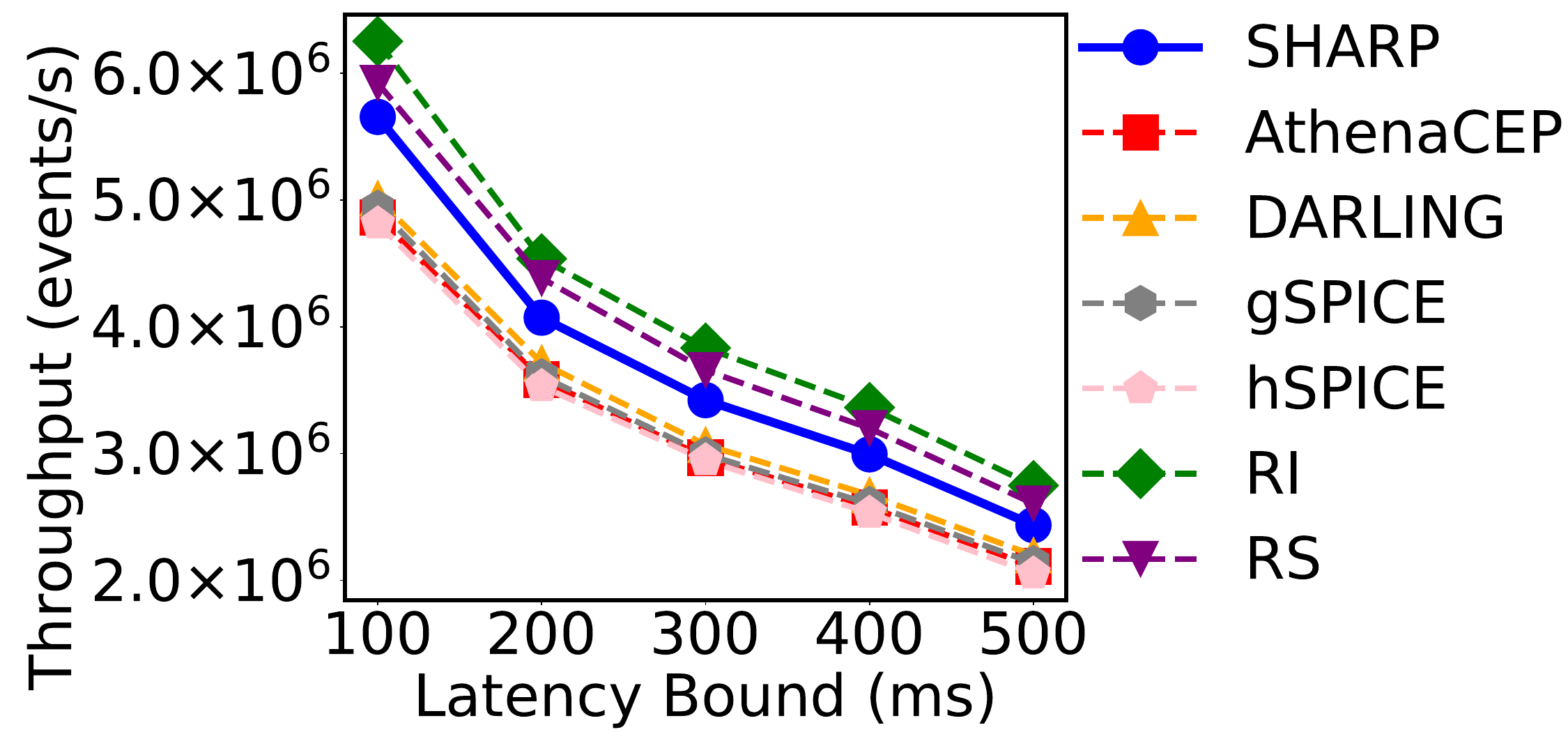}
 		\caption{\footnotesize{Throughput(Strict-contiguity+Consume)}}
 		\label{fig:Strict-contiguity-Consume}
 	\end{subfigure}

 	\caption{Impact of selection / consumption policies (\code{P$_{3}$-P$_4$} over \code{DS1})}
 	\label{fig:policy}
 \end{figure}

The selection and consumption policies (see \S\ref{sec:need4pattern})~\cite{wu2006high,chakravarthy1994snoop,zimmer1999semantics,cugola2010tesla} affect the number of  partial matches, \ie performance, of CEP.
We examine the impact of different policy configurations, using the shared patterns \code{P$_3$} and \code{P$_4$} over \code{Citi\_Bike}~\cite{bike}. 
We consider the selection policies: \code{skip-till-any}, \code{skip-till-next}, and \code{strict-contiguity} and  consumption policies: \code{reuse} and \code{consume} \cite{wu2006high,chakravarthy1994snoop,zimmer1999semantics,cugola2010tesla}. 
As showed in \F\ref{fig:policy}, \sys outperforms baselines in all policy configurations, especially in \code{skip-till-any} (see \F\ref{fig:policy}a-b).
Here, the recall of baselines drops sharply under tight latency bounds.
At 100\unit{ms}, \sys achieves 68.3\% recall, compared  \code{gSPICE}'s 27.9\%, \code{hSPICE}'s 24.8\%, \code{DARLING}'s 18.0\% and  \code{AthenaCEP}'s 20.1\%. 
This is because the configuration of \code{skip-till-next} \code{+}  \code{reuse} (or \code{consume})  produces the most partial matches, which suits \sys's PSD design and cost model to efficiently select the most promising states under strict latency bounds.  \looseness = -1

 \subsection{Integration/Comparison  with \texttt{\small Neo4j-GraphRAG}} \label{sec:neo4j}

We investigate how \sys (as a stand-alone state reduction library) improves the performance of a leading industrial GraphRAG system, \code{Neo4j-GraphRAG}~\cite{neo4j-graphRAG-python}.
To this end, we integrated \sys into \code{Neo4j-GraphRAG}, 
coined as \code{Neo4j-GraphRAG+\sys}.
We integrated \sys into \code{Neo4j-GraphRAG} as \code{Neo4j-GraphRAG+\sys}
\footnote{https://github.com/benyucong/SHARP/tree/master/Graph\%20RAG/neo4j-graphrag-python-hop3
} as follows:
(i) we enforce intermediate-result sharing by merging and jointly executing the three path queries; (ii) we leverage \code{apoc.text.distance} from the Neo4j APOC library~\cite{neo4j_apoc} to approximate the contribution of intermediate results: this function computes textual distance as a proxy for semantic similarity, where smaller values indicate higher relevance and thus higher contribution; (iii) we model the computational overhead using the in-memory footprint of intermediate results; and (iv) we rank intermediate results using the previously defined cost model, and apply \code{LIMIT} based selection following the algorithm in \S4.3, ensuring that the overall execution stays within the specified latency bounds.

To construct the baselines, we extended the default \code{Neo4j-\\GraphRAG} with state sharing and state reduction.
(i) \code{Neo4j-GraphRAG}: the default Neo4j-GraphRAG that executes LLM-generated queries without shared state and state reduction. 
(ii) \code{Neo4j-GraphRAG-Shared}: we enhance the Neo4j-GraphRAG with shared state but without state reduction. 
(iii)  \code{Neo4j-GraphRAG-RS}:  Neo4j-GraphRAG enhanced with random state reduction but without shared state.
(iv) \code{Neo4j-\\GraphRAG-Shared-RS}: Neo4j-GraphRAG with  shared state and random state reduction.
, and compare its performance to four baselines of  \code{Neo4j-GraphRAG}.
We constructed the GraphRAG pipeline using the Qwen3-8B~\cite{qwen3_8b} LLM and evaluated 14,273 \emph{3-hop} path queries from \code{KG-MetaQA}~\cite{Zhang_Dai_Kozareva_Smola_Song_2018}.

\begin{figure}[t]
	\centering
	
	\begin{minipage}{0.9\linewidth}
		\centering
		\includegraphics[width=\linewidth]{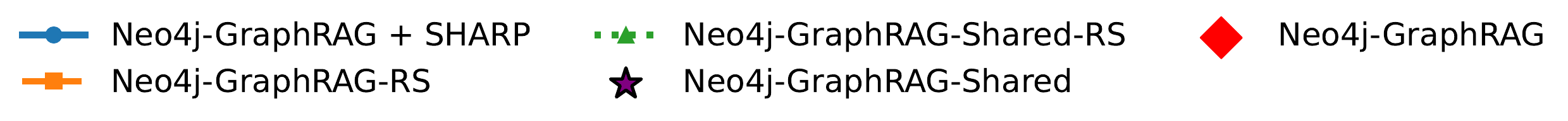}
	\end{minipage}
	
	\vspace{0.2em} 
	
	\begin{subfigure}[b]{0.32\linewidth}
		\centering
		\includegraphics[width=\linewidth]{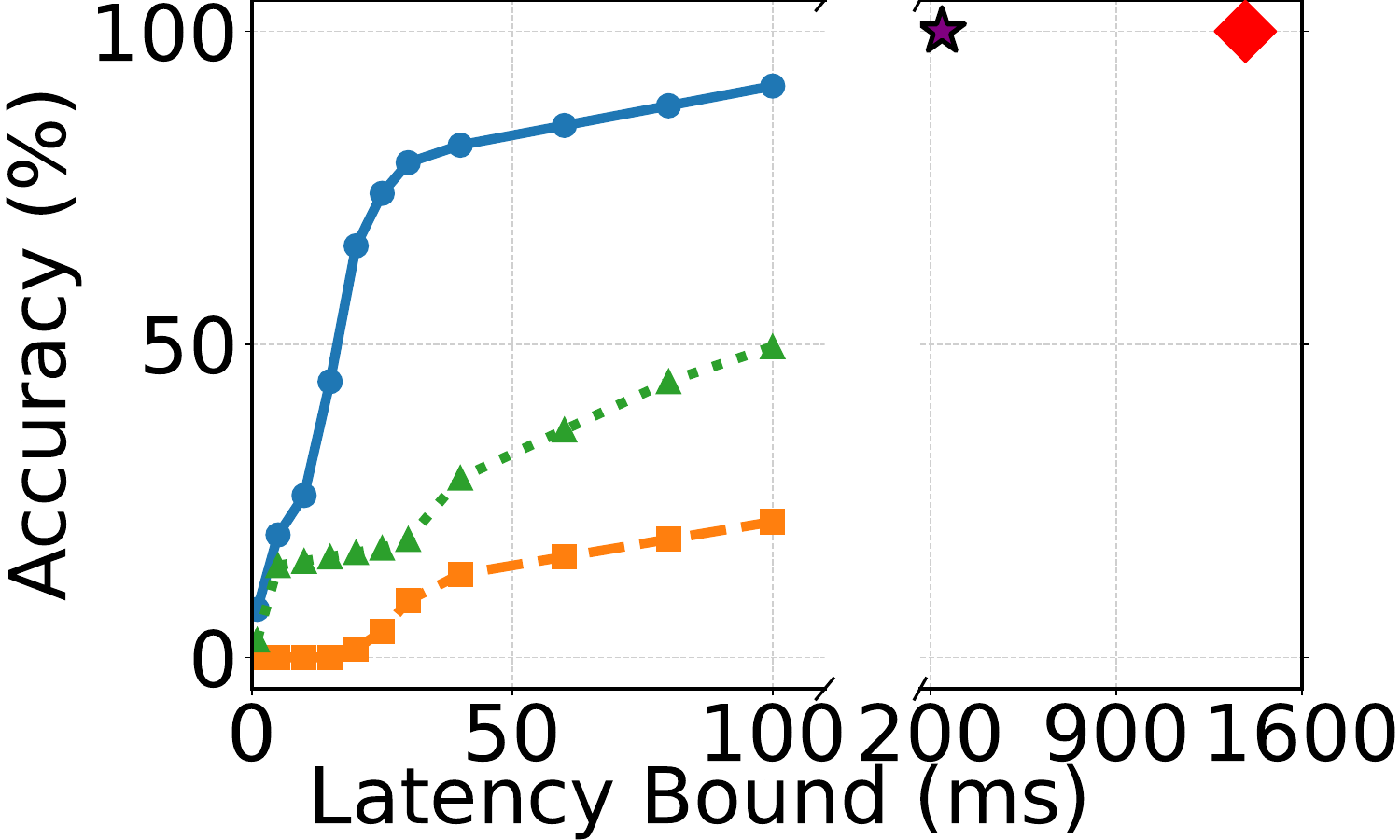}
		\caption{Accuracy}
		\label{fig:3-hop-accuracy}
	\end{subfigure}
	\hfill
	\begin{subfigure}[b]{0.32\linewidth}
		\centering
		\includegraphics[width=\linewidth]{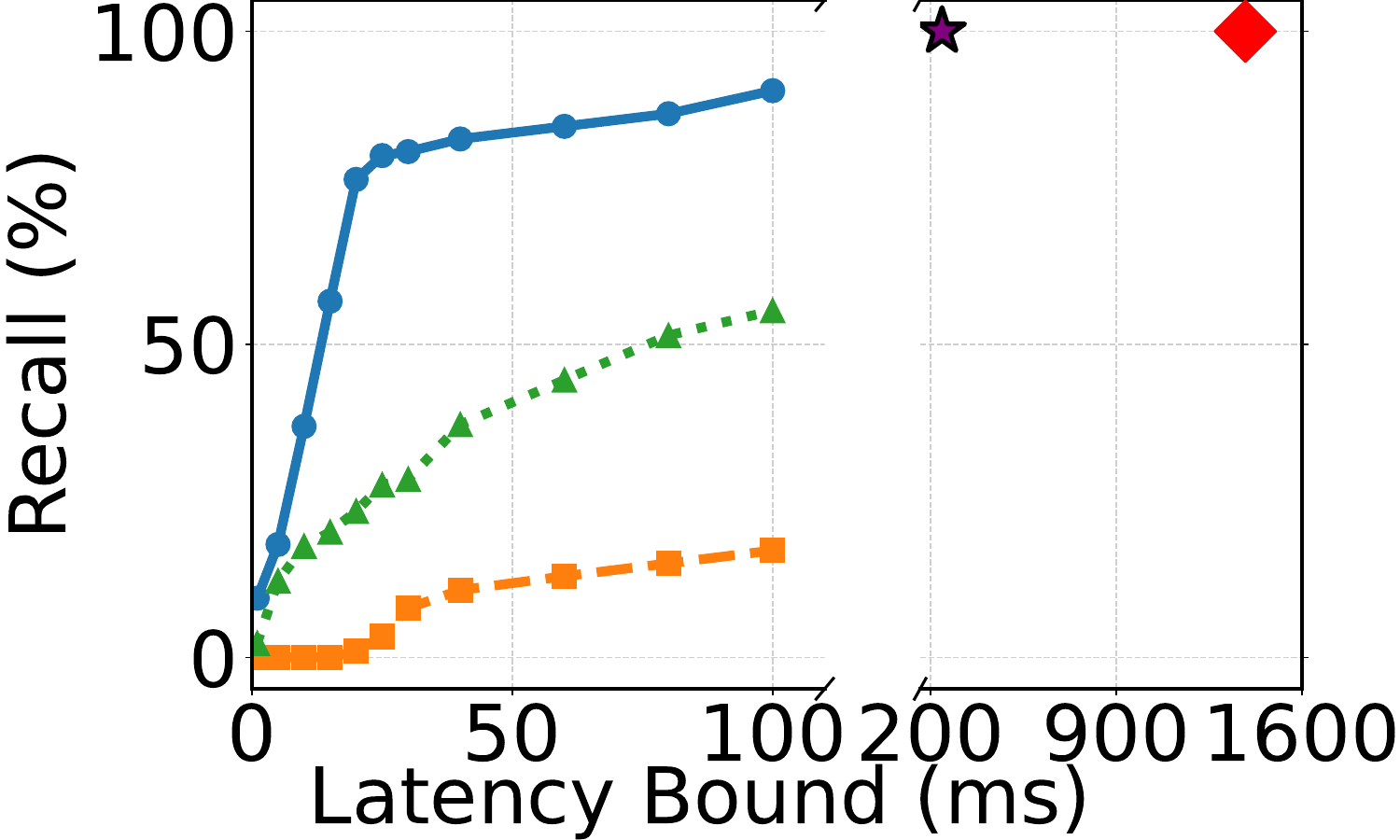}
		\caption{Recall}
		\label{fig:3-hop-recall}
	\end{subfigure}
	\hfill
	\begin{subfigure}[b]{0.32\linewidth}
		\centering
		\includegraphics[width=\linewidth]{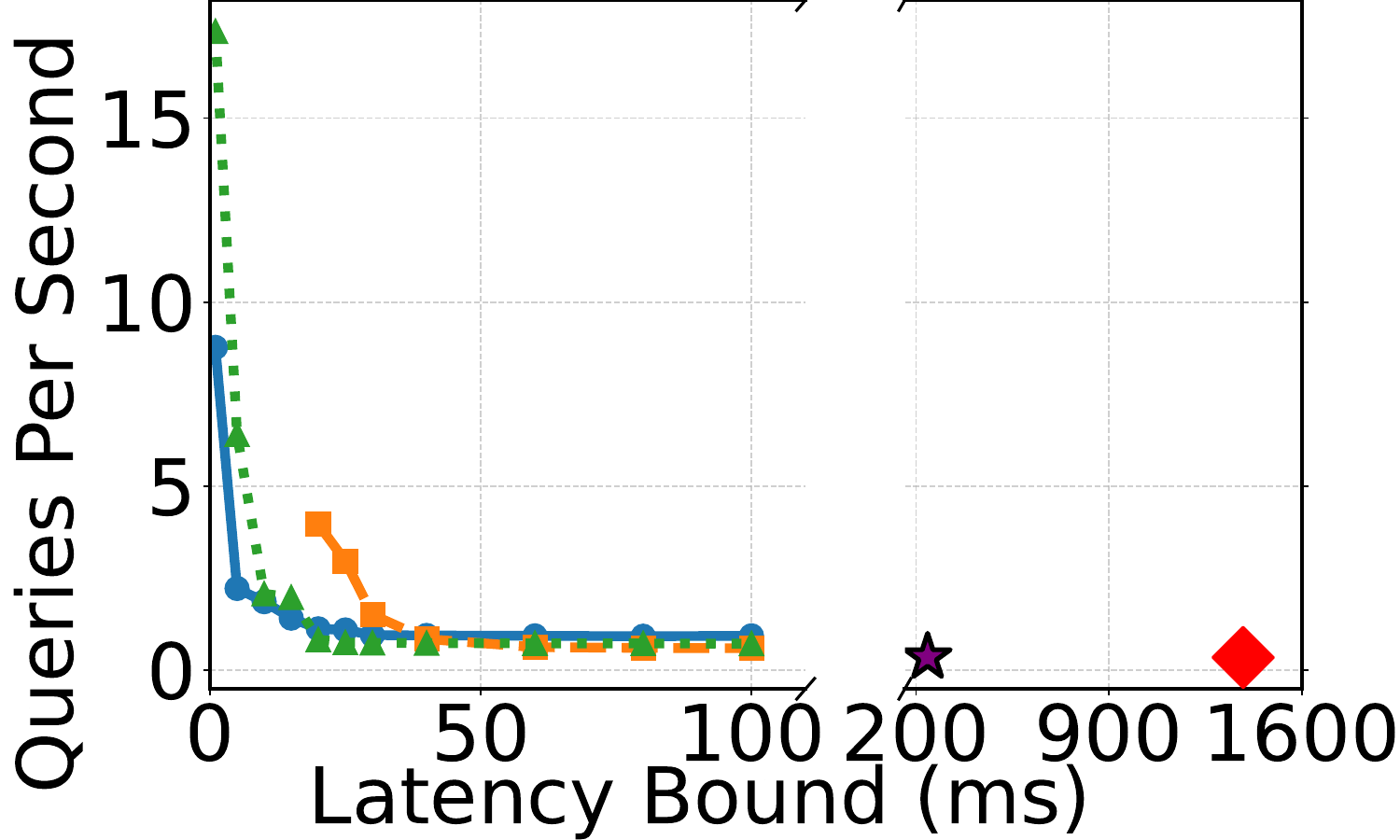}
		\caption{Throughput}
		\label{fig:3-hop-throughput}
	\end{subfigure}
	
	\caption{Performance comparison between \sys-enhanced Neo4j GraphRAG and baselines}
\label{fig:3-hop}
\vspace{-1em}
\end{figure}

\F\ref{fig:3-hop} shows the accuracy, recall, and throughput obtained with all approaches. 
Without state reduction, \code{Neo4j-GraphRAG} achieves 100\% accuracy and recall, but with an unacceptably high latency of 1,429\unit{ms}. \code{Neo4j-GraphRAG+Shared} improves this by sharing state across multiple path queries, reducing the latency to 242\unit{ms}.

We further study the performance of state reduction under latency bounds ranging from 10\unit{ms} to 100\unit{ms}.
\code{Neo4j-GraphRAG+\sys} consistently achieves the highest accuracy, recall and throughput.
At 100\unit{ms} bound,  \code{Neo4j-GraphRAG+\sys} achieves 91.2\% accuracy and 90.8\% recall: $1.84\times$ and $1.63\times$ higher than \code{Neo4j-GraphRAG-}\code{Shared-RS}, $4.21\times$ and $5.30\times$ higher than \code{Neo4j-GraphRAG-RS}. 
The largest margin is around the 40\unit{ms} bound: \code{Neo4j-GraphRAG+\sys} achieves 84.9\% accuracy and 87.3\% recall, compared to 28.7\% and 37.3\% for \texttt{\footnotesize Neo4j-GraphRAG-Shared-RS}, and only 12.3\% and 10.8\% for \code{Neo4j-GraphRAG-RS}.
We conclude that with \sys, \code{Neo4j-GraphRAG} supports tighter latency constraints with little quality loss.\looseness=-1

\subsection{Optimality of \sys's State Selection} \label{sec:optimal}
\begin{figure}[t]
  \centering
  \begin{subfigure}[b]{0.495\linewidth}
    \centering
    \includegraphics[width=\linewidth]{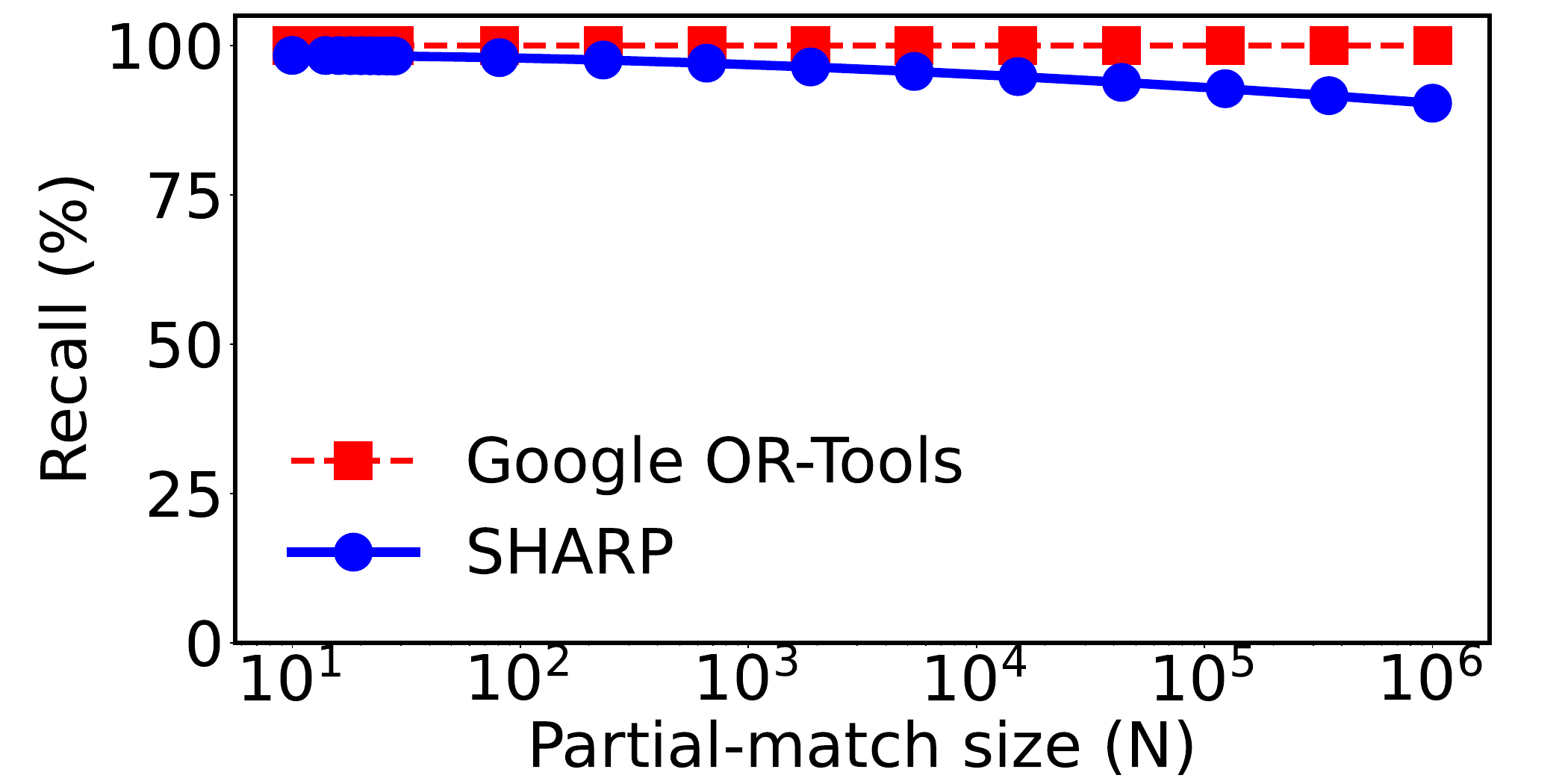}
    \caption{Recall changing with PM size}
    \label{fig:recall_pm}
  \end{subfigure}
  \hfill
  \begin{subfigure}[b]{0.495\linewidth}
    \centering
    \includegraphics[width=\linewidth]{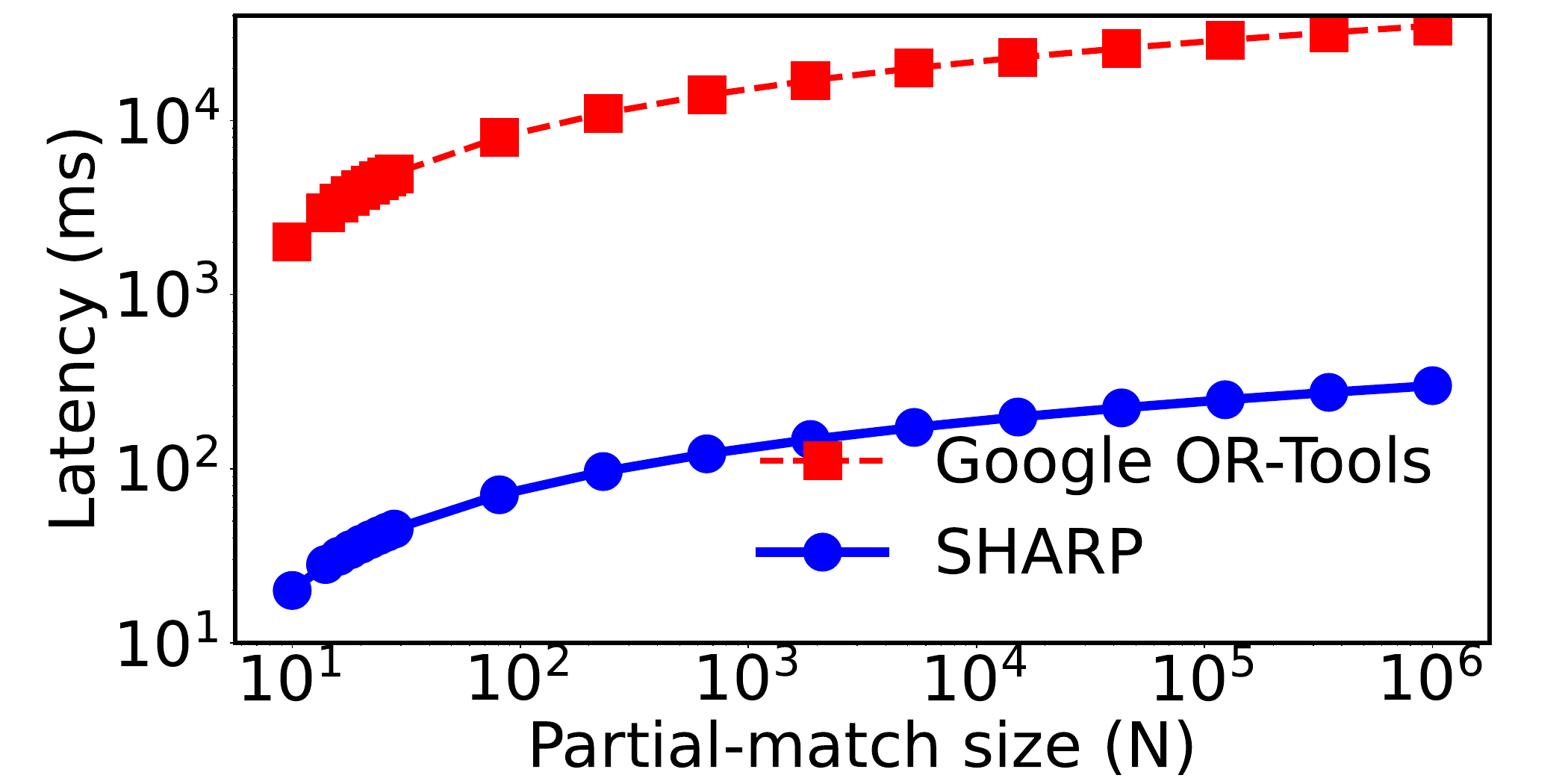}
    \caption{Latency changing with PM size}
    \label{fig:latency_pm}
  \end{subfigure}
  \caption{\sys compared with dynamic programming}
  \label{fig:optimality_pm}
  \vspace{-2em}
\end{figure}

We invesigate the performance of \sys's cost model in state selection compared to the theoretically optimal dynamic programming (DP) oracle. 
To this end, we compare \sys's state selection to a leading industrial DP solver, Google OR-Tools (CP-SAT)~\cite{cpsatlp}, in settings in which the number of partial matches increases from 10 to  $10^6$.
\F\ref{fig:recall_pm} shows the recall and \F\ref{fig:latency_pm} shows the processing latency when selecting the state. 
At $10^2$ partial matches,  \sys achieves 99.3\% recall in selection and 70.9\unit{ms} latency. 
In contrast, CP-SAT keeps 100\% recall but with $3.13\times 10^3$\unit{ms} latency ($44.1\times$ slower than \sys). 
When selecting $10^6$ partial matches,  \sys maintains 90.3\% recall and 313\unit{ms} latency. CP-SAT still keeps 100\% recall but the latency becomes unacceptably high, $3.45\times10^4$\unit{ms} ($110\times$ slower).
The results show that \sys's state selection achieves comparable quality with the theoretically optimal solution at much lower computational overhead (at least two orders of magnitude faster).
In contrast, the theoretically optimal solution is impractical for low-latency constraints due to its high computational complexity.\looseness=-1

\section{Related Work}\label{sec:related}

\mypar{Multi-pattern sharing optimizations} 
Multi-pattern optimization techniques~\cite{zhang2017multi,ray2016scalable,kolchinsky2019real,poppe2018sharon,GRETA2017,poppe2021share,ma2022gloria} reuse the computation of shared sub-patterns and reduce memory footprint.
 SPASS~\cite{ray2016scalable} estimates the benefit of
sharing based on intra- and inter- query correlations.
Sharon~\cite{poppe2018sharon} and HAMLET~\cite{poppe2021share} further
support online aggregation.
While MCEP~\cite{kolchinsky2019real}, GRETA~\cite{GRETA2017} and
GLORIA~\cite{ma2022gloria} allow sharing in \code{Kleene} closure.
These approaches are complementary to \sys, with the focus on improving resource utilization. In contrast, \sys's focal point is best-effort processing to satisfy latency bound. We have integrated above sharing schemes into \sys. \looseness=-1

\mypar{Load shedding} Load shedding techniques discard a set of data elements or partial matches without processing them based on estimated utility~\cite{tatbul2003load, dindar2011efficiently, eSPICE, he2013load, hSPICE, slo2019pspice, chapnik2021darling, zhao2020load}.
Input-based shedding~\cite{eSPICE,hSPICE,chapnik2021darling} drops input data based on their estimated importance of the final results.
In contrast, state-based shedding~\cite{DBLP:conf/icde/Zhao18,slo2019pspice} discards partial matches using utilities based on probabilistic models.
 Hybrid shedding~\cite{zhao2020load}, on the other hand, combines the shedding of input events and partial matches and uses a cost model to balance the trade-offs.
However, \emph{the above load shedding schemes have not  considered the interaction among multiple patterns via shared state}.

\myparr{GraphRAG}~\cite{sun2024surveygraphrag} improves RAG by retrieving data relationships from knowledge graphs~\cite{yin2024graphrag,liu2024graphrag,wang2023knowledgegpt,yang2024hybridrag}.
Neo4j~\cite{neo4j} and Memgraph~\cite{memgraph} accelerate single-query execution through query planning and caching, but they lack the support for sharing state across multiple path queries and best effort processing for strict latency bounds.  These limitations motivate the design of state reduction in \sys to enhance GraphRAG systems.

\mypar{Approximate query processing (AQP)}
AQP estimates the result of  queries~\cite{DBLP:conf/sigmod/ChaudhuriDK17} to fast approximate answers based on sampling~\cite{DBLP:conf/sigmod/0002SSK21} or workload
knowledge~\cite{DBLP:journals/pacmmod/JoT24,DBLP:conf/sigmod/ParkMSW18}. For aggregation queries, sketches~\cite{DBLP:journals/ftdb/CormodeGHJ12} are
employed for efficient, but lossy data stream processing.
 AQP was also explored for sequential pattern matching~\cite{DBLP:journals/pvldb/LiG16a}, focusing on delivering complete matches that deviate from what is
specified in a pattern.
Although AQP aims at best-effort processing, the goal is different from \sys. \sys detects \emph{exact} complete matches that are defined in patterns, not the approximated ones.

\section{Conclusions}\label{sec:conclusions}

We described \sys, a state management library for best-effort processing in shared pattern matching. 
The goal is to satisfy strict latency bounds (specified in applications' SLO), while maximizing the results quality.
\sys proposes the new abstraction of \emph{pattern-sharing degree} (PSD) to capture state sharing schemes and interactions across multiple patterns and a \emph{cost model} to assess the importance and computational overhead of shared state, \ie partial matches.
\sys' \emph{state selector} efficiently selects a set partial matches in a hierarchical manner for further processing. The optimizations of PSD-based bitmap indexing and partial ordering of cost models enable the state selection in constant time.
We have comprehensively evaluated the efficacy of \sys using real-world data in CEP, OLAP and GraphRAG, compare to several SOTA baselines.

\section{Acknowledgment}
This work is funded by Research Council of Finland (grant number 362729),
Business Finland (grant number
169/31/2024).
The authors thank Sukanya Bhowmik and Ahmad Slo for providing the pointers to the baseline implementations of hSPICE and gSPICE.

\balance
\bibliographystyle{ACM-Reference-Format}
\bibliography{sample-base}

@String{Computing = "Computing" }

@String{Computer = "{IEEE} Computer" }

@inproceedings{zhao2020load,
  title={Load shedding for complex event processing: Input-based and state-based techniques},
  author={Zhao, Bo and Hung, Nguyen Quoc Viet and Weidlich, Matthias},
  booktitle={2020 IEEE 36th International Conference on Data Engineering (ICDE)},
  pages={1093--1104},
  year={2020},
  organization={IEEE}
}

@inproceedings{DBLP:conf/sigmod/0002SSK21,
	author       = {Xi Liang and
	Stavros Sintos and
	Zechao Shang and
	Sanjay Krishnan},
	editor       = {Guoliang Li and
	Zhanhuai Li and
	Stratos Idreos and
	Divesh Srivastava},
	title        = {Combining Aggregation and Sampling (Nearly) Optimally for Approximate
	Query Processing},
	booktitle    = {{SIGMOD} '21: International Conference on Management of Data, Virtual
	Event, China, June 20-25, 2021},
	pages        = {1129--1141},
	publisher    = {{ACM}},
	year         = {2021},
	url          = {https://doi.org/10.1145/3448016.3457277},
	doi          = {10.1145/3448016.3457277},
	bibsource    = {dblp computer science bibliography, https://dblp.org}
}

@article{DBLP:journals/ftdb/CormodeGHJ12,
	author       = {Graham Cormode and
	Minos N. Garofalakis and
	Peter J. Haas and
	Chris Jermaine},
	title        = {Synopses for Massive Data: Samples, Histograms, Wavelets, Sketches},
	journal      = {Found. Trends Databases},
	volume       = {4},
	number       = {1-3},
	pages        = {1--294},
	year         = {2012},
	url          = {https://doi.org/10.1561/1900000004},
	doi          = {10.1561/1900000004},
	bibsource    = {dblp computer science bibliography, https://dblp.org}
}

@article{DBLP:journals/pvldb/LiG16a,
	author       = {Zheng Li and
	Tingjian Ge},
	title        = {History is a mirror to the future: Best-effort approximate complex
	event matching with insufficient resources},
	journal      = {Proc. {VLDB} Endow.},
	volume       = {10},
	number       = {4},
	pages        = {397--408},
	year         = {2016},
	url          = {http://www.vldb.org/pvldb/vol10/p397-ge.pdf},
	doi          = {10.14778/3025111.3025121},
	bibsource    = {dblp computer science bibliography, https://dblp.org}
}

@inproceedings{DBLP:conf/sigmod/ParkMSW18,
	author       = {Yongjoo Park and
	Barzan Mozafari and
	Joseph Sorenson and
	Junhao Wang},
	editor       = {Gautam Das and
	Christopher M. Jermaine and
	Philip A. Bernstein},
	title        = {VerdictDB: Universalizing Approximate Query Processing},
	booktitle    = {Proceedings of the 2018 International Conference on Management of
	Data, {SIGMOD} Conference 2018, Houston, TX, USA, June 10-15, 2018},
	pages        = {1461--1476},
	publisher    = {{ACM}},
	year         = {2018},
	url          = {https://doi.org/10.1145/3183713.3196905},
	doi          = {10.1145/3183713.3196905},
	bibsource    = {dblp computer science bibliography, https://dblp.org}
}

@inproceedings{DBLP:conf/sigmod/ChaudhuriDK17,
	author       = {Surajit Chaudhuri and
	Bolin Ding and
	Srikanth Kandula},
	editor       = {Semih Salihoglu and
	Wenchao Zhou and
	Rada Chirkova and
	Jun Yang and
	Dan Suciu},
	title        = {Approximate Query Processing: No Silver Bullet},
	booktitle    = {Proceedings of the 2017 {ACM} International Conference on Management
	of Data, {SIGMOD} Conference 2017, Chicago, IL, USA, May 14-19, 2017},
	pages        = {511--519},
	publisher    = {{ACM}},
	year         = {2017},
	url          = {https://doi.org/10.1145/3035918.3056097},
	doi          = {10.1145/3035918.3056097},
	bibsource    = {dblp computer science bibliography, https://dblp.org}
}

@article{DBLP:journals/pacmmod/JoT24,
	author       = {Saehan Jo and
	Immanuel Trummer},
	title        = {ThalamusDB: Approximate Query Processing on Multi-Modal Data},
	journal      = {Proc. {ACM} Manag. Data},
	volume       = {2},
	number       = {3},
	pages        = {186},
	year         = {2024},
	url          = {https://doi.org/10.1145/3654989},
	doi          = {10.1145/3654989},
	bibsource    = {dblp computer science bibliography, https://dblp.org}
}

@inproceedings{DBLP:conf/icde/Zhao18,
	author       = {Bo Zhao},
	title        = {Complex Event Processing under Constrained Resources by State-Based
	Load Shedding},
	booktitle    = {34th {IEEE} International Conference on Data Engineering, {ICDE} 2018,
	Paris, France, April 16-19, 2018},
	pages        = {1699--1703},
	publisher    = {{IEEE} Computer Society},
	year         = {2018},
	url          = {https://doi.org/10.1109/ICDE.2018.00218},
	doi          = {10.1109/ICDE.2018.00218},
	bibsource    = {dblp computer science bibliography, https://dblp.org}
}

@inproceedings{he2013load,
  author       = {Yeye He and
                  Siddharth Barman and
                  Jeffrey F. Naughton},
  editor       = {Nicole Schweikardt and
                  Vassilis Christophides and
                  Vincent Leroy},
  title        = {On Load Shedding in Complex Event Processing},
  booktitle    = {Proc. 17th International Conference on Database Theory (ICDT), Athens,
                  Greece, March 24-28, 2014},
  pages        = {213--224},
  publisher    = {OpenProceedings.org},
  year         = {2014},
  url          = {https://doi.org/10.5441/002/icdt.2014.23},
  doi          = {10.5441/002/ICDT.2014.23},
  bibsource    = {dblp computer science bibliography, https://dblp.org}
}

@article{chapnik2021darling,
  title={DARLING: data-aware load shedding in complex event processing systems},
  author={Chapnik, Koral and Kolchinsky, Ilya and Schuster, Assaf},
  journal={Proceedings of the VLDB Endowment},
  volume={15},
  number={3},
  pages={541--554},
  year={2021},
  publisher={VLDB Endowment}
}

@inproceedings{kolchinsky2019real,
  title={Real-time multi-pattern detection over event streams},
  author={Kolchinsky, Ilya and Schuster, Assaf},
  booktitle={Proceedings of the 2019 International Conference on Management of Data},
  pages={589--606},
  year={2019}
}

@inproceedings{poppe2021share,
  title={To share, or not to share online event trend aggregation over bursty event streams},
  author={Poppe, Olga and Lei, Chuan and Ma, Lei and Rozet, Allison and Rundensteiner, Elke A},
  booktitle={Proceedings of the 2021 International Conference on Management of Data},
  pages={1452--1464},
  year={2021}
}

@inproceedings{ma2022gloria,
  title={Gloria: Graph-based Sharing Optimizer for Event Trend Aggregation},
  author={Ma, Lei and Lei, Chuan and Poppe, Olga and Rundensteiner, Elke A},
  booktitle={Proceedings of the 2022 International Conference on Management of Data},
  pages={1122--1135},
  year={2022}
}

@misc{bike,
  author       = {},
  title        = {Citi Bike},
  howpublished = {\url{http://www.citibikenyc.com/system-data}},
  year         = {2024},
  note         = {}
}

@inproceedings{nguyen-etal-2024-direct,
    title = "Direct Evaluation of Chain-of-Thought in Multi-hop Reasoning with Knowledge Graphs",
    author = "Nguyen, Thi  and
      Luo, Linhao  and
      Shiri, Fatemeh  and
      Phung, Dinh  and
      Li, Yuan-Fang  and
      Vu, Thuy-Trang  and
      Haffari, Gholamreza",
    editor = "Ku, Lun-Wei  and
      Martins, Andre  and
      Srikumar, Vivek",
    booktitle = "Findings of the Association for Computational Linguistics: ACL 2024",
    month = aug,
    year = "2024",
    address = "Bangkok, Thailand",
    publisher = "Association for Computational Linguistics",
    url = "https://aclanthology.org/2024.findings-acl.168",
    doi = "10.18653/v1/2024.findings-acl.168",
    pages = "2862--2883",
}

@article{Zhang_Dai_Kozareva_Smola_Song_2018, title={Variational Reasoning for Question Answering With Knowledge Graph}, volume={32}, url={https://ojs.aaai.org/index.php/AAAI/article/view/12057}, DOI={10.1609/aaai.v32i1.12057}, abstractNote={ &lt;p&gt; Knowledge graph (KG) is known to be helpful for the task of question answering (QA), since it provides well-structured relational information between entities, and allows one to further infer indirect facts. However, it is challenging to build QA systems which can learn to reason over knowledge graphs based on question-answer pairs alone. First, when people ask questions, their expressions are noisy (for example, typos in texts, or variations in pronunciations), which is non-trivial for the QA system to match those mentioned entities to the knowledge graph. Second, many questions require multi-hop logic reasoning over the knowledge graph to retrieve the answers. To address these challenges, we propose a novel and unified deep learning architecture, and an end-to-end variational learning algorithm which can handle noise in questions, and learn multi-hop reasoning simultaneously. Our method achieves state-of-the-art performance on a recent benchmark dataset in the literature. We also derive a series of new benchmark datasets, including questions for multi-hop reasoning, questions paraphrased by neural translation model, and questions in human voice. Our method yields very promising results on all these challenging datasets. &lt;/p&gt; }, number={1}, journal={Proceedings of the AAAI Conference on Artificial Intelligence}, author={Zhang, Yuyu and Dai, Hanjun and Kozareva, Zornitsa and Smola, Alexander and Song, Le}, year={2018}, month={Apr.} }

@inproceedings{yih-etal-2016-value,
    title = "The Value of Semantic Parse Labeling for Knowledge Base Question Answering",
    author = "Yih, Wen-tau  and
      Richardson, Matthew  and
      Meek, Chris  and
      Chang, Ming-Wei  and
      Suh, Jina",
    editor = "Erk, Katrin  and
      Smith, Noah A.",
    booktitle = "Proceedings of the 54th Annual Meeting of the Association for Computational Linguistics (Volume 2: Short Papers)",
    month = aug,
    year = "2016",
    address = "Berlin, Germany",
    publisher = "Association for Computational Linguistics",
    url = "https://aclanthology.org/P16-2033",
    doi = "10.18653/v1/P16-2033",
    pages = "201--206",
}

@inproceedings{abul2017multiple,
  title={Multiple-query optimization of regular path queries},
  author={Abul-Basher, Zahid},
  booktitle={2017 IEEE 33rd International Conference on Data Engineering (ICDE)},
  pages={1426--1430},
  year={2017},
  organization={IEEE}
}

@inproceedings{ray2016scalable,
  title={Scalable pattern sharing on event streams},
  author={Ray, Medhabi and Lei, Chuan and Rundensteiner, Elke A},
  booktitle={Proceedings of the 2016 international conference on management of data},
  pages={495--510},
  year={2016}
}

@inproceedings{zhang2017multi,
  title={Multi-query optimization for complex event processing in SAP ESP},
  author={Zhang, Shuhao and Vo, Hoang Tam and Dahlmeier, Daniel and He, Bingsheng},
  booktitle={2017 IEEE 33rd International Conference on Data Engineering (ICDE)},
  pages={1213--1224},
  year={2017},
  organization={IEEE}
}

@misc{chicago_data,
  title        = {Crimes - 2001 to Present},
  author       = {City of Chicago},
  howpublished = {\url{https://data.cityofchicago.org/Public-Safety/Crimes-2001-to-Present/ijzp-q8t2}},
  year         = {2024}
}

@techreport{iso19075-5,
  author       = {{ISO/IEC JTC 1/SC 32 Data management and interchange}},
  title        = {{ISO/IEC TR 19075-5:2016 Information technology -- Database languages -- SQL Technical Reports -- Part 5: Row Pattern Recognition in SQL}},
  year         = {2016},
  institution  = {International Organization for Standardization (ISO)},
  url          = {https://www.iso.org/standard/65143.html},
}

@online{findeisen2021,
  author       = {Findeisen, Kasia},
  title        = {Row pattern recognition with MATCH\_RECOGNIZE},
  year         = {2021},
  url          = {https://trino.io/blog/2021/05/19/row_pattern_matching.html},
}

@online{laker2017,
  author       = {Laker, Keith},
  title        = {MATCH\_RECOGNIZE and predicates - everything you need to know},
  year         = {2017},
  url          = {https://blogs.oracle.com/datawarehousing/post/match_recognize-and-predicates-everything-you-need-to-know},
}

@online{paes2019,
  author       = {Paes, Marta},
  title        = {MATCH\_RECOGNIZE: where Flink SQL and Complex Event Processing meet},
  year         = {2019},
  url          = {https://www.ververica.com/blog/match_recognize-where-flink-sql-and-complex-event-processing-meet},
}

@online{alves2019,
  author       = {Alves, Rodrigo},
  title        = {Azure Stream Analytics now supports MATCH\_RECOGNIZE},
  year         = {2019},
  url          = {https://azure.microsoft.com/en-us/blog/azure-stream-analytics-now-supports-match-recognize/},
}

@online{snowflake2021,
  author       = {{Snowflake}},
  title        = {Identifying Sequences of Rows That Match a Pattern},
  year         = {2021},
  url          = {https://docs.snowflake.com/en/user-guide/match-recognize-introduction.html},
}

@inproceedings{wu2006high,
  title={High-performance complex event processing over streams},
  author={Wu, Eugene and Diao, Yanlei and Rizvi, Shariq},
  booktitle={Proceedings of the 2006 ACM SIGMOD international conference on Management of data},
  pages={407--418},
  year={2006}
}

@inproceedings{mei2009zstream,
  title={Zstream: a cost-based query processor for adaptively detecting composite events},
  author={Mei, Yuan and Madden, Samuel},
  booktitle={Proceedings of the 2009 ACM SIGMOD International Conference on Management of data},
  pages={193--206},
  year={2009}
}

@inproceedings{poppe2018sharon,
  title={Sharon: Shared online event sequence aggregation},
  author={Poppe, Olga and Rozet, Allison and Lei, Chuan and Rundensteiner, Elke A and Maier, David},
  booktitle={2018 IEEE 34th International Conference on Data Engineering (ICDE)},
  pages={737--748},
  year={2018},
  organization={IEEE}
}

@article{GRETA2017,
author = {Poppe, Olga and Lei, Chuan and Rundensteiner, Elke A. and Maier, David},
title = {GRETA: graph-based real-time event trend aggregation},
year = {2017},
issue_date = {September 2017},
publisher = {VLDB Endowment},
volume = {11},
number = {1},
issn = {2150-8097},
url = {https://doi.org/10.14778/3151113.3151120},
doi = {10.14778/3151113.3151120},
journal = {Proc. VLDB Endow.},
month = sep,
pages = {80–92},
numpages = {13}
}

@inproceedings{eSPICE,
author = {Slo, Ahmad and Bhowmik, Sukanya and Rothermel, Kurt},
title = {eSPICE: Probabilistic Load Shedding from Input Event Streams in Complex Event Processing},
year = {2019},
isbn = {9781450370097},
publisher = {Association for Computing Machinery},
address = {New York, NY, USA},
url = {https://doi.org/10.1145/3361525.3361548},
doi = {10.1145/3361525.3361548},
booktitle = {Proceedings of the 20th International Middleware Conference},
pages = {215–227},
numpages = {13},
location = {Davis, CA, USA},
series = {Middleware '19}
}

@inproceedings{hSPICE,
author = {Slo, Ahmad and Bhowmik, Sukanya and Rothermel, Kurt},
title = {hSPICE: state-aware event shedding in complex event processing},
year = {2020},
isbn = {9781450380287},
publisher = {Association for Computing Machinery},
address = {New York, NY, USA},
url = {https://doi.org/10.1145/3401025.3401742},
doi = {10.1145/3401025.3401742},
booktitle = {Proceedings of the 14th ACM International Conference on Distributed and Event-Based Systems},
pages = {109–120},
numpages = {12},
location = {Montreal, Quebec, Canada},
series = {DEBS '20}
}

@article{DBLP:journals/pvldb/GeorgeCW16,
	author       = {Lars George and
	Bruno Cadonna and
	Matthias Weidlich},
	title        = {IL-Miner: Instance-Level Discovery of Complex Event Patterns},
	journal      = {Proc. {VLDB} Endow.},
	volume       = {10},
	number       = {1},
	pages        = {25--36},
	year         = {2016},
	url          = {http://www.vldb.org/pvldb/vol10/p25-weidlich.pdf},
	doi          = {10.14778/3015270.3015273},
	bibsource    = {dblp computer science bibliography, https://dblp.org}
}

@inproceedings{dindar2011efficiently,
  title={Efficiently correlating complex events over live and archived data streams},
  author={Dindar, Nihal and Fischer, Peter M and Soner, Merve and Tatbul, Nesime},
  booktitle={Proceedings of the 5th ACM international conference on Distributed event-based system},
  pages={243--254},
  year={2011}
}

@inproceedings{korber2021index,
  title={Index-accelerated pattern matching in event stores},
  author={K{\"o}rber, Michael and Glombiewski, Nikolaus and Seeger, Bernhard},
  booktitle={Proceedings of the 2021 International Conference on Management of Data},
  pages={1023--1036},
  year={2021}
}

@article{zhu2023high,
  title={High-performance row pattern recognition using joins},
  author={Zhu, Erkang and Huang, Silu and Chaudhuri, Surajit},
  journal={Proceedings of the VLDB Endowment},
  volume={16},
  number={5},
  pages={1181--1195},
  year={2023},
  publisher={VLDB Endowment}
}

@inproceedings{WadhwaPRBB19,
  author       = {Sarisht Wadhwa and
                  Anagh Prasad and
                  Sayan Ranu and
                  Amitabha Bagchi and
                  Srikanta Bedathur},
  editor       = {Peter A. Boncz and
                  Stefan Manegold and
                  Anastasia Ailamaki and
                  Amol Deshpande and
                  Tim Kraska},
  title        = {Efficiently Answering Regular Simple Path Queries on Large Labeled
                  Networks},
  booktitle    = {Proceedings of the 2019 International Conference on Management of
                  Data, {SIGMOD} Conference 2019, Amsterdam, The Netherlands, June 30
                  - July 5, 2019},
  pages        = {1463--1480},
  publisher    = {{ACM}},
  year         = {2019},
  url          = {https://doi.org/10.1145/3299869.3319882},
  doi          = {10.1145/3299869.3319882},
  bibsource    = {dblp computer science bibliography, https://dblp.org}
}

@misc{AWSGraphRAG2024,
  author       = {AWS Machine Learning Blog},
  title        = {Improving retrieval-augmented generation accuracy with GraphRAG},
  year         = {2024},
  url          = {https://aws.amazon.com/blogs/machine-learning/improving-retrieval-augmented-generation-accuracy-with-graphrag/},
}

@misc{MicrosoftGraphRAG,
  author       = {Microsoft},
  title        = {GraphRAG - Microsoft Research},
  year         = {2024},
  url          = {https://microsoft.github.io/graphrag/},
}

@misc{Neo4jGraphRAG2024,
  author       = {Neo4j Developer Blog},
  title        = {Knowledge Graph RAG Application},
  year         = {2024},
  url          = {https://neo4j.com/developer-blog/knowledge-graph-rag-application/},
}

@misc{SiemensKnowledgeGraph,
  author       = {Siemens},
  title        = {Artificial Intelligence: Industrial Knowledge Graph},
  year         = {2024},
  url          = {https://www.siemens.com/global/en/company/stories/research-technologies/artificial-intelligence/artificial-intelligence-industrial-knowledge-graph.html},
}

@misc{MicrosoftPrivateData2024,
  author       = {Microsoft Research},
  title        = {GraphRAG: Unlocking LLM Discovery on Narrative Private Data},
  year         = {2024},
  url          = {https://www.microsoft.com/en-us/research/blog/graphrag-unlocking-llm-discovery-on-narrative-private-data/},
}

@misc{Neo4jAutomaker2024,
  author       = {Neo4j},
  title        = {Global Automaker},
  year         = {2024},
  url          = {https://neo4j.com/customer-stories/global-automaker/},
}

@article{lust2012multiobjective,
  title={The multiobjective multidimensional knapsack problem: a survey and a new approach},
  author={Lust, Thibaut and Teghem, Jacques},
  journal={International Transactions in Operational Research},
  volume={19},
  number={4},
  pages={495--520},
  year={2012},
  publisher={Wiley Online Library}
}

@article{da2008core,
  title={Core problems in bi-criteria $\{$0, 1$\}$-knapsack problems},
  author={Da Silva, Carlos Gomes and Cl{\'\i}maco, Jo{\~a}o and Figueira, Jos{\'e} Rui},
  journal={Computers \& Operations Research},
  volume={35},
  number={7},
  pages={2292--2306},
  year={2008},
  publisher={Elsevier}
}

@article{cugola2012processing,
  title={Processing flows of information: From data stream to complex event processing},
  author={Cugola, Gianpaolo and Margara, Alessandro},
  journal={ACM Computing Surveys (CSUR)},
  volume={44},
  number={3},
  pages={1--62},
  year={2012},
  publisher={ACM New York, NY, USA}
}

@article{10.1145/3654935,
author = {Akili, Samira and Purtzel, Steven and Weidlich, Matthias},
title = {DecoPa: Query Decomposition for Parallel Complex Event Processing},
year = {2024},
issue_date = {June 2024},
publisher = {Association for Computing Machinery},
address = {New York, NY, USA},
volume = {2},
number = {3},
url = {https://doi.org/10.1145/3654935},
doi = {10.1145/3654935},
journal = {Proc. ACM Manag. Data},
month = may,
articleno = {132},
numpages = {26}
}

@article{10.1145/3709682,
author = {Sattler, Rebecca and Kleest-Mei\ss{}ner, Sarah and Lange, Steven and Schmid, Markus L. and Schweikardt, Nicole and Weidlich, Matthias},
title = {DISCES: Systematic Discovery of Event Stream Queries},
year = {2025},
issue_date = {February 2025},
publisher = {Association for Computing Machinery},
address = {New York, NY, USA},
volume = {3},
number = {1},
url = {https://doi.org/10.1145/3709682},
doi = {10.1145/3709682},
journal = {Proc. ACM Manag. Data},
month = feb,
articleno = {32},
numpages = {26}
}

@inproceedings{abiteboul1997regular,
  title={Regular path queries with constraints},
  author={Abiteboul, Serge and Vianu, Victor},
  booktitle={Proceedings of the sixteenth ACM SIGACT-SIGMOD-SIGART symposium on Principles of database systems},
  pages={122--133},
  year={1997}
}

@article{angles2017foundations,
  title={Foundations of modern query languages for graph databases},
  author={Angles, Renzo and Arenas, Marcelo and Barcel{\'o}, Pablo and Hogan, Aidan and Reutter, Juan and Vrgo{\v{c}}, Domagoj},
  journal={ACM Computing Surveys (CSUR)},
  volume={50},
  number={5},
  pages={1--40},
  year={2017},
  publisher={ACM New York, NY, USA}
}

@article{liu2020resource,
  title={Resource management and scheduling in distributed stream processing systems: a taxonomy, review, and future directions},
  author={Liu, Xunyun and Buyya, Rajkumar},
  journal={ACM Computing Surveys (CSUR)},
  volume={53},
  number={3},
  pages={1--41},
  year={2020},
  publisher={ACM New York, NY, USA}
}

@inproceedings{zhang2014complexity,
  title={On complexity and optimization of expensive queries in complex event processing},
  author={Zhang, Haopeng and Diao, Yanlei and Immerman, Neil},
  booktitle={Proceedings of the 2014 ACM SIGMOD international conference on Management of data},
  pages={217--228},
  year={2014}
}

@inproceedings{tatbul2003load,
  title={Load shedding in a data stream manager},
  author={Tatbul, Nesime and {\c{C}}etintemel, U{\u{g}}ur and Zdonik, Stan and Cherniack, Mitch and Stonebraker, Michael},
  booktitle={Proceedings 2003 vldb conference},
  pages={309--320},
  year={2003},
  organization={Elsevier}
}

@inproceedings{kim-etal-2023-kg,
    title = "{KG}-{GPT}: A General Framework for Reasoning on Knowledge Graphs Using Large Language Models",
    author = "Kim, Jiho  and
      Kwon, Yeonsu  and
      Jo, Yohan  and
      Choi, Edward",
    editor = "Bouamor, Houda  and
      Pino, Juan  and
      Bali, Kalika",
    booktitle = "Findings of the Association for Computational Linguistics: EMNLP 2023",
    month = dec,
    year = "2023",
    address = "Singapore",
    publisher = "Association for Computational Linguistics",
    url = "https://aclanthology.org/2023.findings-emnlp.631/",
    doi = "10.18653/v1/2023.findings-emnlp.631",
    pages = "9410--9421"
}

@inproceedings{DBLP:conf/iclr/LuoLHP24,
  author       = {Linhao Luo and
                  Yuan{-}Fang Li and
                  Gholamreza Haffari and
                  Shirui Pan},
  title        = {Reasoning on Graphs: Faithful and Interpretable Large Language Model
                  Reasoning},
  booktitle    = {The Twelfth International Conference on Learning Representations,
                  {ICLR} 2024, Vienna, Austria, May 7-11, 2024},
  publisher    = {OpenReview.net},
  year         = {2024},
  url          = {https://openreview.net/forum?id=ZGNWW7xZ6Q},
  bibsource    = {dblp computer science bibliography, https://dblp.org}
}

@inproceedings{abul2016swarmguide,
  title={SwarmGuide: Towards Multiple-Query Optimization in Graph Databases.},
  author={Abul-Basher, Zahid and Yakovets, Nikolay and Godfrey, Parke and Chignell, Mark H},
  booktitle={AMW},
  year={2016}
}

@misc{tan2025pathsovergraph,
      title={Paths-over-Graph: Knowledge Graph Empowered Large Language Model Reasoning}, 
      author={Xingyu Tan and Xiaoyang Wang and Qing Liu and Xiwei Xu and Xin Yuan and Wenjie Zhang},
      year={2025},
      eprint={2410.14211},
      archivePrefix={arXiv},
      primaryClass={cs.CL}
}

@article{DBLP:journals/pvldb/McSherryLSR20,
  author       = {Frank McSherry and
                  Andrea Lattuada and
                  Malte Schwarzkopf and
                  Timothy Roscoe},
  title        = {Shared Arrangements: practical inter-query sharing for streaming dataflows},
  journal      = {Proc. {VLDB} Endow.},
  volume       = {13},
  number       = {10},
  pages        = {1793--1806},
  year         = {2020},
  url          = {http://www.vldb.org/pvldb/vol13/p1793-mcsherry.pdf},
  doi          = {10.14778/3401960.3401974},
  bibsource    = {dblp computer science bibliography, https://dblp.org}
}

@inproceedings{DBLP:conf/iclr/SunXTW0GNSG24,
  author       = {Jiashuo Sun and
                  Chengjin Xu and
                  Lumingyuan Tang and
                  Saizhuo Wang and
                  Chen Lin and
                  Yeyun Gong and
                  Lionel M. Ni and
                  Heung{-}Yeung Shum and
                  Jian Guo},
  title        = {Think-on-Graph: Deep and Responsible Reasoning of Large Language Model
                  on Knowledge Graph},
  booktitle    = {The Twelfth International Conference on Learning Representations,
                  {ICLR} 2024, Vienna, Austria, May 7-11, 2024},
  publisher    = {OpenReview.net},
  year         = {2024},
  url          = {https://openreview.net/forum?id=nnVO1PvbTv},
  bibsource    = {dblp computer science bibliography, https://dblp.org}
}

@inproceedings{slo2023gspice,
  title={gspice: Model-based event shedding in complex event processing},
  author={Slo, Ahmad and Bhowmik, Sukanya and Rothermel, Kurt},
  booktitle={2023 IEEE International Conference on Big Data (BigData)},
  pages={263--270},
  year={2023},
  organization={IEEE}
}

@article{huang2023t,
  title={T-rex: Optimizing pattern search on time series},
  author={Huang, Silu and Zhu, Erkang and Chaudhuri, Surajit and Spiegelberg, Leonhard},
  journal={Proceedings of the ACM on Management of Data},
  volume={1},
  number={2},
  pages={1--26},
  year={2023},
  publisher={ACM New York, NY, USA}
}

@article{mendelzon1995finding,
  title={Finding regular simple paths in graph databases},
  author={Mendelzon, Alberto O and Wood, Peter T},
  journal={SIAM Journal on Computing},
  volume={24},
  number={6},
  pages={1235--1258},
  year={1995},
  publisher={SIAM}
}

@inproceedings{libkin2012regular,
  title={Regular path queries on graphs with data},
  author={Libkin, Leonid and Vrgo{\v{c}}, Domagoj},
  booktitle={Proceedings of the 15th International Conference on Database Theory},
  pages={74--85},
  year={2012}
}

@inproceedings{pacaci2020regular,
  title={Regular path query evaluation on streaming graphs},
  author={Pacaci, Anil and Bonifati, Angela and {\"O}zsu, M Tamer},
  booktitle={Proceedings of the 2020 ACM SIGMOD International Conference on Management of Data},
  pages={1415--1430},
  year={2020}
}

@inproceedings{slo2019pspice,
  title={pspice: Partial match shedding for complex event processing},
  author={Slo, Ahmad and Bhowmik, Sukanya and Flaig, Albert and Rothermel, Kurt},
  booktitle={2019 IEEE International Conference on Big Data (Big Data)},
  pages={372--382},
  year={2019},
  organization={IEEE}
}

@misc{lumi,
  author = {{LUMI Consortium}},
  title = {LUMI Supercomputer},
  howpublished = {\url{https://lumi-supercomputer.eu/}},
}

@misc{triton,
  author        = {{Aalto University, SciComp}},
  title         = {Triton HPC},
  howpublished  = {\url{https://scicomp.aalto.fi/triton/}},
}

@inproceedings{10.1007/978-3-030-27520-4_7,
author = {Nasu, Yuya and Kitagawa, Hiroyuki and Nakabasami, Kosuke},
title = {Efficient Row Pattern Matching Using Pattern Hierarchies for Sequence OLAP},
year = {2019},
isbn = {978-3-030-27519-8},
publisher = {Springer-Verlag},
address = {Berlin, Heidelberg},
url = {https://doi.org/10.1007/978-3-030-27520-4_7},
doi = {10.1007/978-3-030-27520-4_7},
booktitle = {Big Data Analytics and Knowledge Discovery: 21st International Conference, DaWaK 2019, Linz, Austria, August 26–29, 2019, Proceedings},
pages = {89–104},
numpages = {16},
location = {Linz, Austria}
}

@article{chakravarthy1994snoop,
  title={Snoop: An expressive event specification language for active databases},
  author={Chakravarthy, Sharma and Mishra, Deepak},
  journal={Data \& Knowledge Engineering},
  volume={14},
  number={1},
  pages={1--26},
  year={1994},
  publisher={Elsevier}
}

@inproceedings{zimmer1999semantics,
  title={On the semantics of complex events in active database management systems},
  author={Zimmer, Detlef and Unland, Rainer},
  booktitle={Proceedings 15th International Conference on Data Engineering (Cat. No. 99CB36337)},
  pages={392--399},
  year={1999},
  organization={IEEE}
}

@inproceedings{cugola2010tesla,
  title={TESLA: a formally defined event specification language},
  author={Cugola, Gianpaolo and Margara, Alessandro},
  booktitle={Proceedings of the Fourth ACM International Conference on Distributed Event-Based Systems},
  pages={50--61},
  year={2010}
}

@article{heap,
author = {Williams, J.W.J.},
title = {Algorithm 232: Heapsort},
year = {2025},
issue_date = {June 1964},
publisher = {Association for Computing Machinery},
address = {New York, NY, USA},
volume = {7},
number = {6},
issn = {0001-0782},
url = {https://doi.org/10.1145/512274.3734138},
doi = {10.1145/512274.3734138},
journal = {Commun. ACM},
month = may,
pages = {347–348},
numpages = {2}
}

@misc{neo4j-graphRAG-python,
  title        = {neo4j/neo4j-graphrag-python: Neo4j GraphRAG for Python},
  author       = {{Neo4j}},
  howpublished = {\url{https://github.com/neo4j/neo4j-graphrag-python}},
}

@misc{neo4j_apoc,
  author       = {{Neo4j Labs}},
  title        = {APOC: Awesome Procedures On Cypher},
  year         = {2025},
  howpublished = {\url{https://neo4j.com/labs/apoc/}},
  note         = {Accessed: 2025-10-17}
}

@misc{slo-example,
  author       = {Saif Gunja},
  title        = {SLO examples for faster, more reliable apps},
  year         = {2023},
  howpublished = {\url{https://www.dynatrace.com/news/blog/service-level-objective-examples-5-slo-examples/}}
}

@misc{pragmaticsre-slo-examples,
  author       = {Pragmatics RE},
  title        = {SLO Examples},
  year         = {2024},
  howpublished = {\url{https://www.pragmaticsre.com/psre-guide/conclusion/slo-examples}},
}

@article{sajjanar2025real,
  title={Real-Time, Low-Latency Surveillance Using Entropy-Based Adaptive Buffering and MobileNetV2 on Edge Devices},
  author={Sajjanar, Poojashree Chandrashekar Pankaj M},
  journal={arXiv preprint arXiv:2506.14833},
  year={2025}
}

@inproceedings{mishra2024abnormal,
  title={Abnormal Human Behaviour Detection by Surveillance Camera},
  author={Mishra, Manas Ranjan and Meher, Pramod Kumar},
  booktitle={2024 Asia Pacific Conference on Innovation in Technology (APCIT)},
  pages={1--6},
  year={2024},
  organization={IEEE}
}

@misc{google-slo-guide,
  author       = {Google SRE},
  title        = {Implementing SLOs},
  howpublished = {\url{https://sre.google/workbook/implementing-slos/}},
}

@misc{qwen3_8b,
  title        = {Qwen3-8B},
  author       = {Qwen Team},
  year         = {2025},
  howpublished = {\url{https://huggingface.co/Qwen/Qwen3-8B}},
  note         = {Hugging Face repository}
}

@software{cpsatlp,
  title        = {CP-SAT},
  version      = {v9.12},
  author       = {Laurent Perron and Frédéric Didier},
  organization = {Google},
  url          = {https://developers.google.com/optimization/cp/cp_solver/},
  date         = {2025-02-17}
}

@misc{yin2024graphrag,
      title={From Local to Global: A Graph RAG Approach to Query-Focused Summarization}, 
      author={Darren Edge and Ha Trinh and Newman Cheng and Joshua Bradley and Alex Chao and Apurva Mody and Steven Truitt and Dasha Metropolitansky and Robert Osazuwa Ness and Jonathan Larson},
      year={2025},
      eprint={2404.16130},
      archivePrefix={arXiv},
      primaryClass={cs.CL},
      url={https://arxiv.org/abs/2404.16130}, 
}

@misc{sun2024surveygraphrag,
      title={Graph Retrieval-Augmented Generation: A Survey}, 
      author={Boci Peng and Yun Zhu and Yongchao Liu and Xiaohe Bo and Haizhou Shi and Chuntao Hong and Yan Zhang and Siliang Tang},
      year={2024},
      eprint={2408.08921},
      archivePrefix={arXiv},
      primaryClass={cs.AI},
      url={https://arxiv.org/abs/2408.08921}, 
}

@misc{wang2023knowledgegpt,
      title={KnowledGPT: Enhancing Large Language Models with Retrieval and Storage Access on Knowledge Bases}, 
      author={Xintao Wang and Qianwen Yang and Yongting Qiu and Jiaqing Liang and Qianyu He and Zhouhong Gu and Yanghua Xiao and Wei Wang},
      year={2023},
      eprint={2308.11761},
      archivePrefix={arXiv},
      primaryClass={cs.CL},
      url={https://arxiv.org/abs/2308.11761}, 
}

@misc{yang2024hybridrag,
      title={HybridRAG: Integrating Knowledge Graphs and Vector Retrieval Augmented Generation for Efficient Information Extraction}, 
      author={Bhaskarjit Sarmah and Benika Hall and Rohan Rao and Sunil Patel and Stefano Pasquali and Dhagash Mehta},
      year={2024},
      eprint={2408.04948},
      archivePrefix={arXiv},
      primaryClass={cs.CL},
      url={https://arxiv.org/abs/2408.04948}, 
}

@misc{liu2024graphrag,
      title={From Local to Global: A Graph RAG Approach to Query-Focused Summarization}, 
      author={Darren Edge and Ha Trinh and Newman Cheng and Joshua Bradley and Alex Chao and Apurva Mody and Steven Truitt and Dasha Metropolitansky and Robert Osazuwa Ness and Jonathan Larson},
      year={2025},
      eprint={2404.16130},
      archivePrefix={arXiv},
      primaryClass={cs.CL},
      url={https://arxiv.org/abs/2404.16130}, 
}

@software{neo4j,
  author       = {{Neo4j\, Inc.}},
  title        = {Neo4j Graph Database},
  version      = {5.26.0},       
  year         = {2025},          
  publisher    = {Neo4j\, Inc.},
  address      = {San Mateo, CA},
  note         = {ACID‐compliant, disk‐based native graph database:contentReference[oaicite:4]{index=4}},
  url          = {https://neo4j.com}
}

@software{memgraph,
  author       = {{Memgraph\, Ltd.}},
  title        = {Memgraph: In‑Memory Graph Database},
  version      = {3.0},           
  year         = {2025},          
  publisher    = {Memgraph\, Ltd.},
  address      = {Zagreb, Croatia},
  note         = {Open‑source, ACID‐compliant, in‑memory graph database written in C/C++:contentReference[oaicite:5]{index=5}:contentReference[oaicite:6]{index=6}},
  url          = {https://memgraph.com}
}



\end{document}